\documentclass[fleqn,usenatbib]{mnras}  
\usepackage{newtxtext,newtxmath}
\usepackage{xfrac}

\usepackage{ae,aecompl}

\usepackage{graphicx}	
\usepackage{amsmath}	
\usepackage[dvipsnames]{xcolor} 
\usepackage{ulem}
\usepackage{microtype}
\usepackage{bbold}
\usepackage{tikz}

\newcommand{\aref}[1]{\hyperref[#1]{Appendix~\ref{#1}}}

\newcommand{\be}{\begin{equation}}
\newcommand{\ee}{\end{equation}}
\newcommand{\zstart}{z_{\mathrm{start}}}

\newcommand{\fett}[1]{\boldsymbol{#1}}
\newcommand{\dd}{{\rm d}}

\newcommand{\ii}{{\rm{i}}}
\newcommand{\ini}{{\rm{ini}}}
\newcommand{\fb}{f_{\rm b}}
\newcommand{\fc}{f_{\rm c}}
\newcommand{\rb}{{\rm b}}
\newcommand{\rc}{{\rm c}}
\newcommand{\rM}{{\rm m}}
\newcommand{\bc}{{\rm bc}}
\newcommand{\nabx}{\nabla_{\!x}}
\newcommand{\nabq}{\fett{\nabla}_{\!q}}
\newcommand{\nab}{\fett{\nabla}}

\newcommand{\smalltext}[1]{\text{\scriptsize #1}}

\definecolor{lime}{HTML}{A6CE39}
\DeclareRobustCommand{\orcidicon}{
	\begin{tikzpicture}
	\draw[lime, fill=lime] (0,0) 
	circle [radius=0.14] 
	node[white] {{\fontfamily{qag}\selectfont \tiny ID}};
	\draw[white, fill=white] (-0.0625,0.095) 
	circle [radius=0.007];
	\end{tikzpicture}
	\hspace{-2mm}
}

\foreach \x in {A, ..., Z}{%
	\expandafter\xdef\csname orcid\x\endcsname{\noexpand\href{https://orcid.org/\csname orcidauthor\x\endcsname}{\noexpand\orcidicon}}
}

\title[Baryon + CDM initial conditions]{Higher-order initial conditions for mixed baryon-CDM simulations}

\author[O. Hahn, C. Rampf \& C. Uhlemann]{
Oliver Hahn$^{\,{\tiny\orcidA{}}\,\,\,\,\hyperlink{OCA}{1},\hyperlink{VA}{2},\hyperlink{VM}{3}}$\thanks{E-mail: oliver.hahn@univie.ac.at}, Cornelius Rampf$^{\,{\tiny\orcidB{}}\,\,\,\,\hyperlink{OCA}{1}}$, and Cora Uhlemann$^{\,{\tiny\orcidC{}}\,\,\,\,\hyperlink{NUC}{4}}$  
\\
\hypertarget{OCA}$^{1}$Universit\'e C\^ote d'Azur, Observatoire de la C\^ote d'Azur, CNRS, Laboratoire Lagrange, \\\quad\!Boulevard de l'Observatoire, CS 34229, 06304 Nice, France \\
\hypertarget{VA}$^{2}$Department of Astrophysics, University of Vienna, T\"urkenschanzstra{\ss}e 17, 1180 Vienna, Austria\\
\hypertarget{VM}$^{3}$Department of Mathematics, University of Vienna, Oskar-Morgenstern-Platz 1, 1090 Vienna, Austria \\
\hypertarget{NUC}$^{4}$School of Mathematics, Statistics and Physics, Herschel Building, Newcastle University,  \\\quad\!\!Newcastle upon Tyne, NE1 7RU, UK}

\date{Accepted XXX. Received YYY; in original form ZZZ}

\pubyear{2020}

\begin{document}
\label{firstpage}
\pagerange{\pageref{firstpage}--\pageref{lastpage}}
\maketitle

\begin{abstract}
We present a novel approach to generate higher-order initial conditions (ICs) for cosmological simulations that take into account the distinct evolution of baryons and dark matter. We focus on the numerical implementation and the validation of its performance, based on both collisionless $N$-body simulations and full hydrodynamic Eulerian and Lagrangian simulations. We improve in various ways over previous approaches that were limited to first-order Lagrangian perturbation theory (LPT). Specifically, we (1) generalize $n$th-order LPT to multi-fluid systems, allowing 2LPT or 3LPT ICs for two-fluid simulations, (2) employ a novel propagator perturbation theory to set up ICs for Eulerian codes that are fully consistent with 1LPT or 2LPT, (3) demonstrate that our ICs resolve previous problems of two-fluid simulations by using variations in particle masses that eliminate spurious deviations from expected perturbative results, (4) show that the improvements achieved by going to higher-order PT are comparable to those seen for single-fluid ICs, and (5) demonstrate the excellent (i.e., few per cent level) agreement between Eulerian and Lagrangian simulations, once high-quality initial conditions are used. The rigorous development of the underlying perturbation theory is presented in a companion paper. All presented algorithms are implemented in the {\sc Monofonic Music-2} package that we make publicly available.
\end{abstract}



\section{Introduction}
The physics of the cosmic microwave background \citep[e.g.][]{Hu:2002,Durrer:2008} implies that baryons do not trace the distribution of dark matter. Since baryons were tightly coupled to photons prior to recombination, they begin to collapse on sub-horizon scales much later than dark matter. Furthermore, sound waves excited prior to recombination decay away over some finite time and the resulting relative streaming motion  \citep{Tseliakhovich:2010}  leaves an imprint on the formation of the very first cosmic objects and their spatial distribution  \citep[cf.][and many later studies]{Dalal:2010,Greif:2011,Yoo:2011,Fialkov:2012}  on scales correlated with the baryon acoustic oscillation (BAO) feature of the power spectrum, which is one of the most sensitive cosmological measures.  Such correlations, if they carry over to the galaxy populations in lower-redshift surveys, are therefore potentially significant biases in BAO measurements of cosmological parameters \citep[e.g.][]{Slepian:2015,Ahn:2016,Blazek:2016,2018MNRAS.474.2109S,Chen:2019}. 

While the difference in the clustering of baryons and CDM -- the baryon bias -- is small on large scales today \citep[e.g.][]{Angulo:2013}, this difference is more significant at earlier times. These epochs are increasingly within reach of ever more sensitive observations \citep[e.g. the Square Kilometre Array, SKA,][]{Bull:2018lat}. At late times and on small scales, the finite temperature of baryons, along with energy injection from supermassive black holes leads to a de-correlation of baryons and CDM in the mature low-redshift Universe \citep[e.g.][]{Chisari:2019}, complicating  access to cosmological information on those scales.  While the cosmological baryon bias is a robust prediction of our cosmological model, the latest generation of cosmological galaxy formation simulations in large-scale structure context \citep[e.g.][for the Horizon-AGN, Eagle, Illustris-TNG, and Borg-Cube simulations]{Dubois:2014,Schaye:2015,Springel:2018,Emberson:2019} still do not model it due to multiple reasons that we discuss below. At the same time, the precision determination of the matter power spectrum has now reached the level at which baryonic effects should be incorporated in predictions from collisionless `total matter' $N$-body simulations \citep[cf.][]{Schneider:2015,Huang:2019,Schneider:2019,Arico:2020}.

Accurate numerical studies of the rich dynamics of the two-fluid system of collisionless dark matter and collisional baryons from the cosmological perspective have  been limited by a range of problems. This includes the generation of simulation initial conditions (ICs) based on perturbation theory (PT). For single-fluid simulations, the Zel'dovich approximation \citep[ZA,][]{Zeldovich:1970} has been used since the early days of cosmological $N$-body methods to set up ICs \citep{Klypin:1983,Efstathiou:1985}, and it is by now standard to employ second-order Lagrangian perturbation theory (2LPT), e.g. \cite{Crocce:2006}. However, to date, higher-order LPT has not been developed for multi-component systems of baryons and dark matter. Some findings have been obtained in the context of Eulerian PT \citep{Somogyi:2010, Bernardeau:2012} but have had no impact on increasing the accuracy of simulations so far. At the same time, simulations studying the baryon streaming predicted by \cite{Tseliakhovich:2010} impose an {\it ad hoc} relative velocity between baryons and dark matter but do not self-consistently account for the non-linear coupling of such a relative velocity in the fluid PT used to set up the simulation ICs. Such relative motion appears as a decaying mode, not sourced by gravity, and is particularly difficult to tackle in the setting of a rigorous PT for simulation ICs. 

$N$-body simulation ICs typically start by considering an initially (statistically) uniform discrete universe of particles, onto which the cosmological perturbations are imprinted, namely by perturbing the particle positions and velocities. The problem of decaying-mode initial conditions is that they are fundamentally inconsistent with this boundary condition (in the sense that for earlier times, one would approach a more inhomogeneous state).
 In full generality, the dark matter and baryon perturbations computed by Einstein--Boltzmann codes of course contain a multitude of effects that are not captured by standard LPT (due to the absence of physical effects beyond the Newtonian two-fluid model). 

For this reason, simulations attempting to take into account more realistic two-fluid perturbations so far employ only a first-order accurate approach, which simply guarantees that density and velocity power spectra imposed on the $N$-body particles have amplitudes that are in accordance with the linear Einstein--Boltzmann system \citep[e.g.,][for technical details]{Yoshida:2003,Hahn:2011}. It has quickly been noted however that simulations of two fluids (i.e., dark matter and baryons) initialized in this way do not accurately reproduce the relative growth between baryons and CDM even on scales where linear PT should apply \citep{OLeary:2012,Angulo:2013}, unless a much larger force softening is applied to the baryon particles than would be typical in usual $N$-body simulations. This finding has been confirmed in a later analysis also by \cite{Valkenburg:2017}. Very recently, \cite{Bird:2020} have claimed that the large softening can be circumvented by arranging baryon and dark matter $N$-body particles in a more refined way than just on two shifted lattices (on which we will comment later on). In any case, the current state-of-the-art view is that two-fluid $N$-body simulations require a more careful suppression of discreteness effects than single fluid `total matter' $N$-body simulations. 

In this paper, we show how to numerically implement high-order ICs for two-fluid cosmological simulations while minimizing discretization errors. Among other things, we discuss how Lagrangian PT can be used to generate growing-mode initial conditions for multiple cold fluids. Such ICs preclude decaying modes, and therefore any relative velocities between the fluids. While we also provide a numerical procedure that includes the linear effects of relative velocities in LPT ICs, it is currently unclear how such decaying modes can be implemented to achieve consistent higher-order ICs. We defer this aspect to future work.

The growing-mode approach that we focus on in the present paper allows for an initially prescribed scale-dependent baryon density bias which, of course, changes significantly during the non-linear evolution. The essential idea of the growing-mode approach is that, to leading order, the local baryon and CDM fractions are constant in time, and therefore can be absorbed into variations of the masses of Lagrangian fluid elements (these variations are fairly small and constant in time). This simple trick guarantees that the particle realizations of the baryon and CDM fractions are locally compensated to high precision, meaning that the individual density fractions change without changing the total matter density. This solution thus strongly improves over previous simulations of this kind that were plagued by discreteness errors even on large scales. Furthermore, based on this approach, it is possible to essentially apply standard $n$-th order LPT results to generate high-order ICs for two-fluid simulations.

Cosmological hydrodynamical simulations usually come in two broad limits: Lagrangian methods, such as smooth particle hydrodynamics or moving mesh techniques -- e.g. the widely used {\sc Gadget-2/3} \citep{Springel:2005}, {\sc Gasoline} \citep{Wadsley:2017}, {\sc Arepo} \citep{Springel:2010,Weinberger:2020}, {\sc Gizmo} \citep{Hopkins:2015}, and {\sc Swift} \citep{Schaller:2016} codes -- and Eulerian methods which use a spatially fixed mesh that can be dynamically refined -- e.g. the widely used {\sc ART} \citep{Kravtsov:1997}, {\sc Ramses} \citep{Teyssier:2002}, {\sc Enzo} \citep{Bryan:2014}, or {\sc Nyx} \citep{Almgren:2013} codes (but note that a moving mesh can be usually used in both Eulerian or quasi-Lagrangian mode). While the mentioned problems of two-fluid cosmological simulations apply to some degree to all Lagrangian $N$-body, SPH, or ``moving mesh'' simulations, the situation for Eulerian ``fixed mesh'' simulations is arguably even more dire. To achieve comparable accuracy in Eulerian PT, one has to go to significantly higher order than in LPT. Obtaining Eulerian ICs by using LPT-evolved fields in combination with local Lagrangian approximation schemes, as proposed by \cite{Hahn:2011}, introduces gravitational non-Gaussianity in the Eulerian density field of the baryons, but is not a consistent PT approach.

To tackle the problem of providing accurate Eulerian ICs for hydrodynamical simulations, we apply in the present paper the propagator perturbation theory \citep[PPT,][]{Uhlemann:2019,Rampf:2020} for multiple fluids.
 This field-level approach, which accurately evaluates LPT-evolved fields at the Eulerian position, has already been used for forward modelling of the matter distribution for Ly-$\alpha$ forest reconstructions by \cite{Porqueres:2020}, to first order in PPT. {\it Using  here second-order PPT to initialize Eulerian hydrodynamical simulations,} we are able for the first time to achieve ICs for both (Eulerian) baryons and (Lagrangian) dark matter that are on a similar footing regarding their accuracy in fixed-order PT.

The structure of this paper is as follows. First, in Section~\ref{sec:theory_summary}, we provide a concise summary of the main results from \cite{Rampf:2020} as they apply to initial conditions for cosmological simulations of baryons and dark matter. We also quantify the error incurred by neglecting contributions inconsistent with the IC boundary conditions compared to the full cosmological Einstein--Boltzmann solution. In Section~\ref{sec:descr_simulations}, we present the numerical simulations we employ in this work, and describe the summary statistics that we use to quantify them. In Section~\ref{sec:two-fluid-gravity-evolution}, we present results for the non-linear evolution of a collisionless two-fluid $N$-body system evolving under self-gravity. We then extend this analysis to full cosmological hydrodynamics plus $N$-body simulations in Section~\ref{sec:two-fluid-full-evolution}. We summarize our main results and conclude in Section~\ref{sec:summary}.

Throughout this paper, we adopt cosmological parameters consistent with the {\sc Planck2018}+LSS results \citep{Planck:2018a}: $\Omega_{\rm m} = 0.3111$, $\Omega_\Lambda = 0.6889$, $\Omega_{\rm b}  = 0.04897$, $\Omega_{\rm r}=9.139\times10^{-5}$, $h = 0.6766$, $\sigma_8 = 0.8102$ and $n_s = 0.9665$. Einstein--Boltzmann results were computed using the {\sc Class} code\footnote{available from \url{http://class-code.net/}} \citep{Blas:2011}. Our computation of the linear theory growth factor $D_+ =: D$ always includes the background contribution due to relativistic species; see e.g. \cite{Fidler:2017ebh} for details. Note that in this work we use the `fixing' technique of \cite{Angulo:2016} -- in which the modulus of the white noise Fourier modes is set to unity -- in order to suppress the impact of cosmic variance on our results without running large ensembles of simulations, but we consider only single simulations and do not perform the additional `pairing'.


\section{Perturbation theory in a nutshell}
\label{sec:theory_summary}
We consider the evolution of two fluids -- specifically CDM, and baryons for which we assume a negligible temperature (i.e., a Jeans scale much smaller than the scales of interest) -- interacting through gravity in an expanding Universe (parametrized by the cosmic scale factor $a(t)$).  We employ co-moving spatial coordinates $\fett{x} =\fett{r}/a$ and define peculiar velocities with respect to the co-moving expansion with $\fett{v} = \partial_D \fett{x}$, where $D$ is the linear growth time in $\Lambda$CDM which we also use as the time variable (for simplicity of notation we use $D:=D_+$ synonymously); we suppress temporal dependencies whenever there is no confusion. The governing equations of the system of two fluids for component $\alpha\in\{\rb, \rc\}$ in the zero temperature limit are 
\begin{subequations} \label{eqs:fluidsD}
 \begin{align}
& \partial_D \fett{v}_\alpha + \fett{v}_\alpha \cdot \nabla \fett{v}_\alpha = - \frac{3 g}{2D} \big( \fett{v}_\alpha + \nabla \varphi \big) \,,   \label{eq:euler}  \\
& \partial_D \delta_\alpha +\nabla \cdot \big[ (1+\delta_\alpha) \,\fett{v}_\alpha \big] =0 \,, \label{eq:massDalpha} \\
&\nabla^2  \varphi =  \frac 1  D \left( \fb \delta_\rb + \fc \delta_\rc \right) \,,  \label{eq:poisson}
 \end{align}
\end{subequations}
where $f_{\rm b}:=\Omega_{\rm b}/\Omega_{\rm m}$ and $\fc := 1 -\fb$ are respectively the global baryon and CDM mass fractions (neglecting other inhomogeneous contributions), and we have defined
\be 
  g := (D/\partial_t D)^2 a^{-3} = 1 + D^3 \Omega_{\Lambda}/(11\Omega_{\rm m}) + \mathcal{O}(D^6) \,.
\ee 
For detailed derivations of the following equations and results, we kindly refer the reader to the companion paper \cite{Rampf:2020}; for convenience, we provide a brief summary of the key technical steps here as well.

In the following, we first report analytical results in Eulerian coordinates, and discuss the validity and limitation of our approach. Results in Lagrangian coordinates as well as for a semi-classical field-based approach are given in Sections~\ref{sec:Lagrange} and~\ref{sec:ppt} respectively. Details for the numerical initialization of the involved fields with a linear Einstein--Boltzmann solver are provided in Section~\ref{sec:back-scaling}. Our method is fairly distinct from others in the literature, with details given in  Section~\ref{sec:forward}. Finally, in Section~\ref{sec:decaying_mode} we provide ways to effectively include relative velocity effects at linear order.
 
 
\subsection{Analytical findings in Eulerian coordinates}
\label{sec:eulerian_theory}
It is convenient \citep[e.g.][]{Schmidt:2016} to rewrite the set of Eqs.\,\eqref{eqs:fluidsD} in terms of the following ``sum'' and ``difference'' variables 
\begin{subequations} \label{eq:defsumdiff}
\begin{align}
  &\delta_{\rm m} =\fb \delta_{\rm b} +  \fc\,\delta_{\rm c} \,, \qquad & \theta_{\rm m}= \fb \theta_{\rm b} +  \fc\,\theta_{\rm c} \,, \\
 &\delta_{\rm bc} = \delta_{\rm b} - \delta_{\rm c}  \,, \qquad &  \theta_{\rm bc} = \theta_{\rm b} - \theta_{\rm c} \,,
\end{align}
\end{subequations}
where  $\theta_\alpha = \nabla \cdot \fett{v}_\alpha$. 
Formulated in these new variables, the linearized Eqs.\,\eqref{eqs:fluidsD} can be combined to 
\begin{subequations} \label{eqs:linearised2fluid}
\begin{align} 
  &\partial_D  \theta_{\rm m} = -  \frac{3g}{2D} \left( \theta_{\rm m} + \frac {\delta_{\rm m}} D \right) \,, \quad 
       &\partial_D \delta_{\rm m} + \theta_{\rm m} = 0 \,,  \label{eq:matterlinear} \\
 &\partial_D  \theta_{\rm bc}  = - \frac{3g}{2D} \theta_{\rm bc} \,,  
       &\partial_D \delta_{\rm bc} + \theta_{\rm bc} = 0 \,, \label{eq:differencelinear} 
\end{align}
\end{subequations}
which have the {\it only non-decaying solutions}  \citep{Rampf:2020}
\begin{subequations}  \label{eqs:sol4slaving}
\begin{align}
  & \delta_\rM = D\, \nabla^2 \varphi^{\rm ini} \,,   \qquad \hspace{1cm} \theta_\rM = - \nabla^2 \varphi^{\rm ini}  \,, \\
  & \delta_\bc = \delta_\bc^{\rm ini} \,, \qquad  \hspace{1.62cm}  \theta_\bc = 0  \,,  
\end{align}
\end{subequations} 
where here and in the following ``ini'' stands for initial evaluation; see section~\ref{sec:back-scaling} for details how we generate those initial fields. Using the definitions~\eqref{eq:defsumdiff}, these solutions imply for the components~$\alpha \in \{ \rb,\rc\}$ at first order in perturbation theory
\begin{align} \label{eq:solEuler+}
  & \delta_\alpha =D \nabla^2 \varphi^{\rm ini}  +\delta_\alpha^{\rm ini}   \,, \qquad  &\theta_\alpha = - \nabla^2 \varphi^{\rm ini}  \,, 
\end{align}
where we have defined  $\delta_\rb^{\rm ini} = \fc \delta_\bc^\ini$ and  $\delta_\rc^{\rm ini} = -\fb \delta_\bc^\ini$ which, from here on, are sometimes called the compensated constant modes (since $f_\rb \delta_\rb^\ini + f_\rc \delta_{\rc}^\ini =0$). Here it is crucial to note that Eqs.\,\eqref{eqs:linearised2fluid} remain regular for arbitrarily short times if and only if $\delta_{\rm m} \to 0$ and $\theta_\rM \to - \nabla^2 \varphi^\ini$ for $D\to 0$, which by virtue of the definitions~\eqref{eq:defsumdiff} implies an initially non-vanishing $\delta_\bc^\ini$ as well as non-vanishing~$\delta_\alpha^\ini$. Indeed, it can easily be verified that terms such as $\delta_{\rm m}/D$ or $\theta_\bc/D$ appearing in~\eqref{eqs:linearised2fluid} would otherwise imply quasi-singular behaviour for $D \to 0$ (which represents the unperturbed initial state). As mentioned in detail in section~\ref{sec:back-scaling}, being able to initialize the evolution at $D=0$ simplifies the boundary analysis tremendously: While only growing-modes are naturally selected at $D=0$, the evolution of baryons and CDM can be effectively decoupled from the full multi-fluid evolution which is governed by the relativistic Einstein--Boltzmann system. 

We remark that for initializing single-fluid simulations, the described procedure is standard and here applied to the two-fluid case. We must leave for future work how decaying modes can be consistently incorporated in such schemes (see however Section~\ref{sec:decaying_mode}), as they are by definition inconsistent with a homogeneous initial state on which the perturbations are imposed.

Before concluding this section, we report for completeness the higher-order results in our present model. For this we begin with the {\it Ans\"atze} 
\begin{subequations}
\begin{align}
  & \delta_\rM =  \sum_{n=1}^\infty  \delta_\rM^{(n)}(\fett{x})\,D^n \,,  \qquad \quad \, \theta_\rM = - \sum_{n=1}^\infty \theta_\rM^{(n)}(\fett{x}) \,D^{n-1}   \,, \\
 &  \delta_\bc =  \sum_{n=1}^\infty  \delta_\bc^{(n)}(\fett{x})\,D^{n-1}\,, \qquad\, \theta_\bc =0 \,,
\end{align}
where $\delta_\rM^{(n)}$ and $\theta_\rM^{(n)}$ are coefficients that can easily be determined from the known recursion relations in perturbation theory \citep[see e.g.][]{Bernardeau:2002,Taruya:2018}, while the analysis of~\cite{Rampf:2020} revealed the following recursion relation for the difference density, 
\be \label{eq:recdeltabc}
   \delta_\bc^{(n)} = \frac{1}{n-1} \sum_{0< s< n} \fett{\nabla} \cdot \left[ \delta_\bc^{(s)} \nabla^{-2} \fett{\nabla} \theta_\rM^{(n-s)} \right]  \,,
\ee
\end{subequations}
for $n>1$, and $\delta_\bc^{(1)}=\delta_\bc^\ini$ for $n=1$. From these recursive relations it is clear that solutions for $\delta_\rb$ and $\delta_\rc$ can be easily determined to arbitrarily high orders; for explicit solutions up to third order, see Appendix B in the companion paper.


\begin{figure}
	\centering
	\includegraphics[type=pdf, ext=.pdf, read=.pdf, width=\columnwidth]{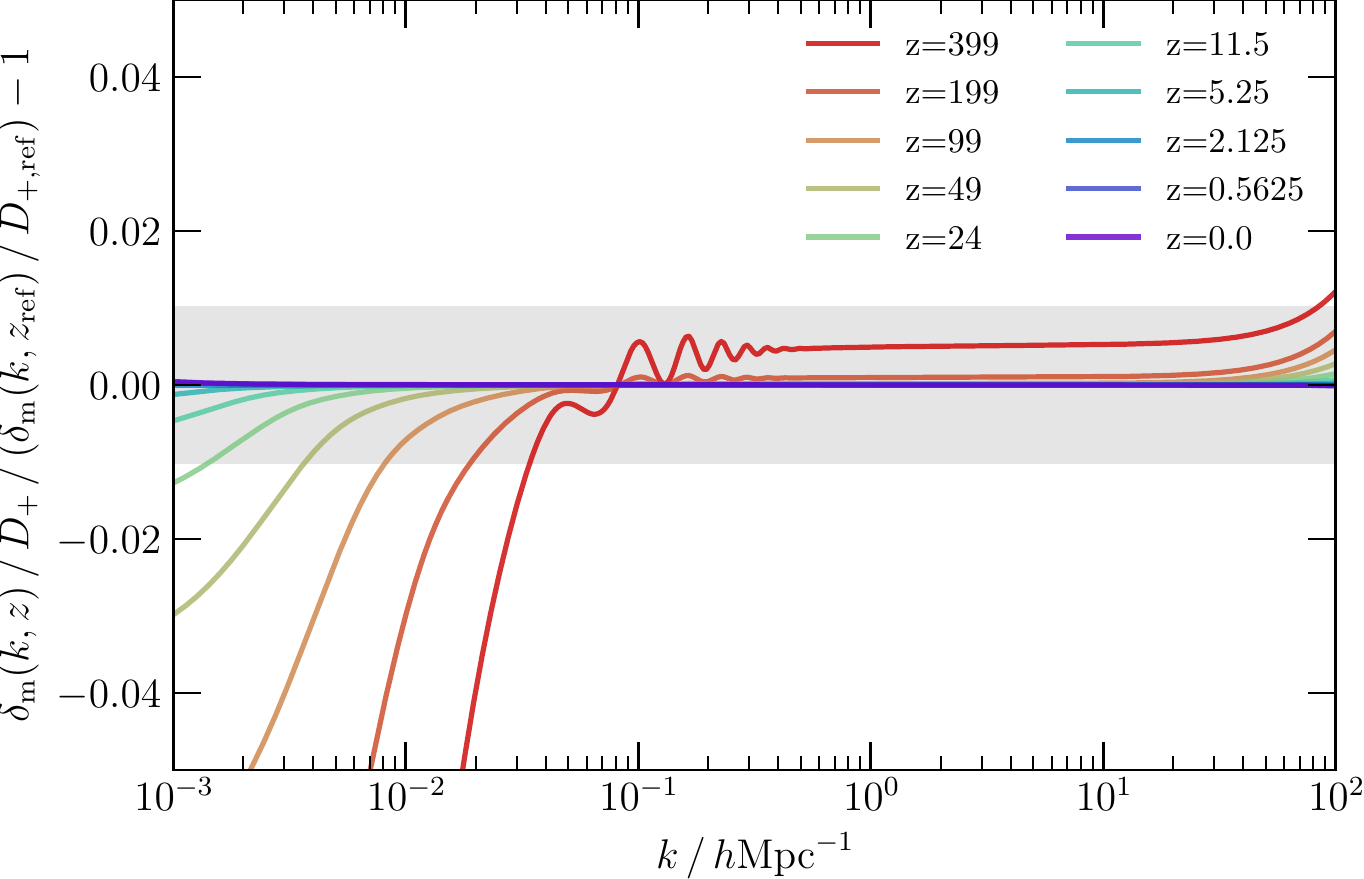}
	\caption{Residual scale-dependent evolution of the total matter density $\delta_{\rm m}$ from redshift 399 to 0 from {\sc Class}, relative to linear growth with $D_+$ and the total matter density amplitude at the reference redshift $z=2.125$. After $z\lesssim100$, evolution on small scales (i.e. $k\gtrsim 0.01\,h\,{\rm Mpc}^{-1}$) is consistent with a purely growing mode $\delta_{\rm}\propto D_+(z)$ at much better than one per cent, while residual evolution due to relativistic effects remains on larger (i.e. horizon-scale) scales, and due to finite baron pressure on small scales $k\gtrsim10^{2}h{\rm Mpc}^{-1}$. Note that the evolution of $D_+$ takes a non-zero $\Omega_{\rm r}$ into account. 	\label{fig:delta_m_evolution}}
\vspace{0.5cm}
	\centering
	\includegraphics[type=pdf, ext=.pdf, read=.pdf, width=\columnwidth]{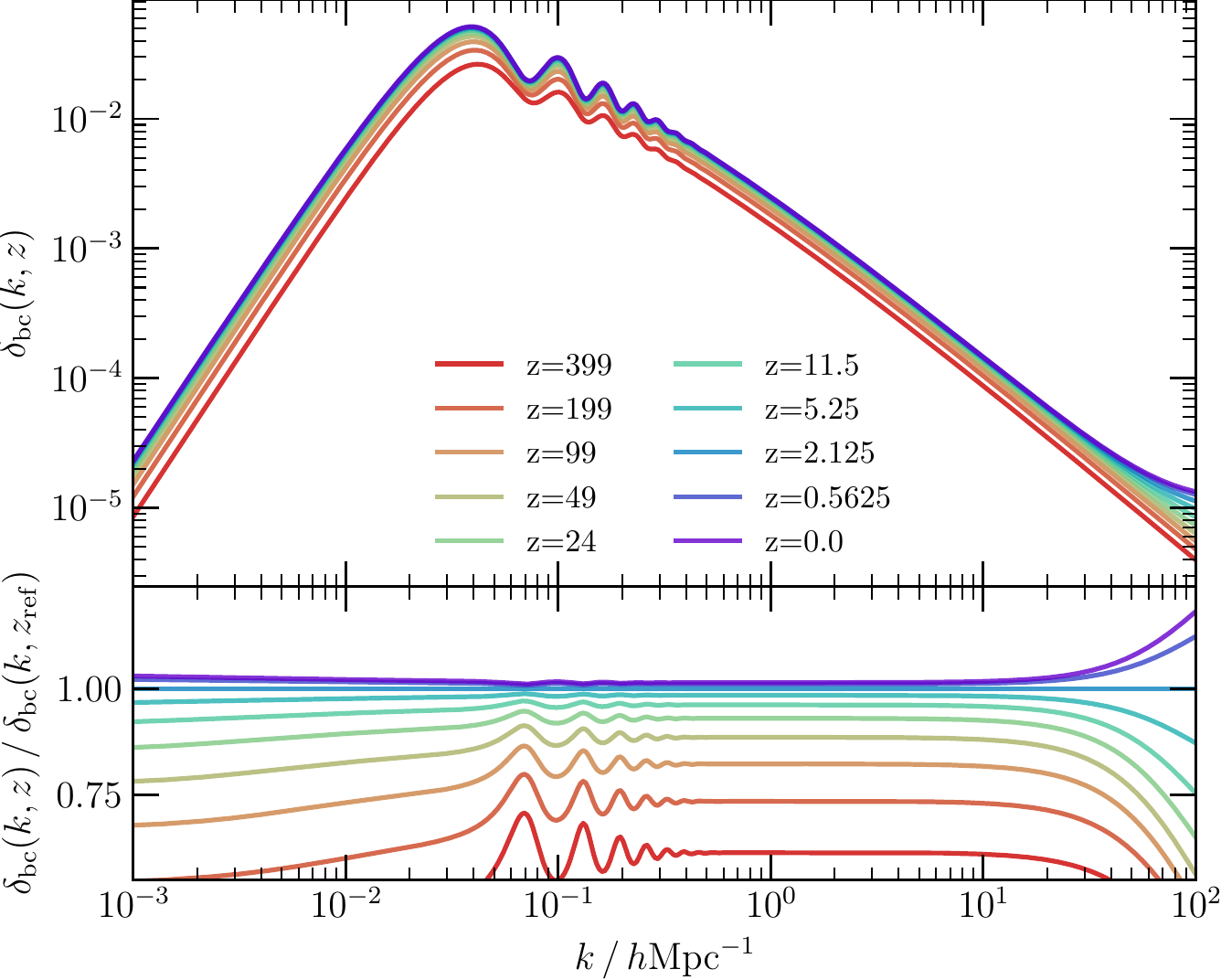}
	\caption{Scale-dependent evolution of the density difference $\delta_{\rm bc}$ from redshift 399 to zero (top panel), and also relative to the density difference at the reference redshift $z_{\rm ref}=2.125$ (bottom panel). The evolution is consistent with a constant mode only at low redshift, roughly $\lesssim 15$~per cent at $z\lesssim50$. At higher redshifts, there is a significant contribution due to an additional decaying mode and evolution of the horizon and Jeans scale. On the smallest scales, $k> 20\,h\,{\rm Mpc}^{-1}$, the impact of Jeans damping is visible.
	\label{fig:delta_bc_evolution}}
\end{figure}

\subsection{Validation of approximations with {\sc Class}} 

It is imperative to test the approximations we had to make in order to obtain consistent perturbative results, which effectively ignores decaying modes (see however further below), a finite sound speed of baryons, as well as couplings to relativistic fluid species -- except the zeroth-order coupling through the background evolution which we do include. To begin with, assuming the validity of  Eqs.\,(\ref{eqs:fluidsD}) implies focusing on a subset of the full set of equations solved in linear Einstein--Boltzmann solvers such as  {\sc Camb} \citep{Lewis2000} or {\sc Class} \citep{Blas:2011}, meaning that any solution based on these equations is already necessarily an approximation. While, clearly, neglecting decaying modes is an important simplification (and one that enables us to carry out the higher-order perturbation theory in the first place), it captures the leading effect of the two-fluid system in the late Universe, namely a spatially varying baryon fraction. 

To check the validity of the restriction to the growing  mode and the compensated constant mode, we show in Fig.~\ref{fig:delta_m_evolution} the evolution of the total matter overdensity amplitude $\delta_{\rm m}$ from redshift 399 to zero, scaled by the linear growth factor $D_+$ and divided by the amplitude at the reference redshift $z_{\rm ref}=2.125$. 

Clearly, inside the cosmological horizon and for scales $k\lesssim 10^{-2}h\,{\rm Mpc}^{-1}$ decaying modes have an impact of less than one per cent at all times of interest for ICs for simulations of the late-time Universe. On scales of the horizon and larger, one clearly sees the relativistic effects of horizon growth, as well as at very early times ($z\gtrsim 199$) the effect of radiation drag due to residual ionisation on the position of the BAO feature. Horizon scale relativistic effects are well understood and can be easily rectified \citep[see e.g.][]{2017MNRAS.466L..68B,2017JCAP...06..043F,2017MNRAS.466.3244Z}. On the smallest scales one sees the impact of the time-evolution of the baryon temperature dependent Jeans scale (which in the $D_+$-scaled solution is fixed to its co-moving value at the reference redshift). 
 
In Fig.~\ref{fig:delta_bc_evolution}, we show similar results for the two-fluid case: The top panel shows the evolution of the compensated mode $\delta_{\rm bc}$ between redshifts $z=399$ and zero, while the bottom panel displays specifically the impact of ignoring decaying modes and baryonic pressure in its evolution (i.e., assuming $\delta_{\rm bc}$ to be constant in time). Evidently,  ignoring decaying modes in the evolution of $\delta_{\rm bc}$ is justified at fairly late times ($z \lesssim 10$) on almost all scales, while on very small scales ($k\gtrsim 20\,h{\rm Mpc}^{-1}$) the impact of the evolution of the Jeans scale due to the evolution of the baryon temperature becomes visible. The dominant scale-dependent evolution is again due to a shift of the BAO feature at high redshift, as well as a weak horizon-scale evolution. Assuming that $\delta_{\rm bc}$ is constant in time thus introduces an almost scale-independent error that is increasingly larger at high redshift. We consider it thus important to tune the reference redshift for the IC generation (see Section~\ref{sec:back-scaling}) well in order to capture most accurately the time of interest (i.e., for a Lyman-$\alpha$ forest simulation, e.g., it is arguably more accurate to use a reference redshift of $z_{\rm ref}\sim2.5$ rather than zero).

\begin{figure}
	\centering
	\includegraphics[type=pdf, ext=.pdf, read=.pdf, width=\columnwidth]{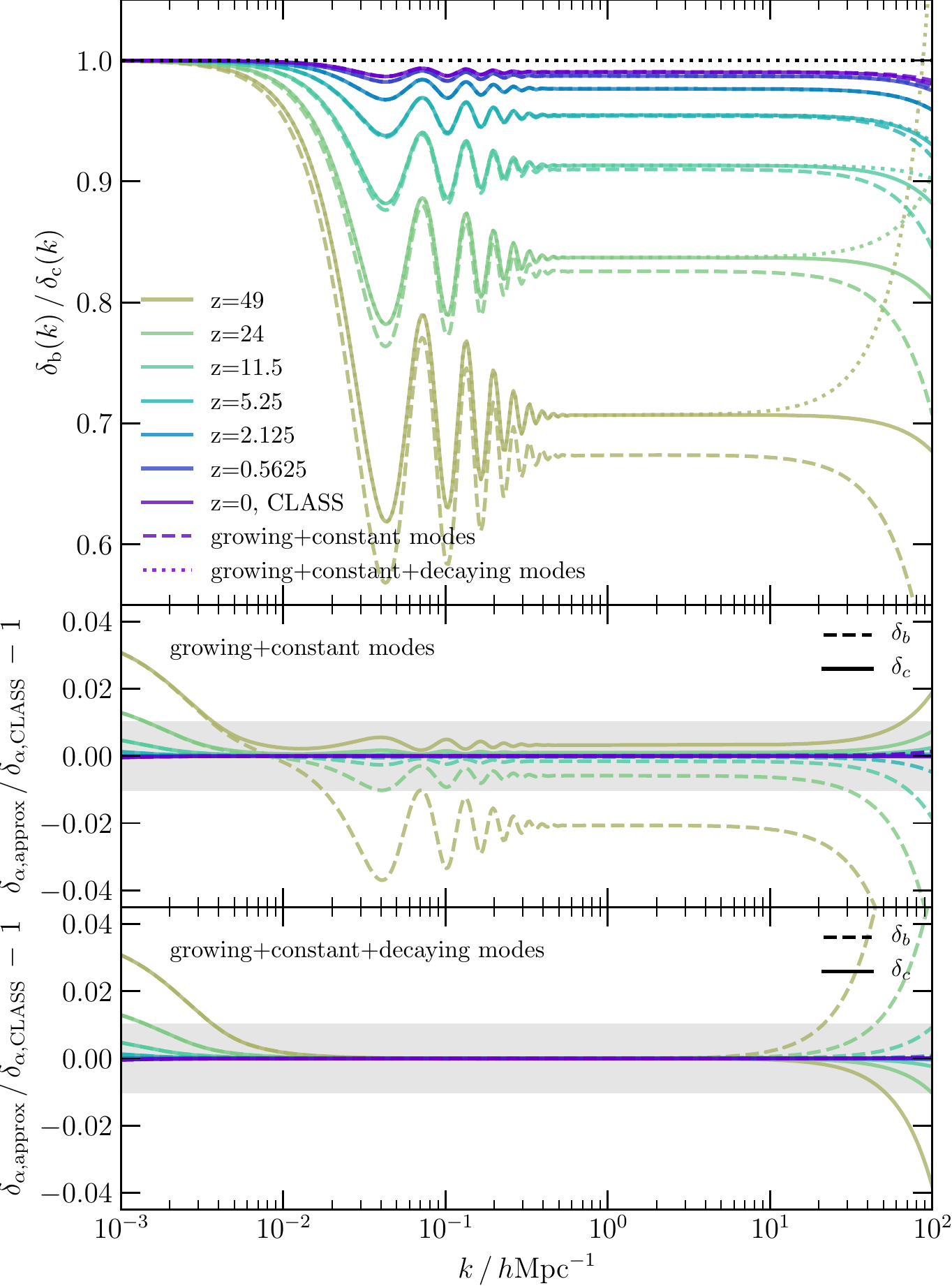}
	\caption{Accuracy of the two- and three-mode approximation for the amplitude of baryon and CDM density perturbations. The top panel shows the ratio of the scale-dependent amplitude of baryon to CDM density perturbations as obtained from {\sc Class} (solid lines) and  using the growing + constant mode approximation (dashed lines; cf.\ Eq.\,\ref{eq:twomode}), and also including the decaying relative velocity mode (dotted lines; cf.\ Eq.\,\ref{eq:threemode}). The middle (bottom) panel shows the scale-dependent fractional difference in each component between the two-mode (three-mode) approximation and the {\sc Class} amplitude for CDM (solid lines) and baryons (dashed lines). The shaded gray area indicates one per cent deviation. The error in each component is sub per cent for redshifts $z\lesssim 24$ on scales smaller than the horizon.   The error on large-scales ($k\lesssim 0.01 h\,{\rm Mpc}^{-1}$) is due to neglecting relativistic effects, the error on small scales is due to neglecting finite temperature effects.}
	\label{fig:slaved_approx_evolution}
\end{figure}

In Fig.~\ref{fig:slaved_approx_evolution}, we show how well the obtained linear baryon and CDM spectra agree with the full multi-physics evolution. Specifically, we consider two approaches, the first being a two-mode approximation (growing+constant) that forms the basis of our higher order PT, and which obeys eq.~(\ref{eq:solEuler+}), i.e.,
\be \label{eq:twomode}
 \delta_\alpha =D \nabla^2 \varphi^{\rm ini}  +\delta_\alpha^{\rm ini} \,,
\ee
with initial fields obtained as detailed in Section~\ref{sec:back-scaling}. The second approach we consider is based on a three-mode approximation (growing+constant+decaying), in which the decaying relative velocity mode is also included (cf.\ Section~\ref{sec:decaying_mode} for generating the additional initial field), i.e., we have then
\be \label{eq:threemode}
 \delta_\alpha =D \nabla^2 \varphi^{\rm ini}  +\delta_\alpha^{\rm ini} + 2 (D^{-1/2}-1)\,\theta_\alpha^{\rm ini},
\ee
with $\delta_{\rm b}^{\rm ini}=\fc \delta_{\rm bc}^{\rm ini}$, $\theta_{\rm b}^{\rm ini}=\fc\delta_{\rm bc}^{\rm ini}$, and $\delta_{\rm c}^{\rm ini}=-f_{\rm b}\delta_{\rm bc}^{\rm ini}$, $\theta_{\rm c}^{\rm ini}=-f_{\rm b}\theta_{\rm bc}^{\rm ini}$.
 In the top panel, we compare the evolution of $\delta_{\rm b}(k)/\delta_{\rm c}(k)$ from $z=49$ to zero as obtained from {\sc Class} (solid lines) including the full linear physics to the two-mode (dashed) and three-mode (dotted) approximations. The relative differences are shown in the bottom panels of the figure. By construction, the solutions coincide at the reference time $z_{\rm ref}=2.125$. When neglecting the decaying relative velocity mode, the relative error is sub-percent for all times $z\lesssim24$ on scales $k\lesssim 10^{2}h\,{\rm Mpc}^{-1}$, but increases rapidly at earlier times. This behaviour is improved when the decaying mode is included, leading to sub-percent agreement over more than four magnitudes in scale at $z\lesssim24$. At late times, the largest error clearly arises from the Jeans scale due to its strong evolution over time, and for simulations aiming at these scales, a more refined discussion, possibly also including temperature fluctuations \citep[e.g.][]{Naoz:2005,Naoz:2011}, might be necessary. Note that the evolution shown here does not include the impact of reionization on the Jeans scale as this is usually captured in the non-linear simulation directly. Since reionization raises the baryon temperature significantly, the exact evolution of these small scales in the IC backscaling process for simulations that do not explicitly resolve the formation of the reionizing objects is of limited interest.

Clearly, there is some improvement if the decaying mode is included in the linear evolution, however we do not yet know how to include it self-consistently in a non-linear PT (but see Section~\ref{sec:decaying_mode} for a workaround). This means that the impact of streaming velocities \citep[cf.][]{Tseliakhovich:2010} cannot be self-consistently included yet. This is arguably the largest drawback and we have to leave the inclusion of a suppression of baryon perturbations on small scales due the relative motion (which is a decaying mode coupling to the baryon density at second order) for future work.  Since this effect is most prominent on the smallest scales (on dimensional grounds it must be close to the baryon Jeans scale), we can safely assume that the approach we present here is accurate for large-scale simulations that do not include the formation of the very first baryonic objects close to the Jeans scale.


\subsection{Lagrangian-coordinates approach}\label{sec:Lagrange}

Introducing the Lagrangian map for both fluid components $\alpha \in\{{\rm b,c}\}$ with $\fett{q} \mapsto \fett{x}^\alpha(\fett{q}, D) = \fett{q} + \fett{\xi}^\alpha (\fett{q},D)$ with corresponding displacement $\fett{\xi}^\alpha$, equations~\eqref{eqs:fluidsD} can be easily transformed into Lagrangian space. For purely growing-mode flows in perturbation theory the Lagrangian equations of motion take the particularly simple form \citep{Rampf:2020} 
\begin{subequations}
\begin{align}
 &\fett{\nabla}_{\!x} \cdot \left( \partial_D^2  +\frac{3g}{2D} \partial_D   \right)  \fett{\xi}^\alpha(\fett{q},D) = - \frac{3g}{2D} \nabx^2 \varphi   \,,  \label{eq:evoxialpha}\\
 &\nabx^2 \varphi = \delta_\rM(\fett{q},D) /D  \,,  \label{eq:PoissDeltaM}
\end{align}
\end{subequations}
where the Poisson source is expressed in terms of
\begin{subequations}
\be \label{eq:massM}
  \delta_\rM(\fett{q},D) = \frac{1}{\det[\fett{\nabla}_q \fett{x}^\rM]}-1 \,,
\ee 
where 
\be
 \fett{x}^\rM -\fett{q} \equiv \fett{\xi}^\rM = \fb \fett{\xi}^\rb + \fc \fett{\xi}^\rc
\ee 
is the displacement of the combined (or, centre-of-mass) matter fluid. The perturbative solution of the combined matter displacement is well-known, which is usually formulated in terms of the following power series
\be \label{eq:xisingle}
   \fett{\xi}^\rM(\fett{q},D) = \sum_{n=1}^\infty \fett{\xi}^{\rM(n)}(\fett{q})\,D^n \,,
\ee
\end{subequations}
with the first-order coefficient $\fett{\xi}^{\rM(1)}= - \fett{\nabla} \varphi^\ini$ denoting the Zel'dovich approximation, and with the second-order coefficient $\fett{\xi}^{\rM(2)}= - \fett{\nabla} \varphi_2$ with $\nabla^2 \varphi_2 = \frac{3}{14} (\varphi_{,ll}^{\rm ini} \, \varphi_{,mm}^{\rm ini} - \varphi_{,lm}^{\rm ini} \, \varphi_{,lm}^{\rm ini}) $. Explicit recursions relations for $\fett{\xi}^{\rM(n)}$ are given by \cite{Rampf:2012b}, \cite{Zheligovsky:2014}, and \cite{Matsubara:2015ipa}.

Since the combined matter displacement $\fett{\xi}^\rM$ is known to all orders, the Poisson equation~\eqref{eq:PoissDeltaM} can be easily determined by virtue of~\eqref{eq:massM}. At the same time, having determined the Poisson source allows us to solve the evolution equation~\eqref{eq:evoxialpha} for the component displacement;
the solution is  particularly simple and reads for the growing modes, for $\alpha\in\{\rb,\rc\}$~\citep{Rampf:2020}
\be \label{eq:xialpha}
     \fett{\xi}^\alpha(\fett{q},D) =  \sum_{n=1}^\infty \fett{\xi}^{\rM(n)}(\fett{q})\,D^n \,,
\ee
which, crucially, must be supplemented with the mass conservation law (cf.\ with Eq.\,\ref{eq:massM})
\be
  \delta_\alpha = \frac{1 + \delta_\alpha^\ini(\fett{q})}{\det[ \nabq \fett{x}^\alpha]} - 1 \,. \label{eq:massalpha} 
\ee
To be specific, although the perturbative solutions for $\fett{\xi}^\alpha$ formally agree with the one of $\fett{\xi}^\rM$, the initial density perturbation~$\delta_\alpha^\ini$ appearing in \eqref{eq:massalpha} must be taken into account since, as mentioned above, only the inclusion of~$\delta_\alpha^\ini$ guarantees the regularity of solutions for arbitrarily short times. 

Recently it has been mathematically proven by \cite{Zheligovsky:2014} and \cite{Rampf:2015} that the LPT series for the single fluid converges for realistic random initial conditions in the growing mode, at least for sufficiently short times. Even more recently, numerical convergence studies to very high perturbation orders revealed that LPT converges until the instance of first shell-crossing (and even beyond, although LPT then ceases to be physically correct; \citealt{Rampf:2020b}). These findings have direct relevance for the considered two-fluid model with growing mode initial conditions. Indeed, since Eqs.\,\eqref{eq:xisingle} and \eqref{eq:xialpha} coincide, it is clear that the observed convergence behaviour for the single fluid directly carries over to the two-fluid model. This is not surprising as the considered model boils down to transporting initial density perturbations along a shared fluid flow (see the companion paper), which comes with the benefit that for two-fluid growing-mode initial conditions, $n$LPT is guranteed to provide more accurate refinements at successively higher orders (provided shell-crossing has not yet occurred).

We remark that when generating initial conditions for the baryon and CDM fluids, $\delta_\alpha^\ini(\fett{q})$ can be explicitly taken into account by varying the particle masses. Further details are provided below (see Eq.\,\ref{eq:pertmass}).

We also remark that alternatively to the above LPT method for two fluids, one may also incorporate $\delta_\alpha^\ini(\fett{q})$ by perturbing the initial positions; see Section~5.4 of~\cite{Rampf:2020} for details. To second order the resulting displacement takes the form  
\be \label{eq:pertdisp}
  \fett{\xi}^\alpha_{\rm pert} =  D\,\fett{\xi}^{\rm m(1)} + D^2 \fett{\xi}^{\rm m(2)} 
    - \nabla^{-2} \fett{\nabla} \delta_{\alpha}^{\rm ini} +  \fett{F}^{(2)}(\delta_\alpha^{\rm ini}) \,,
\ee
where $\fett{F}^{(2)}$ is a vector-valued function that depends quadratically on $\delta_{\alpha}^{\rm ini}$  and can be read off from Eq.~(56) in \cite{Rampf:2020}. While the corresponding mass conservation then simplifies to
$1+\delta_\alpha = 1/\det [\mathbb{1}+\nabq \fett{\xi}^\alpha_{\rm pert} ]$, we find that spurious discretization errors are excited (see discussion below and Fig.~\ref{fig:delta_r_nbody}), and thus we do not recommend this avenue for the present context.

\paragraph*{Pre-initial conditions for numerical implementation.}

Since these LPT results directly translate to initial conditions for Lagrangian methods, such as $N$-body, we can simply (pre-)initialize a set of $N$ particles with the positions and velocities starting from a discrete set of locations $\fett{q}_{i=1\dots N}$. In this work, we always place the particles initially on a simple cubic (SC) Bravais lattice, so that initial particle positions coincide with the uniform 3D grid on which we computed the velocity and displacement fields using Fourier methods \citep[cf.][]{Michaux:2020}. 

For Lagrangian hydrodynamics codes, such as {\sc Gadget} or {\sc Arepo} which we discuss below, also the baryon fluid elements need to be set up using LPT. A dilemma arises if one wants to construct a force-free set up of the initial unperturbed CDM+baryon fluid. In the $\delta_{\rm bc}=0$ case this is possible, two individual SC lattices, each carrying $N$ particles, shifted by half a cell diagonal, with particle masses $\bar{m}_\alpha =  m_p \Omega_\alpha/\Omega_\rM$ achieve this ($m_p$ is $M_{\rm box}/N$). The result corresponds to the CsCl crystal structure. One can in principle also use any other diatomic crystal structure, such as two shifted face centred cubic (FCC) lattices corresponding to either a NaCl or a Zincblende crystal, depending on the relative shift vector. To evaluate the perturbation fields at the shifted locations, we use a simple Fourier shift of the field\footnote{Using that for a field shifted by $\fett{x}_0$, i.e. $g:=f(\fett{x}-\fett{x}_0)$, the Fourier transforms obey $\hat{g}(\fett{k}) = \exp\left[ -{\rm i}\fett{k}\cdot\fett{x}_0\right]\, \hat{f}(\fett{k})$.} in the inverse direction and evaluate at cell centres.

As discussed above, a non-zero $\delta_{\rm bc}$ can be realized in two ways: either by using the total mass LPT displacements $\fett{\xi}^\rM(\fett{q},D)$ for all species and perturbing the individual particle masses
\be \label{eq:pertmass}
m_\alpha(\fett{q}) = \bar{m}_\alpha \, \left(1+\delta^{\rm ini}_\alpha(\fett{q})\right)\,, \qquad \bar{m}_\alpha:= \Omega_\alpha\,/\,\Omega_{\rm m}\,,
\ee
or by absorbing this perturbation into a perturbed displacement, i.e., applying Eq.\,\eqref{eq:pertdisp}. Either case leads to discretization errors, however we find that only with a perturbed displacement that this error has a spurious growing mode, while for the perturbed masses, the discreteness errors are confined to small scales only. For this reason we adopt the perturbed mass approach in most parts of this paper. We present an analysis of the impact of mass vs. displacement perturbations on the power spectrum in Section~\ref{sec:two-fluid-gravity-evolution}. Even in `forward' simulations (see Section~\ref{sec:compensated-collisionless} for details) it would seem preferable to use perturbed masses instead of displacements to set up the compensated perturbations.

Note that we do not consider the proposed solution of~\cite{Bird:2020} in this article, which uses glass pre-initial conditions~\citep[cf.][]{White:1996} for baryon particles and an SC lattice for DM particles. While seemingly also solving the spurious growth problem, this approach appears to introduce significant additional noise on small scales compared to a Bravais lattice, so that we see no advantage over our approach.

A potential concern in multi-mass collisionless simulations is the evolution towards mass segregation of $N$-body particles in equipartitioned systems \citep{Binney:2008} due to spurious collisional relaxation. We therefore want to emphasize that the mass perturbations introduced by Eq.~(\ref{eq:pertmass}) are {\it small}, {\it independent of the starting redshift} (in the fastest growing approximation), and {\it vary on rather large scales}. For the set-up we investigate later, i.e. a $250\,h^{-1}{\rm Mpc}$ box with $2\times512^3$ particles, the relative fractional variation ($1\sigma$) in particle mass is $\sim2.96\times10^{-3}$ for the CDM particles and $\sim1.69\times10^{-2}$ for the baryon fluid elements.%
\footnote{Note that the mass perturbations have amplitudes of $\sigma_{\rm m_\rb}=\bar{m}_\rb \fc \sigma_{\rm bc}$ and $\sigma_{\rm m_\rc}=\bar{m}_\rc f_\rb \sigma_{\rm bc}$ where $\sigma_{\rm bc}^2 = (2\pi^2)^{-1}\int_0^{k_{\rm max}} {\rm d}k\, k^2P_{\rm bc}(k)$. If we assume a late-time baryon Jeans scale of order $k_{\rm J}\sim 100\, h{\rm Mpc}^{-1}$ as our $k_{\rm max}$, then $\sigma_{\rm bc}\sim0.026$ for our cosmology. So even at higher resolution, the mass perturbation amounts to at best a few per cent. It would increase of course  beyond the (evolving) baryon Jeans scale, but finite temperature effects are beyond the scope of our study here.} 
In addition, this variation of a few per cent is spatially correlated with a pronounced peak at the BAO scale (cf. Fig.~\ref{fig:delta_bc_evolution}), meaning that smaller-scale non-linear regions will always have less variation among their particle masses (which can be seen by eye e.g. in the bottom left panel of Fig.~\ref{fig:ppt_example} where we show the spatial behaviour of $\delta_{\rm bc}$). Note that furthermore the relative variation in each species is significantly smaller than the difference in particle masses between baryons and CDM in these simulations (which is of order $\Omega_{\rm b}/\Omega_{\rm c}\simeq 1/5.4$). With mass differences at the sub-per-cent level for CDM particles, the relaxation time can therefore safely be expected to be much longer than that due to spurious scattering between `stars' and CDM particles \citep[cf.][]{Ludlow:2020}.

\subsection{Propagator perturbation theory}\label{sec:ppt}

In contrast to Lagrangian methods, cosmological hydrodynamic codes based on Eulerian hydrodynamics, such as the finite volume codes {\sc Ramses} \citep{Teyssier:2002}, {\sc Enzo} \citep{Bryan:2014}, or {\sc Nyx} \citep{Almgren:2013}, need to start the baryon evolution from the Eulerian density and momentum fields, given at fixed locations discretized in Eulerian space. A possibility to obtain such fields consistent with LPT is by interpolating the fluid elements back to Eulerian grid cells, incurring however the problem of high quality conservative interpolation. Here we follow an alternative approach by using propagator perturbation theory (PPT), as proposed by \cite{Uhlemann:2019} and extended to two fluids in \cite{Rampf:2020}, which is able to yield Eulerian density and momentum fields consistent with LPT without {\it ad-hoc} interpolation \citep[see also][]{Porqueres:2020}.

In the following, we briefly summarize essential equations together with relevant results; further technical details are provided in the companion paper.

\paragraph*{Analytical findings in PPT.} \ The central aspect of PPT is to solve for the wavefunction  $\psi_\alpha$ of the fluid components $\alpha \in \{ \rb, \rc\}$ whose time evolution is given by the Schr\"odinger equation 
\begin{align} 
\label{eq:Schroedi2fluid}
 \ii \hbar \partial_D \psi_\alpha &= \ - \frac{\hbar^2}{2} \nabla_x^2 \psi_\alpha + V_{\rm eff} \,\psi_\alpha  \,,
\end{align}
where, $V_{\rm eff}$ is an ``effective'' gravitational potential defined in relation to the fluid equations~\eqref{eqs:fluidsD}. In PPT, $V_{\rm eff}$ is treated as an external potential determined by standard perturbation theory. The evolution is expressed through the propagator  $K(\fett{q},\fett{x};D)$  that propagates the initial wave function (defined at $D=0$)
\be
 \psi_\alpha^\ini(\fett{q}) =  \sqrt{1+ \delta_\alpha^{\rm ini}(\fett{q})} \, \exp \left[ \frac{\rm i}{\hbar} \varphi^{\rm ini}(\fett{q}) \right] \label{eq:psi_ini} 
\ee  
to the current state at time $D$ and position $\fett{x}$, i.e., 
\be
\label{eq:PsiKernel}
  \psi_\alpha(\fett{x};D) =  \int \! \dd^3 q \, K(\fett{q},\fett{x};D) \,  \psi_\alpha^{\rm ini}(\fett{q}) \,.
\ee
  At leading order $V_{\rm eff} \equiv 0$, and the solution of the resulting potential-free Schr\"odinger equation~\eqref{eq:Schroedi2fluid}  is readily obtained from the ``free propagator'' 
\begin{align}
  &K_{\rm free}(\fett{q},\fett{x};D) =  (2 \uppi \ii \hbar D)^{-3/2} \,\exp \left[ \ii (\fett{x}-\fett{q})^2 /(2\hbar D)\right] \,, \label{eq:freep}
\end{align}
where the prefactor guarantees that Eq.\,\eqref{eq:PsiKernel} returns $\psi_\alpha^\ini$ for $D \to 0$.

At next-to-leading order, dubbed 2PPT, a time-independent $V_{\rm eff}$ becomes relevant and is given by the expression
\be
  \nabla^2 V_{\rm eff}  = \frac 3 7 \left(\varphi_{,ll}^{\rm ini} \, \varphi_{,mm}^{\rm ini} - \varphi_{,lm}^{\rm ini} \, \varphi_{,lm}^{\rm ini} \right) \,.
\ee
 As shown in the companion paper, the 2PPT propagator reads
\begin{align}
\label{eq:KNLOsol}
  K (\text{\small $\fett{q},\fett{x};D$}) = K_{\rm free}(\text{\small $\fett{q},\fett{x};D$}) \, \exp\left[-\frac{\ii D}{2\hbar}\left(V_{\rm eff}(\fett{q}) + V_{\rm eff}(\fett{x}) \right)\right] \,.
\end{align}
The semiclassical limits of the free and 2PPT propagators return, respectively, the classical Zel'dovich approximation and the second-order improvement 2LPT. \cite{Uhlemann:2019} have shown that the 2PPT results are in fact more accurate than 2LPT since additional symmetries are preserved due to the underlying Hamiltonian structure of~\eqref{eq:Schroedi2fluid}. Notably, no spurious higher-order vorticity is excited.

\begin{figure}
	\centering
	\includegraphics[type=pdf, ext=.pdf, read=.pdf, width=\columnwidth]{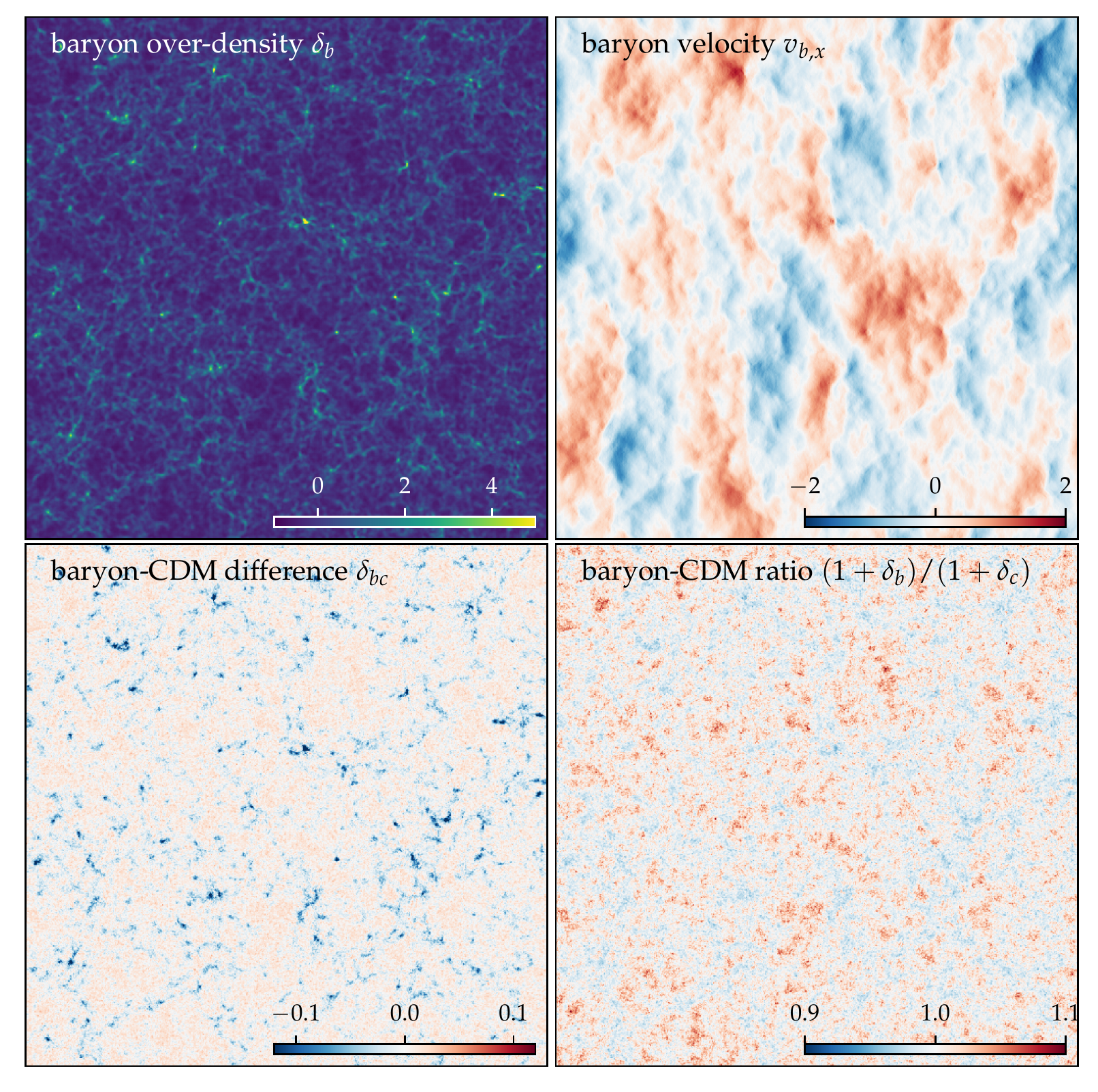}
	\caption{Eulerian fields at $z=8$ obtained with 2PPT as described in Section~\ref{sec:ppt}: the baryon overdensity $\delta_{\rm b}$ (top left), x-component of the baryon peculiar velocity field $v_{{\rm b},x}$ (top right), the compensated density difference $\delta_{\rm bc}$ (bottom left), and the ratio of baryon to CDM density fluctuations (bottom right). We show an $x$-$y$-slice through the highest density point ($\delta_{{\rm b,max}}\simeq13$, north of the centre of the image) for a box of side-length $250\,h^{-1}{\rm Mpc}$ computed using a resolution of $512^3$. }
	\label{fig:ppt_example}
\end{figure}

Having obtained numerical solutions for the wave function (see the following paragraph for details), the desired Eulerian fields, e.g., the density $\rho_\alpha = 1 + \delta_\alpha$ and the momentum density field $\fett{\pi}_\alpha = \rho_\alpha \fett{v}_\alpha $ for each species, are
\begin{subequations}
\begin{align}
\rho_\alpha(\fett{x},a) &= \psi_\alpha\,\overline{\psi}_\alpha,\quad\textrm{and} \\
\fett{\pi}_\alpha(\fett{x},a)&= \frac{{\rm i}\hbar}{2} \left( \psi_\alpha \fett{\nabla} \overline{\psi}_\alpha - \overline{\psi}_\alpha \fett{\nabla} \psi_\alpha \right),
\end{align}
\end{subequations}
where an overline denotes complex conjugation. In principle, one could also extract an effective temperature from the next higher moment, but we will neglect finite temperature effects here altogether and always assume the cold limit on the PT side.

 In Figure~\ref{fig:ppt_example} we show the baryon density, velocity $\fett{v}_\alpha = \fett{\pi}_\alpha / \rho_\alpha$, and the 2PPT density difference $\delta_\bc$ for a $L=250\,h^{-1}{\rm Mpc}$ box with $512^3$ resolution elements at $z=8$ (which is much later than the time we would initialize a simulation and was just chosen for illustrative purposes). For further numerical tests of PPT in the single-fluid case we refer to~\cite{Uhlemann:2019}.

\paragraph*{Numerical implementation of PPT.} \ Numerically, the expression for the free propagator~\eqref{eq:freep} is most conveniently evaluated using a discrete Fourier transform (DFT), since the cyclic convolution with the propagator becomes a simple multiplication in Fourier space. Let us therefore assume without change of notation that all spatial coordinates, $\fett{x}$ and $\fett{q}$, refer to positions on a discrete regular grid with spacing $\Delta$, whenever we refer to the numerical implementation. Then, the equivalent statement of~\eqref{eq:freep} at the operational level can be executed using the ``drift'' operator $\hat{\rm D}$, defined through
\begin{align}
\psi_\alpha(\fett{x},a) &= \hat{\rm D} \, \psi_\alpha^{\rm ini}\\
&=:\underset{\fett{k}\to\fett{x}}{\rm DFT}^{-1}\left\{ \exp\left[ -{\rm i} \hbar D_+(a)\, \frac{k^2}{2}  \right]\, \underset{\fett{q}\to\fett{k}}{\rm DFT}\left\{ \psi_\alpha^{\rm ini}(\fett{q}) \right\}\right\} \nonumber,
\end{align}
where $\fett{k}$ denotes a discrete wave vector and $k$ its modulus. Similarly, to incorporate the aforementioned 2PPT correction, one introduces the ``kick'' operator 
\be
\hat{\rm K} := \exp\left[ -\frac{{\rm i}}{\hbar} \frac{D_+(a)}{2} V_{\rm eff} \right]  
\ee
in real space, which corresponds to a half `time step' in $D_+$. The final 2PPT operator evolution equation is given by the single-step leap frog
\be
 \psi_\alpha(\fett{x};a) = \hat{\rm K} \, \hat{\rm D} \, \hat{\rm K} \, \psi_\alpha^{\rm ini} \,.
\ee
It can be effectively evaluated by performing the drift step in Fourier space and the kick steps in regular space. 

\begin{figure}
	\centering
	\includegraphics[type=pdf, ext=.pdf, read=.pdf, width=\columnwidth]{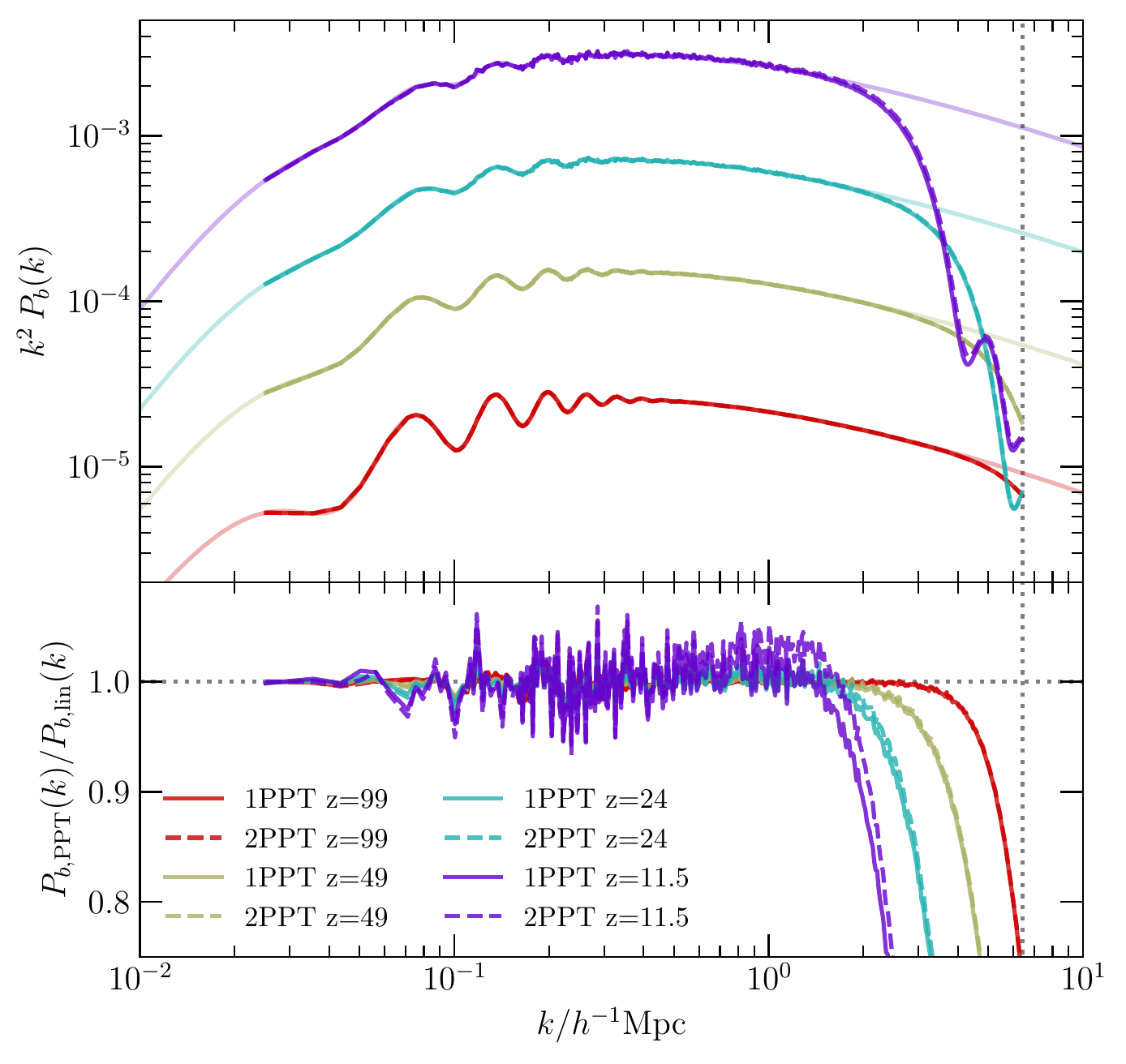}
	
	\caption{Evolution of the baryon power spectrum $P_{\rm b}(k,z)$ in `growing mode' Eulerian linear theory and in PPT (top panel), and the ratio of the two (bottom panel) for a $250\,h^{-1}{\rm Mpc}$ box with $512^3$ resolution.  The finite $\hbar$, which is set by numerical resolution (see Eq.\,\ref{eq:hbar}), introduces an evolving scale beyond which power is sharply suppressed due to effectively coarse grained dynamics. }
	\label{fig:power_ppt}
\end{figure}

\paragraph*{The $\fett{\hbar}$-parameter.} Finally, for numerical implementations of PPT, one chooses a finite $\hbar$ that is as small as possible in order to be closest to the semi-classical limit. Since we evaluate the propagator using a DFT, the smallest numerically possible $\hbar$ is determined by the Nyquist--Shannon sampling theorem, which requires that the phase in adjacent sampling points changes by at most $\uppi$. This implies
\be  \label{eq:hbar}
\hbar \ge  \frac{1}{\uppi} \,\, \max_{\fett{q},d} \,\left| \varphi^{\rm ini}(\fett{q}) - \varphi^{\rm ini}(\fett{q}+\Delta\, \hat{\fett{e}}_d)\right|,
\ee
where $\hat{\fett{e}}_d$ is the Cartesian basis vector for the $d$-th dimension and $\Delta$ the grid spacing -- the expression thus runs over all points and considers the (three) neighbours in three dimensions. We determine~$\hbar$ once we have generated the initial field $\varphi^{\rm ini}$ from the input power spectrum, and it depends thus explicitly both on the form of the perturbation spectrum realized in the simulation volume and the grid spacing $\Delta$.

The numerically finite value of $\hbar$ has of course a consequence, namely it acts as an effective coarse-graining scale of the LPT dynamics over phase space cells of size $\hbar$. Since $\hbar$ is determined mostly by the resolution and more weakly by the shape of the perturbation spectrum, the resolution sets the effective temperature of PPT. This manifests itself as a `Jeans'-like suppression of power on the smallest scales, similar to what is observed in PT for axion-like particles \citep[cf. e.g.][]{Guth:2015}, but note that the scale related to $\hbar/m$ has a different time dependence in PPT than in the axion-like case (here $k_{\rm cut}\propto a^{-1/2}=(1+z)^{1/2}$). In Figure~\ref{fig:power_ppt}, we show the effect on the baryon power spectrum. In the top panel, we show the power spectrum from linear Eulerian PT, restricted to our `growing mode' model, in comparison to the first- and second-order PPT results, measured numerically at different times $z=99,49,24$ and $11.5$. The effective Jeans smoothing is clearly visible as a sharp power suppression on small scales that increases for later starting times. In the bottom panel, the ratio between Eulerian PT and PPT spectra is shown for a more quantitative comparison. Note that the power spectrum does however not capture the significant amount of non-Gaussianity that is already present in the fields at the later times. As we show below in Section~\ref{sec:evolution_nongaussian}, since the suppression affects scales of 2-3 cells only, and the full non-Gaussian character of LPT is mapped to the Eulerian grid, baryon simulations initialized with PPT evolve quite consistently with those initialized with LPT.

\subsection{Generating the initial fields -- backscaling Einstein--Boltzmann}\label{sec:back-scaling}

Traditionally, to generate  first-order initial conditions for two-fluid numerical studies for baryons and CDM, one takes the respective fluid variables from a linear Einstein--Boltzmann code at the time when the simulation is to be initialized \citep[see e.g.][]{Yoshida:2003,Hahn:2011,Angulo:2013,Valkenburg:2017,Bird:2020}. By contrast, in simulations for single-matter fields, it is very common to employ the so-called backscaling procedure, which effectively takes the Boltzmann code from very late times, usually around~$z \simeq 0$ (but note that we use $z=2.125$ as the pivot redshift in this work), and rescales the respective gravitational potential $\varphi$ such that the initialized particle configuration at $z_\ini$ has the correct density amplitude. It is important to realize that in a (fictitious) universe with zero radiation content, both approaches reproduce the same initial matter, baryon and CDM power spectrum. However, the two approaches disagree in a realistic Universe, due to the nontrivial evolution of relativistic species which mostly impact the largest scales.

Here, we adopt the backscaling procedure to allow for the initialization of two fluids. This has the advantage that the evolved large-scale power spectra agree, by definition, with the corresponding predictions in general relativity. In addition, also the finite temperature of the baryons is partially included implicitly through the baryon transfer function, determined at the reference time, just not its adiabatic evolution under compression and expansion.

Due to the choice of used boundary conditions~\eqref{eqs:sol4slaving}, which effectively set the decaying modes for the two fluids to zero,  only two fields need to be specified initially. One of those fields is the total matter field $\delta_\rM$ which relates to the associated gravitational potential according to $\nabla^2 \tilde \varphi(a) = \delta_\rM(a)/a$, where $\tilde \varphi = D_+ \varphi/a$. Since growing-mode initial conditions are obtained from the output of an Einstein-Boltzmann code at sufficiently late times $a_{\rm ref}$, one can write  $\delta_\rM^{\rm code}(a_{\rm ref}) = C_+(\fett{x}) D_+(a_{\rm ref})$. Using these relations, the initial gravitational potential at $a=0$  is \citep[cf.][]{Michaux:2020}
\be
\label{eq:phi_ini}
   \varphi^\ini = \frac{\nabla^{-2} \delta_\rM^{\rm code}(a_{\rm ref})}{D_+(a_{\rm ref})} \lim_{a \to 0} \frac{D_+(a)}{a} \,.
\ee
In the present paper, we choose $a_{\rm ref} =0.32$, in accordance with our choice of reference redshift $z_{\rm ref}=2.125$. The other initial field that should be prescribed is the linear difference $\delta_\bc$ which, in the absence of decaying modes, is constant in time. Thus, the amplitude $\delta_\bc$ does not need to be rescaled, and can instead be directly extracted from a Boltzmann code at $a=a_{\rm ref}$
\be
\label{eq:deltabc_ini}
   \delta_\bc^\ini = \delta_\bc^{\rm code}(a_{\rm ref}) \,.
\ee
 Of course, having specified both $\varphi^\ini$ and $\delta_\bc^{\rm ini}$ initially also yields the initial fields for $\delta_\rb$ and $\delta_\rc$, as well as $\theta_\rb$ and $\theta_\rc$, by virtue of the definitions~\eqref{eq:defsumdiff} and boundary conditions~\eqref{eqs:sol4slaving}.

\subsection{Relation to the forward approach}\label{sec:forward}

Previous studies modelling two-fluid dynamics in $N$-body simulations \citep[e.g.][]{Yoshida:2003,Angulo:2013,Valkenburg:2017,Bird:2020} all (to our knowledge) rely on the forward approach, where the Lagrangian displacement and velocity fields are initialized directly with the output of the linear Einstein--Boltzmann code at time $z_{\rm start}$ as
\be \label{eq:1lpt_forward}
 \begin{aligned}
   &\fett{x}_\alpha(\fett{q}, \zstart) = \fett{q}-\nabla^{-2}\fett{\nabla}\delta_\alpha^{\rm code}(\zstart) \,,  \\
   &\fett{v}_\alpha(\fett{q}, \zstart) = \nabla^{-2}\fett{\nabla}\theta_\alpha^{\rm code}(z_{\rm start}) \,. 
  \end{aligned}
\ee
Within the two-fluid picture, the four code input fields $\delta_\alpha^{\rm code}$ and $\theta_\alpha^{\rm code}$ may be expressed in terms of the standard growing and decaying modes. Putting aside the decaying modes for the  moment (justified at sufficiently late times), one can approximate~\eqref{eq:1lpt_forward} with
\be  \label{eq:1lpt_forward2}
 \begin{aligned}
  &\fett{x}_\alpha(\fett{q}, \zstart) \approx \fett{q}- D(z_{\rm start}) \,\fett{\nabla}\varphi^\ini - \nabla^{-2} \nab  \delta_\alpha^\ini\,, \\ 
  &\fett{v}_\alpha(\fett{q}, \zstart) \approx   \dot D(z_{\rm start}) \fett{\nabla} \varphi^\ini \,.
 \end{aligned}
\ee
where we have used Eq.\,\eqref{eq:solEuler+} to express the growing modes in terms of the input fields in our backscaling approach. Thus, in the forward approach, the initial perturbations $\delta_\alpha^{\rm ini}$ are effectively included in~\eqref{eq:1lpt_forward2} along with the other modes. 
Hence, the forward approach is very close in spirit to the `displacement perturbation' approach presented in Eq.\,(\ref{eq:pertdisp}), and, therefore, comes with fairly similar numerical challenges as discussed e.g.\ in detail by \cite{Angulo:2013,Bird:2020}. In Section~\ref{sec:two-fluid-gravity-evolution}, and in particular in~\ref{sec:compensated-collisionless}, we present a detailed comparison between `displacement perturbed' and `mass perturbed'~ICs.


\subsection{Including the decaying relative velocity mode at first order}
\label{sec:decaying_mode}

The two-fluid perturbation theory presented above neglects all decaying modes. This leads to small but noticeable differences between the growing mode PT and the evolution in e.g. {\sc Class}, as shown in Fig.~\ref{fig:slaved_approx_evolution}. While it is still unclear how to incorporate these decaying modes rigorously in higher order LPT/PPT, we note that it is quite straightforward to include the relative velocity $v_{\rm bc}$ between baryons and CDM at linear order in LPT. Given the difference fields $\delta_{\rm bc}(\fett{q};\,a_{\rm ref})$ and $\theta_{\rm bc}(\fett{q};\,a_{\rm ref})$ from the Einstein-Boltzmann code at the reference time $a_{\rm ref}$, one can modify the mass perturbation and initial particle velocity by rescaling the decaying relative velocity mode from $a_{\rm ref}$ to the starting time $a_{\rm start}$ of the simulation as
\be
\label{eq:decayingmass}
\begin{aligned}
m_\alpha(\fett{q};\,a_{\rm start}) &\to m_\alpha(\fett{q};\,a_{\rm start})  \\
& +  2  \bar{m}_\alpha  \left[  \left(\frac{D_+(a_{\rm ref})}{D_+(a_{\rm start})}\right)^{1/2} - 1 \right] \theta_{\alpha}^{\rm ini}(\fett{q}) \\
\boldsymbol{v}_\alpha(\boldsymbol{q};\,a_{\rm start}) &\to \boldsymbol{v}_\alpha(\boldsymbol{q};\,a_{\rm start})  \\
& +  \left(\frac{D_+(a_{\rm ref})}{D_+(a_{\rm start})}\right)^{1/2} {\nabla^{-2}}\boldsymbol{\nabla}\theta_{\alpha}^{\rm ini}(\fett{q}).
\end{aligned}
\ee
This allows for a first-order correction to the $n$LPT two-fluid ICs which restores the agreement between {\sc Class} and LPT ICs. We note however that it is not part of a rigorous perturbative framework, and it is yet unclear how to incorporate a similar fix in PPT. We demonstrate below in Section~\ref{sec:convergence_ratio_bc_power} that the inclusion improves, as expected, the agreement with the linear {\sc Class} calculation at high redshift. It appears therefore that, whenever possible, this mode should be included, even though a self-consistent higher order PT is not known to include its non-linear coupling.


\section{Employed simulation set-up and summary statistics}
\label{sec:descr_simulations}
In this section, we briefly summarize the simulation codes as well as the simulations we use in this work. We also discuss the technicalities of the analyses we perform on the simulations.

\begin{table}
\begin{center}
\begin{tabular}{llll}
PT & $z_{\rm start}$ & $N_{\rm part}$ &code  \\
\hline
1LPT & 49 & $2\times512^3$ &{\sc Gadget-2}$^{1,\ast,\bigtriangleup}$\\
2LPT & 49 & $2\times512^3$ &{\sc Gadget-2}$^{1}$\\
2LPT & 24 & $2\times512^3$ &{\sc Gadget-2}$^{1,\bigtriangleup}$\\
3LPT & 24 & $2\times256^3$ &{\sc Gadget-2}$^{1}$\\
3LPT & 24 & $2\times512^3$ &{\sc Gadget-2}$^{1}$\\
3LPT & 24 & $2\times1024^3$ &{\sc Gadget-2}$^{1}$\\
\hline
1LPT + 1PPT& 49 & $2\times512^3$ &{\sc Ramses}$^{2,\dagger}$\\
2LPT + 2PPT& 24 & $2\times512^3$ &{\sc Ramses}$^{2,\ddagger}$\\
1LPT & 49 & $2\times512^3$ &{\sc Arepo}$^{2,\dagger}$\\
2LPT & 24 & $2\times512^3$ &{\sc Arepo}$^{2,\ddagger}$\\
\hline
\end{tabular}
\end{center}
\caption{Simulations marked with superscript `${1}$' are used in Section~\ref{sec:two-fluid-gravity-evolution} for the study of the purely gravitational, cold and  collisionless evolution of our two-fluid ICs, those with `${2}$' in Section~\ref{sec:two-fluid-full-evolution} for full $N$-body plus collisional hydrodynamics simulations. For the run marked with `${\ast}$', we ran also with ICs with perturbed initial positions (using Eq.\,\eqref{eq:pertdisp} to first order); all others ICs use perturbed masses (see Eq.\,\eqref{eq:pertmass}) which is our preferred method; and for those marked with '${\bigtriangleup}$', we also ran ICs with the decaying relative velocity mode included at linear order (using Eq.~\ref{eq:decayingmass}).  We also refer to the 1LPT/PPT hydro runs `${\dagger}$' as `leading order' (LO) and the 2LPT/PPT runs `${\ddagger}$' as `next-to-leading-order' (NLO). All simulations represent a cosmological volume of side length $250\, h^{-1}{\rm Mpc}$.\label{tab:sims}}
\end{table}

\subsection{Simulation methods}
In order to compare the performance of Lagrangian and Eulerian cosmological hydrodynamics codes as well as the impact of the collisional nature of baryons vs.\ the effect of gravity alone, we use a multitude of cosmological simulation codes in this work. Specifically, we use the Tree-SPH code {\sc Gadget-2} \citep{Springel:2005} for all gravity-only simulations, in which we do not use the SPH part but evolve both species as collisionless zero-temperature fluids.

For more realistic baryon+CDM simulations that evolve baryons hydrodynamically, we use the finite volume code {\sc Ramses} \citep{Teyssier:2002}, as well as the moving mesh code {\sc Arepo} \citep{Springel:2010,Weinberger:2020}, respectively, to evolve our initial conditions. Note that this present choice of simulation codes is fairly arbitrary, and an increasingly larger set of codes is becoming freely available to the community.  A more stringent code comparison of the results that we sketch in the following sections, that includes other codes, is certainly desirable at some point in the future. Any details of the two codes we use beyond the Lagrangian-Eulerian distinction of the hydrodynamic scheme are not very important for this paper. 

At early times, we are in a regime where the finite temperature of the baryons is negligibly small, and pressure effects become important only after shell-crossing and the related formation of shocks and caustics \citep[e.g.][]{Shandarin:1989}. We decidedly do not include additional physics such as radiative cooling or even astrophysical processes such as star formation or energy injection, and switch off UV and other backgrounds. A Lagrangian method therefore has the trivial advantage of (in principle) solving the cold non-linear advection problem with self-gravity more accurately than a Eulerian method prior to shell crossing\footnote{This statement is strictly speaking not correct since pseudo-spectral Eulerian would have also negligible (possibly even superior) advection errors. Pseudo-spectral methods are however not used in cosmological simulations due to their lack of adaptivity and poor convergence at singularities.}.  We list all simulations employed in this work in Table~\ref{tab:sims}. Our motivation to consider both a Eulerian and a Lagrangian code was to validate the performance of $n$PPT against $n$LPT, differences between {\sc Ramses} and {\sc Arepo} are of secondary interest to us here.

\paragraph*{Collisionless fluids with {\sc Gadget-2}.} \ Before considering collisional simulations of the baryons (i.e., in the hydrodynamic limit), we will study the purely gravitational, collisionless evolution using a two component $N$-body simulation. For these simulations we use {\sc Gadget-2} and treat both baryons and CDM as $N$-body particles (i.e., we do not use the Smoothed Particle Hydrodynamics (SPH) part of the code). {\sc Gadget-2} uses a tree-PM approach to solve for self-gravity and we employ the code specific parameters listed in Table~\ref{tab:arepo_params}.
 
\paragraph*{Eulerian baryons with {\sc Ramses}.} \  {\sc Ramses} is based on a second order MUSCL \citep{VanLeer:1979} finite volume scheme to solve the equations of ideal hydrodynamics. For the evolution of the collisionless dark matter, {\sc Ramses} employs an adaptive particle mesh scheme. We adopt the usual quasi-Lagrangian refinement strategy in which refinements are triggered by the number of $N$-body particles and a gas mass threshold based on the initial average baryon mass per cell. In order to achieve a more accurate large-scale integration, we refine the base grid level already when a cell exceeds 4 times its initial (cosmic average) mass, all higher levels at the default threshold of 8 times. All accuracy-related code specific parameters are listed in Table~\ref{tab:ramses_params}. 
 
\paragraph*{Lagrangian baryons with {\sc Arepo}.} \ The moving mesh code  {\sc Arepo} is strictly speaking not a fully Lagrangian method, since the mesh does not strictly follow the Lagrangian tracers, and fluxes between cells are taken into account. Prior to shell crossing we are in the advection dominated regime, and we preferentially probe here the Lagrangian aspect of this approach, while arguably the late time deeply non-linear evolution might be more similar to a Eulerian finite volume method. We list all accuracy-related code specific parameters in Table~\ref{tab:arepo_params}.

\subsection{Analysis of simulations -- power spectra}
\label{sec:definition_powerspectrum}
In this article, we analyse the density statistics of the baryon-CDM two-fluid system mainly through the isotropic (auto) power spectrum $P_X(k)$ defined as
\be
\left\langle \delta_X(\fett{k}) \, \delta_X(\fett{k}') \right\rangle = (2\uppi)^3 \delta_{\rm D}^{(3)}(\fett{k}+\fett{k}^\prime)\,P_X(k),
\ee
where $X\in\{{\rm b},{\rm c},{\rm bc},{\rm m}\}$, we have 
 $k:=\|\fett{k}\|$, and $\delta_{\rm D}^{(3)}$ is the three-dimensional Dirac delta. Numerically, we compute all power spectra using DFTs based on the mass distribution on a regular mesh. 

\paragraph*{Density fields for particles.} \ If the density field is represented by Lagrangian elements (i.e., particles or moving cells), we employ a `cloud-in-cell' \citep[CIC, cf.][]{Hockney:1981} interpolation to a regular grid. To accurately estimate the power spectrum, we use the interlacing technique proposed by \cite{Sefusatti:2016} along with deconvolution with the CIC assignment kernel. We always employ twice the resolution in the DFT mesh compared to the particle resolution, i.e., for $N^3$ particles, we compute DFTs of size $(2N)^3$, to resolve the particle grid itself. Note that we do not correct for shot noise.

\paragraph*{Density fields for finite volume cells.} \ For the finite volume {\sc Ramses} simulations, the baryon density is given as a volume average on the adaptively refined oct-tree mesh. In order to evaluate the density field on a regular grid at the same resolution as the particles, it is necessary to deal with cells that are larger than the grid on which one desires to compute the power spectrum. For those cells, that are at a coarser resolution, we use the slope-limited piecewise linear reconstruction used also during the actual {\sc Ramses} simulation to `refine' coarse cells to the target resolution. We found that a deconvolution with the cell volume average is necessary to achieve an estimate of the power spectrum that is relatively independent of the resolution used for its estimation (just as with the interlacing and deconvolution in the case of the particles). The volume average is represented by the convolution with the `nearest-grid-point'  \citep[NGP, cf.][]{Hockney:1981} kernel
\be
W_{\smalltext{NGP}} = \left( 2\uppi \right) ^{3/2}\prod_{i\in\{x,y,z\}}\frac{\sin \left( \frac{\uppi}{2} \, {k_i}\,/\,{k_\smalltext{Ny}}\right)}{{ k_i }\,/\,{ k_\smalltext{Ny}}},
\ee
where $k_{\smalltext{Ny}}$ is the grid Nyquist wave number. Note also that $W_{\smalltext{CIC}} = W_\smalltext{NGP}^2$ for the kernel used to deconvolve the CIC particle projection.

\subsection{Analysis of simulations -- bispectra}
\label{sec:definition_bispectrum}
To capture the growth of non-Gaussianity in the baryon-CDM two-fluid system, we also consider the (isotropic) component bispectrum $B_X(k_1,k_2,k_3)$, defined by
\be
\left\langle \delta_X(\fett{k}_1) \delta_X(\fett{k}_2) \delta_X(\fett{k}_3)\right\rangle =\ (2\uppi)^3 \delta_{\rm D}^{(3)}(\fett{k}_1+\fett{k}_2+\fett{k}_3)\,B_X(k_1,k_2,k_3),
\ee
with $X\in\{\rb,\rc\}$ (but one could also consider $\{\bc,\rM\}$ as well, of course). To simplify the discussion, we only focus on equilateral bispectra here, i.e., where $k:=k_1=k_2=k_3$. We use the {\sc Python} package {\sc Bskit} \citep{Foreman:2019}, to numerically compute the bispectrum from the same three-dimensional component density fields as the power spectra described in the previous subsection (i.e., we perform our own CIC deconvolution for the particle density fields and NGP deconvolution for the finite volume density field). {\sc BSkit} is based on the ``Scoccimarro estimator'' for the bispectrum \citep[cf.][]{Scoccimarro:2000,Sefusatti:2016,Tomlinson:2019}.
 
\subsection{Analysis of simulations -- cumulants}
\label{sec:definition_cumulants}
In addition to the bispectra, to quantify the amount of non-Gaussianity present in the simulation, we also consider directly the third and fourth cumulants (i.e., skewness and kurtosis) of the density field, which we define as the dimensionless quantities
\begin{subequations}
\label{eq:cumulants}
\begin{align}
C^\alpha_3 &:= \langle \delta_\alpha^3 \rangle_{\rm s} \, / \, \langle \delta_\alpha^2 \rangle_{\rm s}^{3/2}\\
C^\alpha_4 &:=  \langle \delta_\alpha^4 \rangle_{\rm s} \, / \,  \langle \delta_\alpha^2 \rangle_{\rm s}^{2} \,- \,3,
\end{align}
\end{subequations}
where $\langle \cdot \rangle_{\rm s}$ is the volume average of the respective field, filtered with a top hat filter of scale $R_{\rm s}$. The skewness is related to the bispectrum through
\begin{align}
 \langle \delta_\alpha^3 \rangle_{\rm s} = \int \dd^3k_1 \dd^3k_2 &\dd^3k_3 \, B_\alpha(k_1,k_2,k_3) \, \delta_{\rm D}^{(3)}(\fett{k}_1+\fett{k}_2+\fett{k}_3)\notag\\
 &\times W(k_1R_{\rm s}) \, W(k_2R_{\rm s}) \, W(k_3R_{\rm s}) \,,
\end{align}
where $W$ is the Fourier kernel of the spherical top-hat smoothing window. Note that we have by definition $\langle \delta \rangle_{\rm s} = 0$. The smoothed density fields are obtained from the same mesh of size $(2N)^3$ as the power and bispectra, and we also perform the deconvolution with the CIC or NGP filter, as described in Section~\ref{sec:definition_powerspectrum}, prior to applying the top hat filter.


\section{Results I: Purely Gravitational Evolution in Lagrangian Simulations}
\label{sec:two-fluid-gravity-evolution}
Before we discuss the performance of our initial conditions evolving the baryons as collisional, it is worth to investigate first the purely gravitational collisionless evolution as a first generalization step of a single fluid cold $N$-body simulation to two fluids. We focus exclusively on the evolution of the power spectrum in this section. In these $2N$-body simulations, we use identical numbers of baryon and CDM particles with a softening of gravitational forces of $1/20$th of the spacing of the initial unperturbed particle lattice, which corresponds to a typical, even slightly conservative, choice in single fluid collisionless $N$-body simulations. The simulations that we use for the analysis in this section are listed in Table~\ref{tab:sims}.

\begin{figure}
	\centering
	\includegraphics[type=pdf, ext=.pdf, read=.pdf, width=\columnwidth]{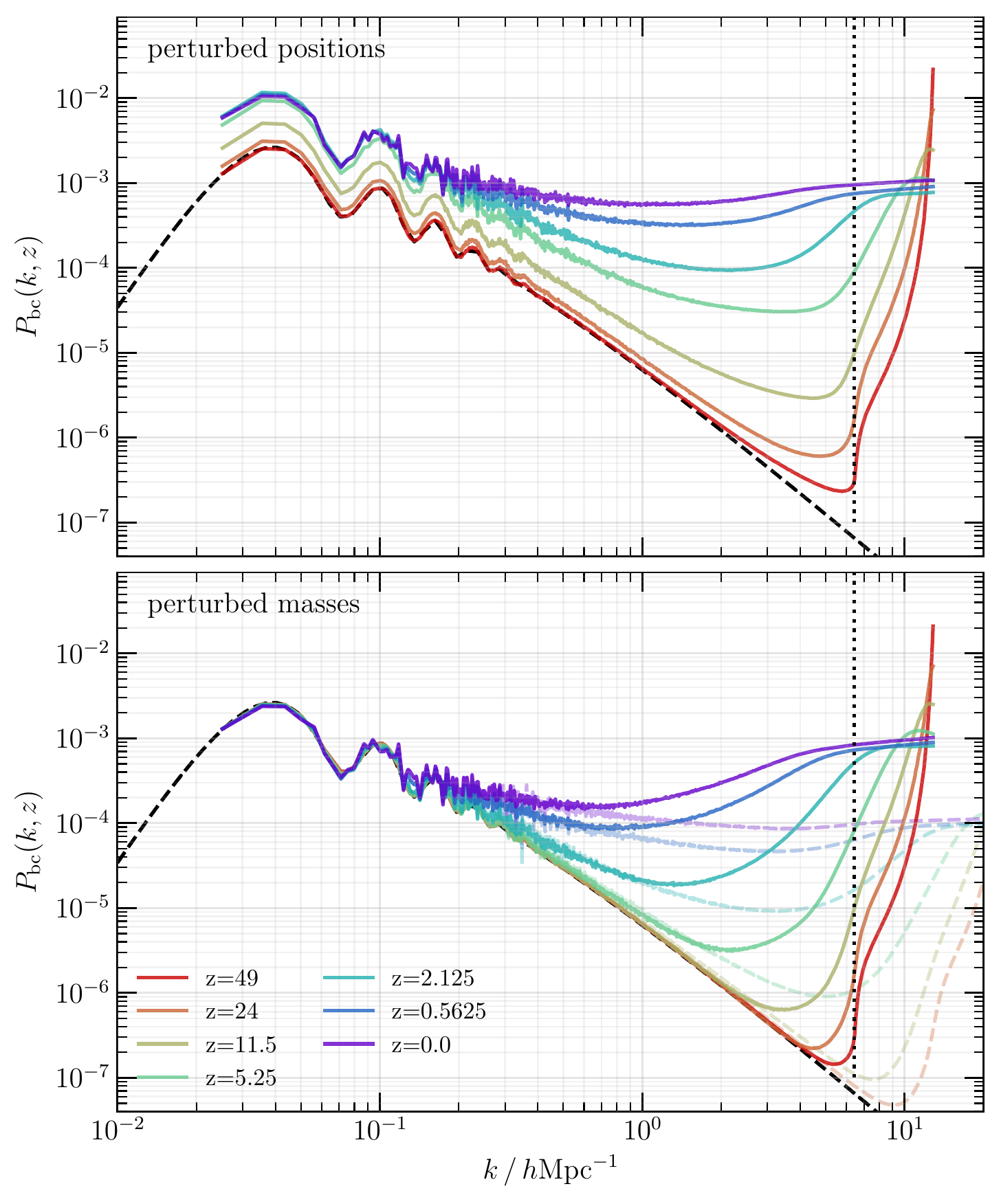}
	
	\caption{Evolution of the baryon-CDM density difference power spectrum $P_{\rm bc}$ in a gravity-only $N$-body simulations where both baryon and CDM are simulated with $N$-body particles with small softening. Simulation results in solid lines (faint dashed lines) with $2 \times 512^3$ ($2 \times 1024^3$) particles initialized at $z_{\rm start}= 49$ ($z_{\rm start}= 24$) using 1LPT (3LPT) ICs, by suitably backscaling the amplitudes from {\sc Class} at the reference redshift $z_{\rm ref}=2.125$. Those backscaled power spectra are shown as black dashed lines, before applying the particle realization. The vertical dotted line indicates the particle Nyquist wave number. {\bf Top panel:} simulation results where $\delta_{\rm bc}^\ini$ is absorbed into particle displacements (using Eq.\,\eqref{eq:pertdisp}). We observe the well-known spurious growing mode due to discreteness errors in the particle discretization of the fluids (the discrete $\delta_\bc$ is not compensated at the discretization scale, if not sufficiently smoothed by gravitational softening, and therefore grows).
{\bf Bottom panel:} Same as above but now $\delta_{\rm bc}^\ini$ is absorbed into perturbed particle masses (using Eq.\,\eqref{eq:pertmass}). Here, discretization errors are strongly suppressed, and the late-time power asymptotes to a constant noise spectrum beyond $k_{\rm Ny}$. These errors are furthermore vastly reduced when $2\times1024^3$ particles and 3LPT ICs are used.}
	\label{fig:delta_r_nbody}
\end{figure}

\subsection{Preservation of the compensated mode at linear scales} 
\label{sec:compensated-collisionless}
A first test is the preservation of the compensated mode, which we assume to be constant in time at linear order and thus, it must be preserved in the absence of numerical errors in the linear regime (i.e., early times and/or large scales). It is well known by now that this is not easy to achieve due to discreteness errors. Previous studies have not used our restriction to just two modes, but have instead followed the ``forward approach'' where the output of the Einstein--Boltzmann code is directly taken at the starting redshift (cf.\ the discussion in Section~\ref{sec:forward}). Without either a gravitational softening of the order of the mean particle separation, or an adaptive softening that arguably achieves this more optimally \citep{OLeary:2012,Angulo:2013}, the relative amplitudes between baryons and CDM do not evolve correctly on any scale, even in the linear regime \citep[cf.\ also][]{Valkenburg:2017}. This is because, effectively, $\delta_{\rm bc}$ is not compensated at the particle level leading to a slowly growing discrete mode.  The low particle-per-force resolution of single-fluid cosmological simulations is usually only possible due to the cold initial conditions, but is known to deviate from the fluid limit on small scales \citep[cf.][]{Joyce:2005,Marcos:2006cn,Joyce:2007}. Alternatively, \cite{Bird:2020} report that using a mix of grid and glass pre-initial conditions can also suppress the spurious growth of the compensated mode. To our knowledge, there is no theoretical understanding what exactly causes the spurious growing mode, and the exact influence of particle pre-initial conditions on it.

In Figure~\ref{fig:delta_r_nbody}, we demonstrate the evolution of discreteness effects in the $\delta_\bc$ power spectrum for the two ways of setting up ICs. In the top panel, we show $P_{\rm bc}(k)$ between $z=49$ and $z=0$ for a simulation with $2 \times 512^3$ particles initialized with two-fluid 1LPT where the initial density perturbation $\delta_\alpha^{\rm (ini)}$ was  incorporated into initial particle displacements; as mentioned above and in section~\ref{sec:forward}, this approach is implicitly followed in previous two-fluid studies.  We see that $P_{\rm bc}$ evolves on {\it all} scales due to numerical errors  -- with the strongest deviations growing at the particle Nyquist wave number $k_{\rm Ny}$ (indicated by a dotted vertical line) -- and linear perturbation theory is {\it not} recovered even on the largest scales. While we show results only for 1LPT and at starting time $z_{\rm start}=49$, this result is almost independent of the used LPT order. Since the error is driven by a spurious growing mode, its amplitude is to first order simply determined by the starting time -- with earlier starts leading to a larger error. It is also insensitive to the specific choice of softening length, as long as it is appreciably smaller than the mean particle separations. Once the small scales have shell-crossed and collapsed, at $z\lesssim 2.5$, the spurious growth is slowed. 

In stark contrast, the situation improves dramatically when the initial density perturbation $\delta_\alpha^{\rm (ini)}$ is incorporated by perturbing the initial particle masses (using Eq.\,\ref{eq:pertmass}), instead of adding initial displacements. This result is shown in the lower panel of Figure~\ref{fig:delta_r_nbody} in solid lines with the same particle load and resolution as above, while the results from the higher resolution simulation ($2\times1024^3$ particles, 3LPT with $z_{\rm start}=24$) are shown as faint dashed lines. In this case, $P_{\rm bc}$ is exactly constant on large scales and exhibits non-linear growth at intermediate scales, as expected.  On the smallest scales, we observe that in all cases the solution asymptotes to a constant power spectrum for $k\gtrsim k_{\rm Ny}$ with a scaling inversely proportional to the particle number \citep[as is expected for shot noise in the power spectrum, e.g.][]{Colombi:2009}. We expect that this behaviour close to the particle Nyquist wavenumber is thus a combination of residual discreteness errors and shot noise contributions to the measured power spectrum. Note that we have not corrected in any way for shot noise.

Our results compare favourably with those of \cite{Bird:2020}, who appear to find a stronger discrete evolution on small scales in their mixed glass+SC approach (cf.\ their Figure~5). Similarly, agreement on large scales can also be achieved using adaptive softening for baryons \citep[e.g.][]{Angulo:2013}, which however leads to an artificial suppression of non-linear growth on small scales.

\begin{figure}
	\centering
	\includegraphics[type=pdf, ext=.pdf, read=.pdf, width=\columnwidth]{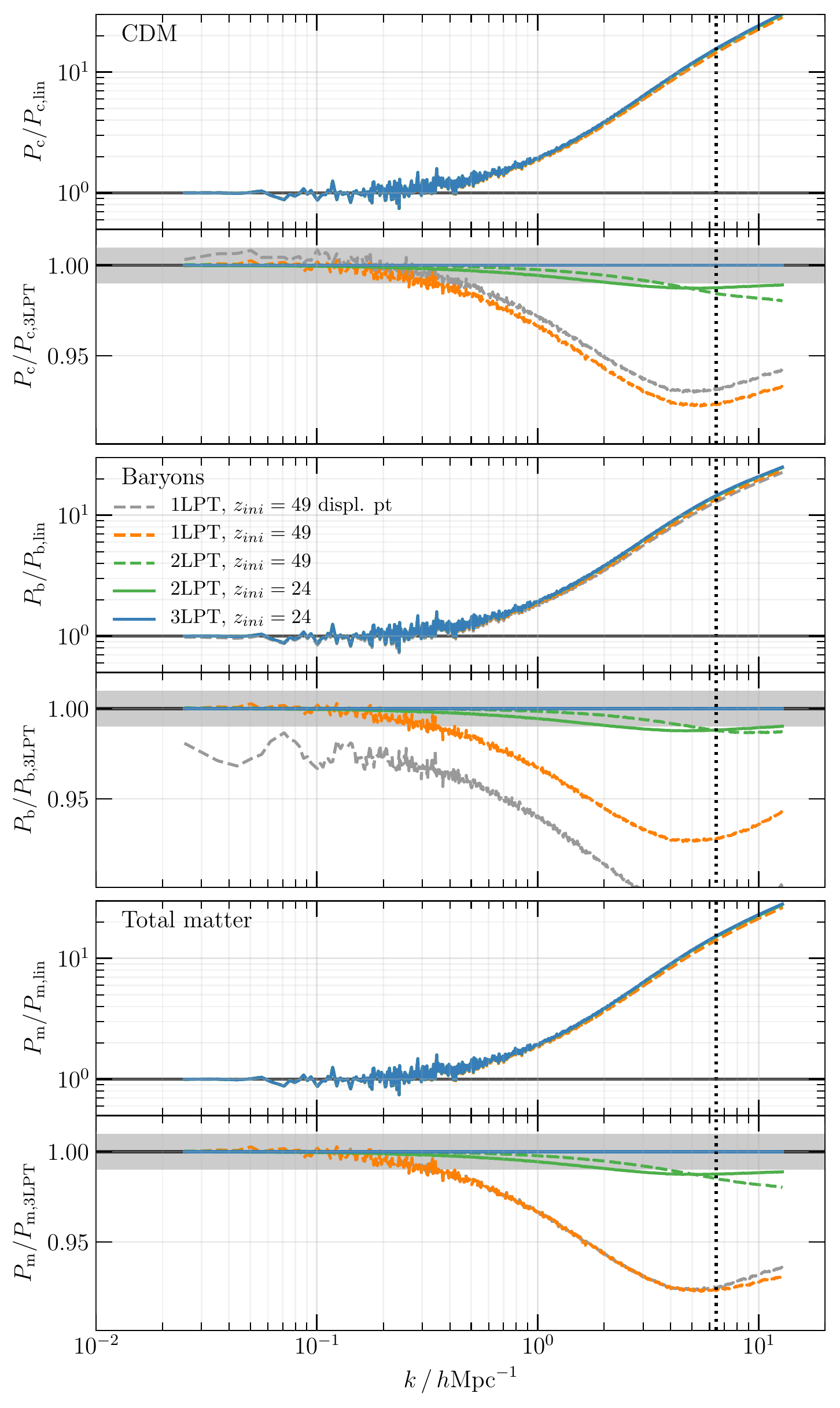}
	
	\caption{CDM, baryon and total matter power spectra at $z=2.125$ for different orders of LPT and starting redshifts in gravity-only simulations. The panels show the CDM power spectrum relative to the linear {\sc CLASS} solution (top) and relative to a reference run using 3LPT and $z_{\rm start}=24$ (second from top), as well as the same for the baryon power spectrum (third and fourth from top), and for the total matter power spectrum (bottom two panels). Line styles represent different combinations of LPT and starting time: 1LPT (orange), 2LPT (green), 3LPT (blue), $z_{\rm start}=24$ (solid) and $z_{\rm start}=49$ (dashed). All simulations use perturbed masses. For comparison we also show the result of a simulation using displacement perturbations in gray dashed. The vertical dotted line indicates the particle Nyquist wave number, and the shaded area indicates a one per cent deviation.}
	\label{fig:power_convergence_nbody}
\end{figure}

\subsection{Impact of the order of perturbation theory} 
In a next step, we test the impact of the order of $n$LPT employed in setting up the ICs on each of the fluid components and on the combined total matter field. To this end, we ran simulations initialized with 1LPT and 2LPT at $z_{\rm start}=49$, and using 2LPT and 3LPT at $z_{\rm start}=24$ using perturbed masses. In addition, we also use one run with perturbed displacements instead of masses, initialized with 1LPT at $z_{\rm start}=49$. We show the results at $z=2.125$, a time of relevance for Lyman-$\alpha$ studies. At our resolution, this also coincides with the onset of stronger non-linear features in the power spectrum, so that one is still probing also the perturbative regime here. The results of this study are shown in Figure~\ref{fig:power_convergence_nbody} for the power spectrum $P_{\rm c}$ of CDM, $P_{\rm b}$ of baryon, and $P_{\rm m}$ of the total matter perturbations (top to bottom panels). Each panel is subdivided in two, showing the respective power spectrum amplitude relative to the linear {\sc Class} prediction, and relative to a reference $N$-body simulation. We use the 3LPT, $z_{\rm start}=24$ run as the reference here.

 As already discussed in the previous section, without large softening, the run that uses perturbed displacements to absorb the compensated density perturbation (grey lines) shows the wrong growth in each fluid component. The deviation is larger in the baryons than in CDM (arguably due to the particle mass difference), however there is no error in the total matter spectrum.  In the case of perturbed masses, for each component, essentially the same discreteness errors arise as in the single fluid case. The results we find for the runs with perturbed masses are consistent with those of \cite{Michaux:2020} for the single fluid case. 
Essentially, starting later with higher-order LPT is preferable, but for our set-up all runs that use at least 2LPT show errors at less than two per cent level at all wave numbers up to the particle Nyquist wave number. In contrast, 1LPT is, as already shown 15 years ago by e.g. \cite{Crocce:2006}, not particularly accurate. As \cite{Michaux:2020} have argued, in principle lower order LPT can be rectified by earlier starts, but then quickly discreteness errors become dominant over LPT-truncation errors, so that this is practically not an option when the goal is to be economic, i.e., to push to the highest wave numbers with the lowest possible number of particles. These discreteness errors can of course be suppressed by resorting to a larger force softening, which however also comes at the price of suppressing power on small scales since the force there is no longer Newtonian.

\begin{figure}
	\centering
	\includegraphics[type=pdf, ext=.pdf, read=.pdf, width=\columnwidth]{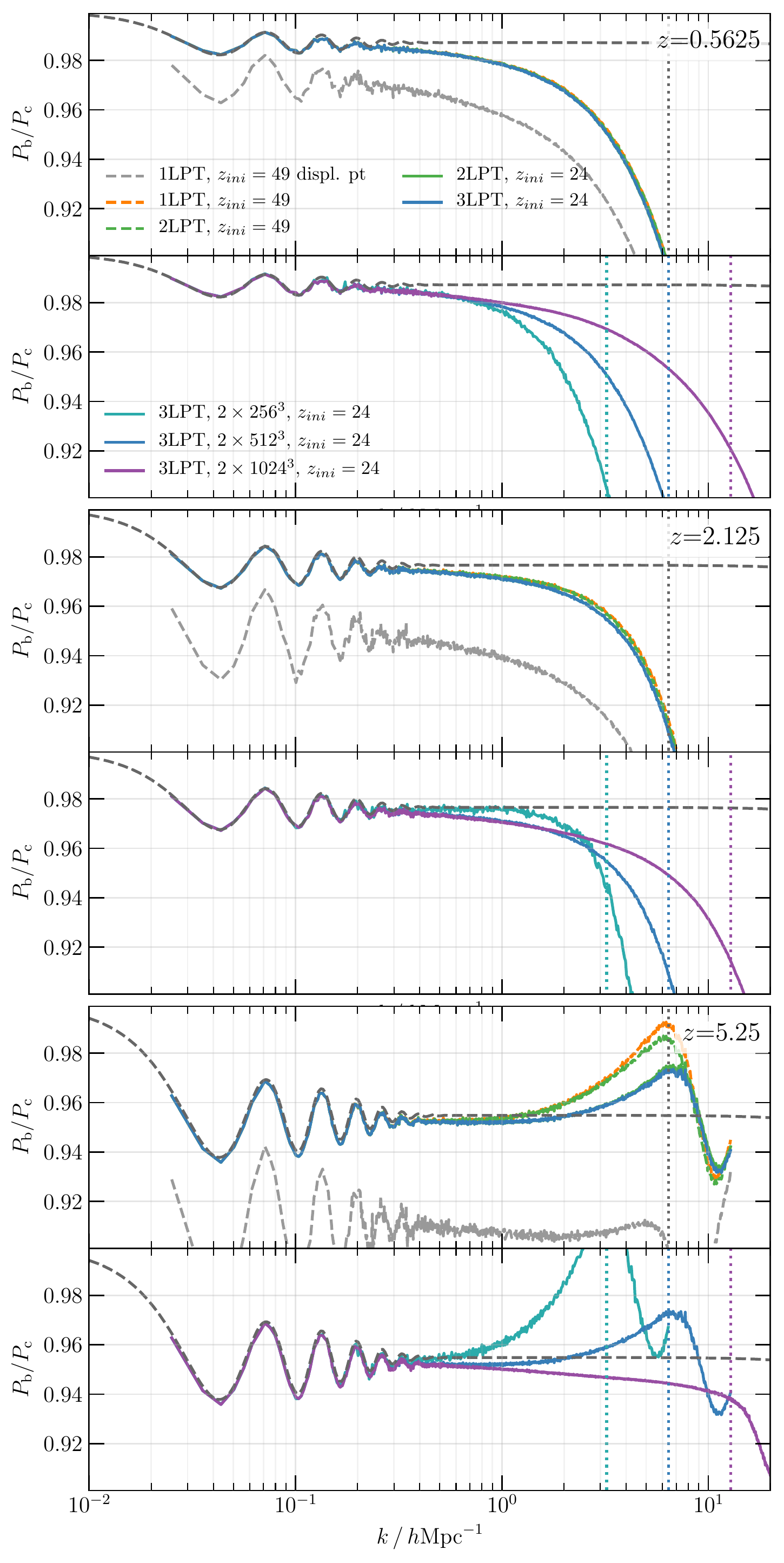}
	
	\caption{Purely gravitational evolution of the relative baryon-CDM power $P_{\rm b}(k) / P_{\rm c}(k)$ from $z= 5.25$ (bottom two panels) to $z= 2.125$ (middle panels) to $z=0.5625$ (top panels). For each redshift, we show the dependence on the order of LPT and starting redshift in the upper panels (colour and line styles have same meaning as in Figure~\ref{fig:power_convergence_nbody}, and the dependence on the particle resolution in the lower panels (light blue corresponding to $2\times 256^3$, blue to $2\times 512^3$, purple to $2\times 1024^3$). The dark grey dashed line indicates the {\sc Class} prediction. The dotted vertical lines indicate the particle Nyquist wave number. }
	\label{fig:power_ratio_evolution_nbody}
\end{figure}

\subsection{Convergence of the baryon-CDM ratio and impact of the decaying relative velocity mode} 
\label{sec:convergence_ratio_bc_power}
Finally, we also study the ratio of CDM to baryon power spectra, $P_{\rm b} / P_{\rm c}$, as a function of the order of LPT, starting time, and numerical resolution. We present the results in Figure~\ref{fig:power_ratio_evolution_nbody} at redshifts $z=0.5625$, $z=2.125$ and $z=5.25$ (top to bottom panels). Each panel is again sub-divided in two, the upper one showing the effect of variations of the order of LPT and starting time $z_{\rm start}$ of the simulation, the lower showing the effect of varying the number of particles used in the simulation. Starting our discussion with the earliest time $z=5.25$, one notices a pronounced peak at the particle Nyquist wave number whose amplitude strongly depends on the starting time of the simulation, and the resolution, but also to a much lesser degree on the order of LPT used. It is clear that this peak arises as a consequence of discreteness errors during the early phases of the simulation. Notably, starting the simulation at $z=24$ instead of $z=49$ reduces this discrete error by almost a factor of two at fixed order of LPT (2LPT in this case). The improvement of 3LPT over 2LPT is minor compared to the discreteness error, but it would be expected that one could push for a later start of the simulation with 3LPT, thus further reducing this specific error \citep[cf. also][]{Michaux:2020}. 

At intermediate redshift, $z=2.125$, and the resolution we consider here, the pronounced peak at the particle Nyquist wave number has disappeared and a physical suppression of $P_{\rm b}/P_{\rm c}$ due to non-linearities becomes visible at small scales. This non-linear suppression is slightly, but visibly, stronger for higher-order LPT, for which one expects more accurate non-linear growth. The difference between the different resolutions at fixed order of LPT is much less pronounced than at higher redshift, but it is clear that the low resolution run does not capture the non-linear suppression yet at this redshift. Note that this suppression has been predicted in perturbation theory by \cite{Somogyi:2010} and has also previously been measured in exactly this type of collisionless two-fluid simulations by \cite{Angulo:2013} (who however used the ``forward'' approach to set up their ICs). The suppression essentially means that fluctuations in the baryon fraction become locked once a given scale collapses. 

Finally, at the lowest redshift we consider, $z=0.5625$, the situation is similar as at $z\sim2$. The differences due to the order of LPT have further decreased, as was already reported by \cite{Michaux:2020} -- essentially non-linearities transport power from larger to smaller scales, and differences in LPT are always smaller at larger scales. The resolution of the simulation still plays an important role in setting the suppression of the power spectrum ratio on small scales. It appears as if at this late time, the suppression is converged at scales $\lesssim k_{\rm Ny}/3$.

\begin{figure}
	\centering
	\includegraphics[type=pdf, ext=.pdf, read=.pdf, width=\columnwidth]{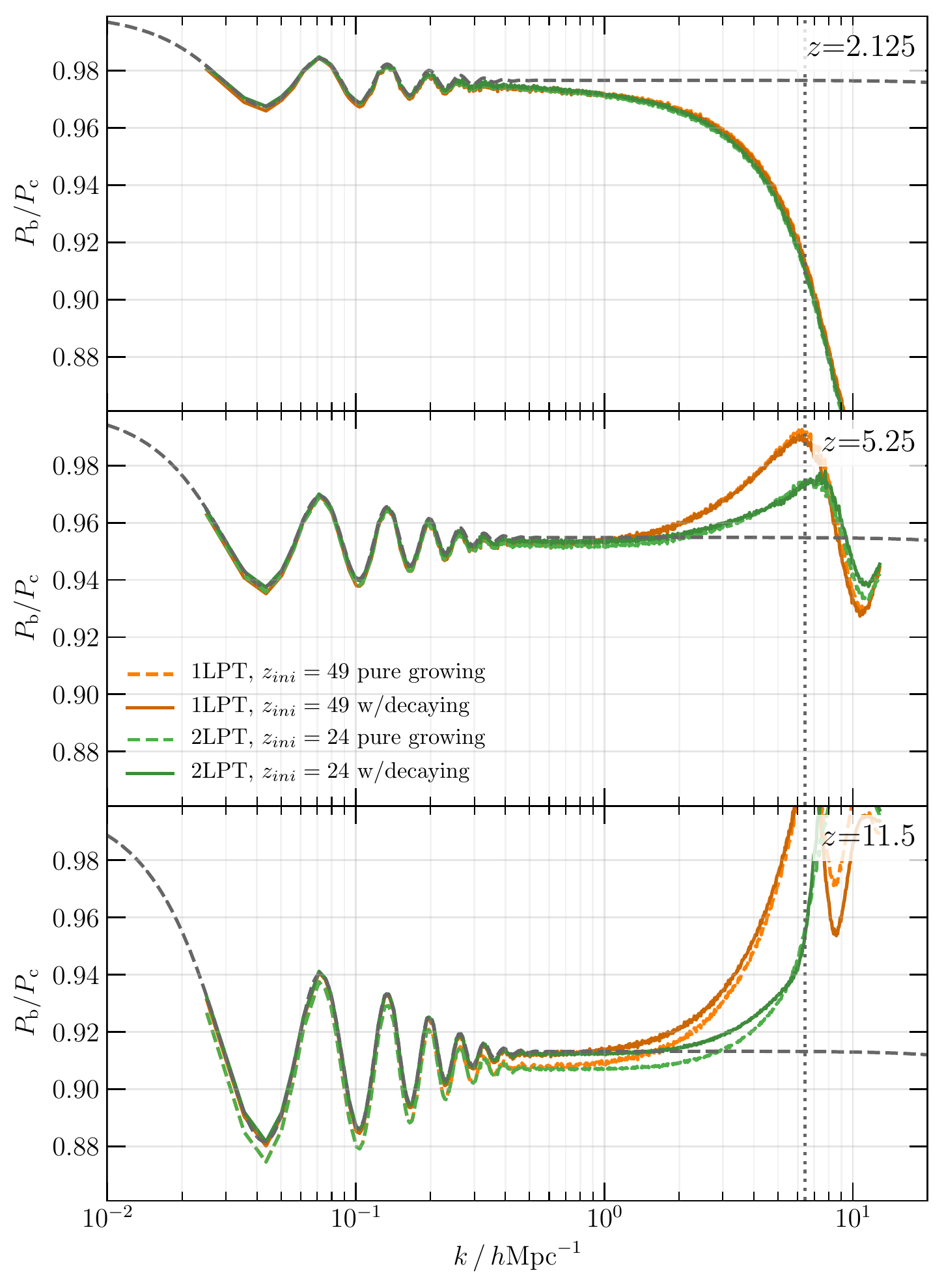}
	
	\caption{Effect of the decaying relative velocity mode on the purely gravitational evolution of the relative baryon-CDM power $P_{\rm b}(k) / P_{\rm c}(k)$ from $z= 11.5$ (bottom panel) to $z=5.25$ (middle panel) to $z= 2.125$ (top panel). For each redshift, we show lines obtained with 1LPT and $z_{\rm start}=49$ (orange) and with 2LPT and $z_{\rm start}=24$ (green) for ICs that include growing modes only (dashed, brighter hues) and that also include the relative velocity decaying mode (solid, darker hues). The grey dashed line indicates the {\sc Class} prediction. The dotted vertical lines indicate the particle Nyquist wave number. Including the relative velocity mode improves the agreement at high redshift but has only sub-percent effect on the power spectrum on these scales at low $z$.}
	\label{fig:power_ratio_wdecay}
\end{figure}
 
Last but not least, we also investigate the impact of including the linear decaying relative-velocity mode, as discussed in Section~\ref{sec:decaying_mode}. The impact on the ratio of baryon to CDM power spectra between redshifts 11.5 and 2.125 is shown in Figure~\ref{fig:power_ratio_wdecay}. We show the ratio of baryon to CDM power spectra with and without the inclusion of the relative velocity mode at linear order. It is obvious that the inclusion of the mode clearly improves the agreement at high redshift ($z\gtrsim5$). At late times ($z\lesssim5$), the effect is however sub-per-cent on the scales we investigate here. Naturally, the relative velocity mode will have a much stronger effect if scales smaller than those probed here are investigated, where the baryon streaming can have a significant effect on the growth of structures. Since the inclusion of the decaying mode is only carried along at first order, one expects however a strong dependence on the starting redshift since non-linear effects due to the relative velocity mode are not captured in the ICs. In order to improve agreement with the linear evolution, it should however be included when possible (i.e., for particle simulations, since it is not clear yet how to include this mode in the PPT framework).



\section{Results II: Mixed CDM + Baryon Simulations}
\label{sec:two-fluid-full-evolution}
After having considered the collisionless, purely gravitational evolution in the previous section, we now turn to fully hydrodynamic simulations using the Eulerian code {\sc Ramses} and the moving mesh {\sc Arepo} codes. The simulations analysed in this section are listed in Table~\ref{tab:sims}. We first present the evolution of density power spectra and compare the results between the two codes and with the purely gravitational evolution. We then analyse in detail the improvements brought about by higher-order PT, which is known to be more prominent in higher-order correlations \citep[e.g.][]{Munshi:1994}.

\begin{figure*}
	\centering
	\includegraphics[type=pdf, ext=.pdf, read=.pdf, width=\columnwidth]{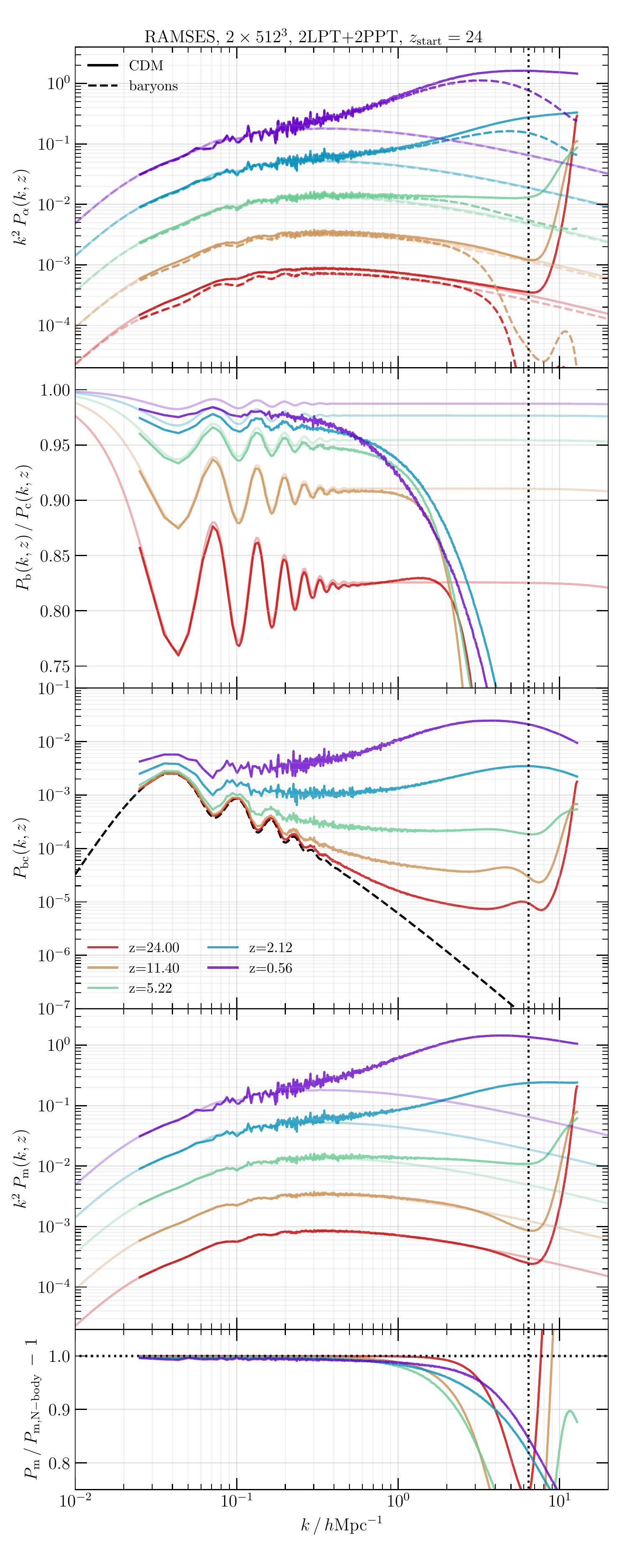}
	\includegraphics[type=pdf, ext=.pdf, read=.pdf, width=\columnwidth]{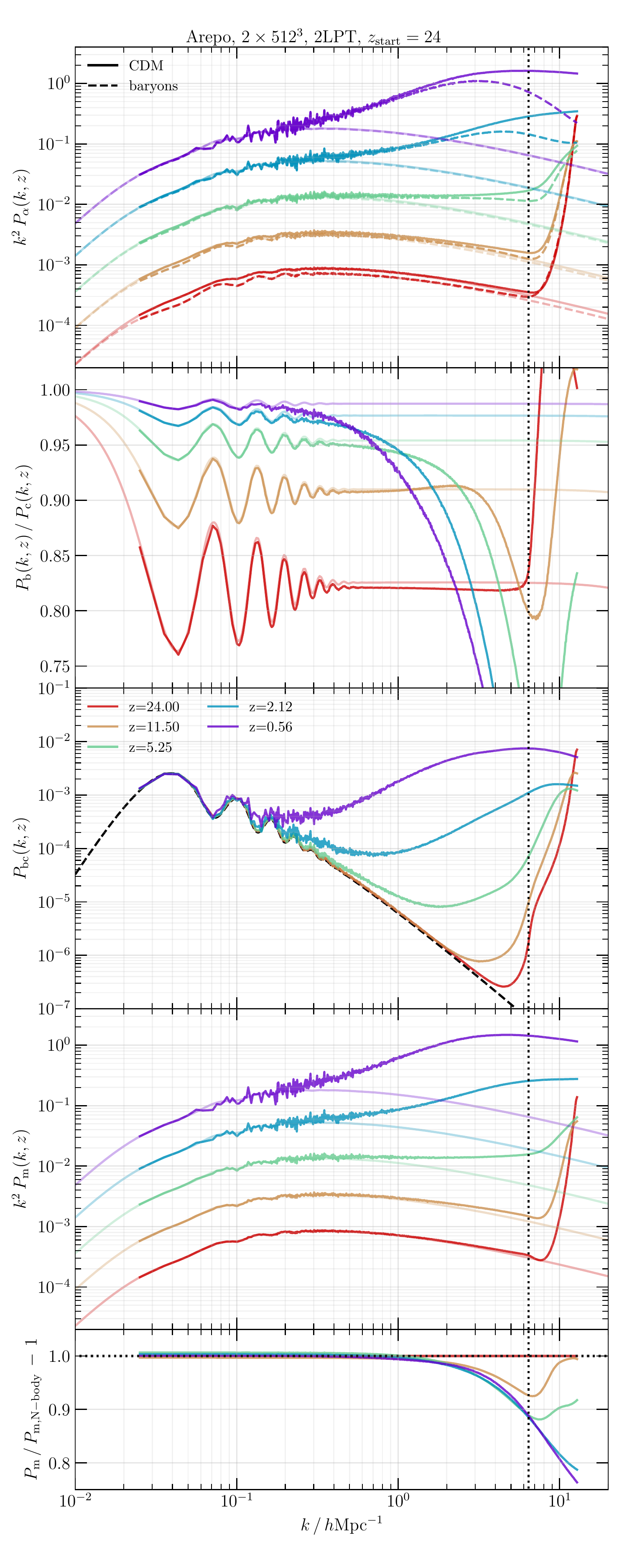}
	\caption{Comparison of the evolution of a mixed CDM+baryon fluid in two commonly used cosmological hydrodynamical codes: {\sc Ramses} (left panels) and {\sc Arepo} (right panels). From top to bottom, the figure shows the evolution of the individual baryon and CDM matter power spectra, $P_{\rm b}$ and $P_{\rm c}$,  the ratio of these two spectra, $P_{\rm b}/P_{\rm c}$, the spectrum of the difference, $P_{\rm bc}$, the total matter power spectrum $P_{\rm m}$, and the ratio of the simulation matter power spectrum to that from a two-fluid collisionless $N$-body run. Light lines indicate the linear `growing-mode' PT results at the precise output times of the snapshots. The vertical dotted line indicates the particle Nyquist wave number. Note that the initial suppression of baryon power on small scales in the {\sc Ramses} simulation is due to the PPT approach (cf. Fig.~\ref{fig:power_ppt}).}
	\label{fig:stats_ramses_arepo}
\end{figure*}


\subsection{Power spectrum evolution}
\paragraph*{Evolution of component spectra.} \ In Figure~\ref{fig:stats_ramses_arepo}, we show the evolution of various density power spectra obtained with the {\sc Ramses} (left panels) and {\sc Arepo} (right panels) codes for initial conditions for the same $250\,h^{-1}{\rm Mpc}$ box as in the previous section. Both simulations use the same initial number of resolution elements. The top-most panels show the evolution of the component, i.e. baryon and CDM, power spectra $P_\alpha(k)$ between the initial time $z=24$ and the final output we considered at $z\simeq0.56$, followed by the evolution of the ratio $P_\rb/P_\rc$ in the second panel from top. We observe that at the initial time $z=24$, the 2PPT initial conditions used for the baryons with {\sc Ramses} are slightly smoother than the Lagrangian field used for {\sc Arepo}, with a suppression of baryon power on scales of about 2-3 top grid cells (see also Figure~\ref{fig:power_ppt} and the related discussion, where we showed how this depends on starting time and resolution). This suppression persists during the quasi-linear stages of the evolution, but we find that once non-linear structure has formed at $z\lesssim 2.5$, the non-linear spectra obtained with {\sc Ramses} and {\sc Arepo} agree very well. In the power ratio $P_{\rm b}/P_{\rm c}$, we observe about 1 per cent deviation of the {\sc Ramses} results from linear theory on large scales, while {\sc Arepo} follows the linear theory perfectly on large scales. This is arguably due to advection errors in {\sc Ramses} causing a slight diffusion that affects even large scales, but note that this error is at the per cent level only. 

\paragraph*{Evolution of the difference spectrum.} \ As expected in the presence of collisional processes, the power spectrum of the baryon-CDM density difference  $P_{\rm bc}$, which is shown in the third panel from the top in Figure~\ref{fig:stats_ramses_arepo}, shows a larger amplitude on small scales compared to the evolution in the collisionless simulation shown in Figure~\ref{fig:delta_r_nbody}. Collisional processes, leading to an isotropic pressure rather than anisotropic stress after shell-crossing \citep[cf.][]{Buehlmann:2019}, as well as entropy production act to decouple the baryon evolution from the collisionless CDM evolution. Similarly to the collisionless simulations, numerical errors are also particularly pronounced in $P_{\rm bc}$, the initial 2-3 grid-scale power suppression in PPT/{\sc Ramses} at $z=24$ is visible as a deviation from  linear theory peaking at the root-grid Nyquist wave number. During quasi linear evolution, the overall level of error very close to the Nyquist scale is similar for both codes, but extending to larger scales in {\sc Ramses}. Also the per cent level growth error at late times compared to linear theory visible for {\sc Ramses} in the $P_{\rm b}/P_{\rm c}$ ratio is visible here as a late time spurious growth. 

\paragraph*{Total matter spectrum and baryon response.} \ Finally, in the second panel from the bottom of Figure~\ref{fig:stats_ramses_arepo}, we show the evolution of the total matter spectrum together with its relative deviation from the total matter power spectrum obtained from the purely collisionless two-fluid evolution. 
While on large scales, an effective pressure arising from shell-crossed shocks plays no role, and the collisional simulations agree perfectly  with the collisionless simulation (i.e., within expected numerical errors), we see that at late times $z\lesssim2.5$, both collisional codes predict a fairly rapid suppression in the total matter spectrum at scales $k\gtrsim 1\,h{\rm Mpc}^{-1}$. 
To compare the collisionless and collisional power spectra, which were output by the respective codes at slightly different snapshot times, we simply rescaled to the output times of {\sc Ramses} and {\sc Arepo} using the linear theory growth factor $D_+$. We note that, at this resolution (which is fairly low compared to state-of-the-art galaxy formation simulations), {\sc Ramses} predicts a slightly larger suppression $\sim 15$ per cent at $k_\smalltext{Ny}$ at $z=0.56$ compared to {\sc Arepo} with $\sim 10$ per cent at $k_\smalltext{Ny}$ which is possibly due to the smoother ICs and/or advection errors. The suppression shape predicted by the {\sc Arepo} simulation has a nearly universal shape across all redshifts. We caution that we expect that the precise evolution of the power suppression depends crucially on additional physics such as cooling, UV backgrounds and AGN feedback in more realistic simulations that attempt to model also astrophysical processes. In that sense it is somewhat surprising that the suppression we observe here is quantitatively not all too different at low $z$ from the range found across state-of-the-art galaxy formation simulations \citep[cf.][in particular their Figure~3]{Chisari:2019}. A suppression that is stronger or affects larger scales, as has been observed in some simulations and is included in recent baryon response models applied to collisionless simulations \citep[cf. e.g.][]{Huang:2019,Schneider:2019,Arico:2020} clearly requires substantial injection of energy into the baryons beyond just offsetting radiative cooling losses. Due to the absence of additional physics, the physical suppression scale in our simulations is set by gravity alone.


\begin{figure}
	\centering
	\includegraphics[type=pdf, ext=.pdf, read=.pdf, width=0.95\columnwidth]{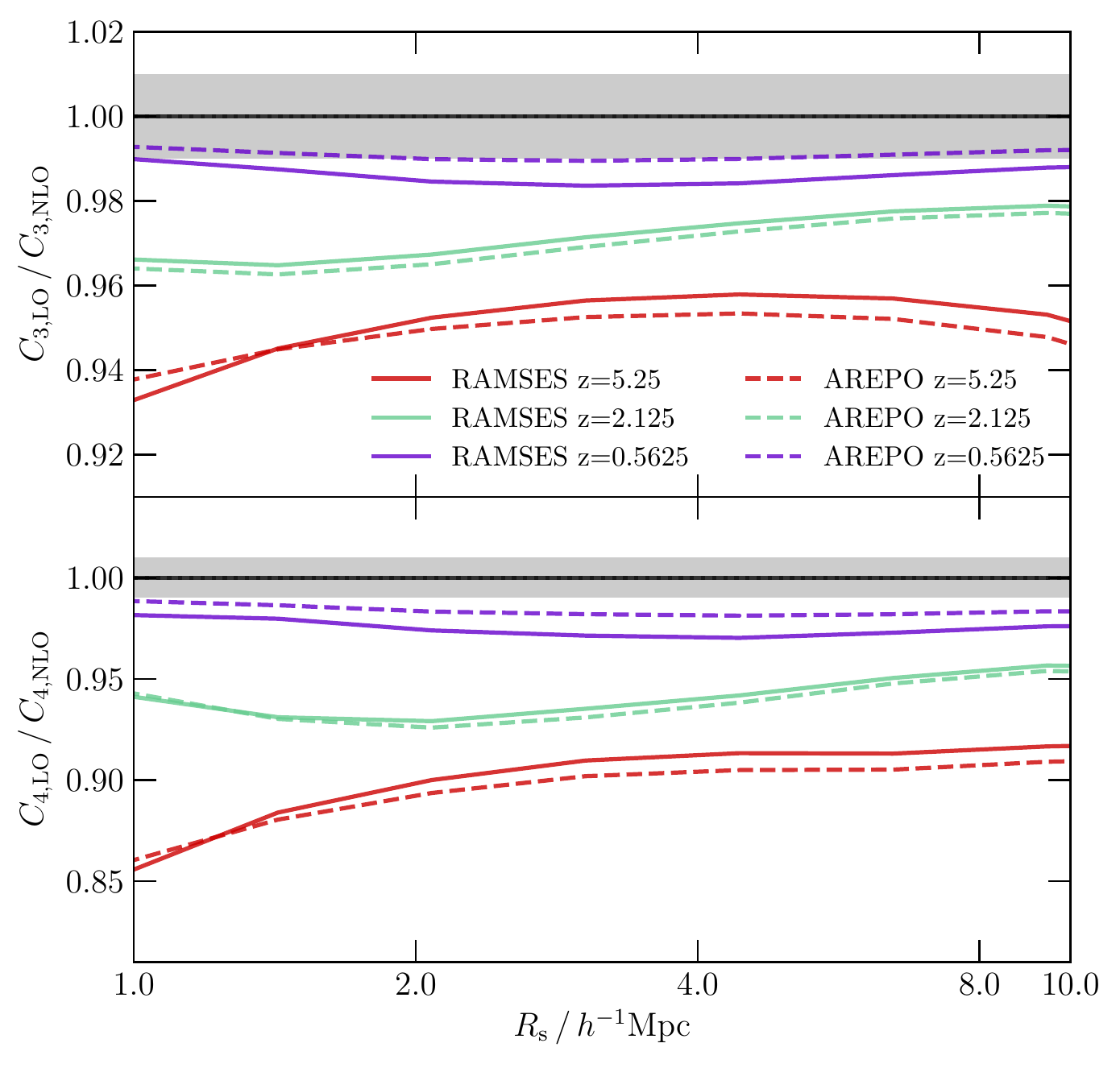}
	\caption{\label{fig:stats_cumulants} Dependence of the third and fourth cumulant, $C_3$ (top panel) and $C_4$ (bottom panel) of the baryon density field on the order of the PT used to set-up the initial conditions. We show the ratio of the cumulants obtained from LO simulations (i.e., 1LPT/1PPT, initialized at $z_{\rm start}=49$) to those obtained from NLO simulations (i.e., 2LPT/2PPT, initialized at $z_{\rm start}=24$) as a function of the scale of the applied top hat filter $R_\smalltext{TH}$. The results show the transient behaviour well known for collisionless simulations also for hydrodynamic simulations: lower order PT leads to an underestimation of non-Gaussianity, particularly at higher redshift. The improvement brought about by going from LO to NLO is similar at all scales for Lagrangian and Eulerian simulations.}
	\vspace{1cm}
	\includegraphics[type=pdf, ext=.pdf, read=.pdf, width=0.95\columnwidth]{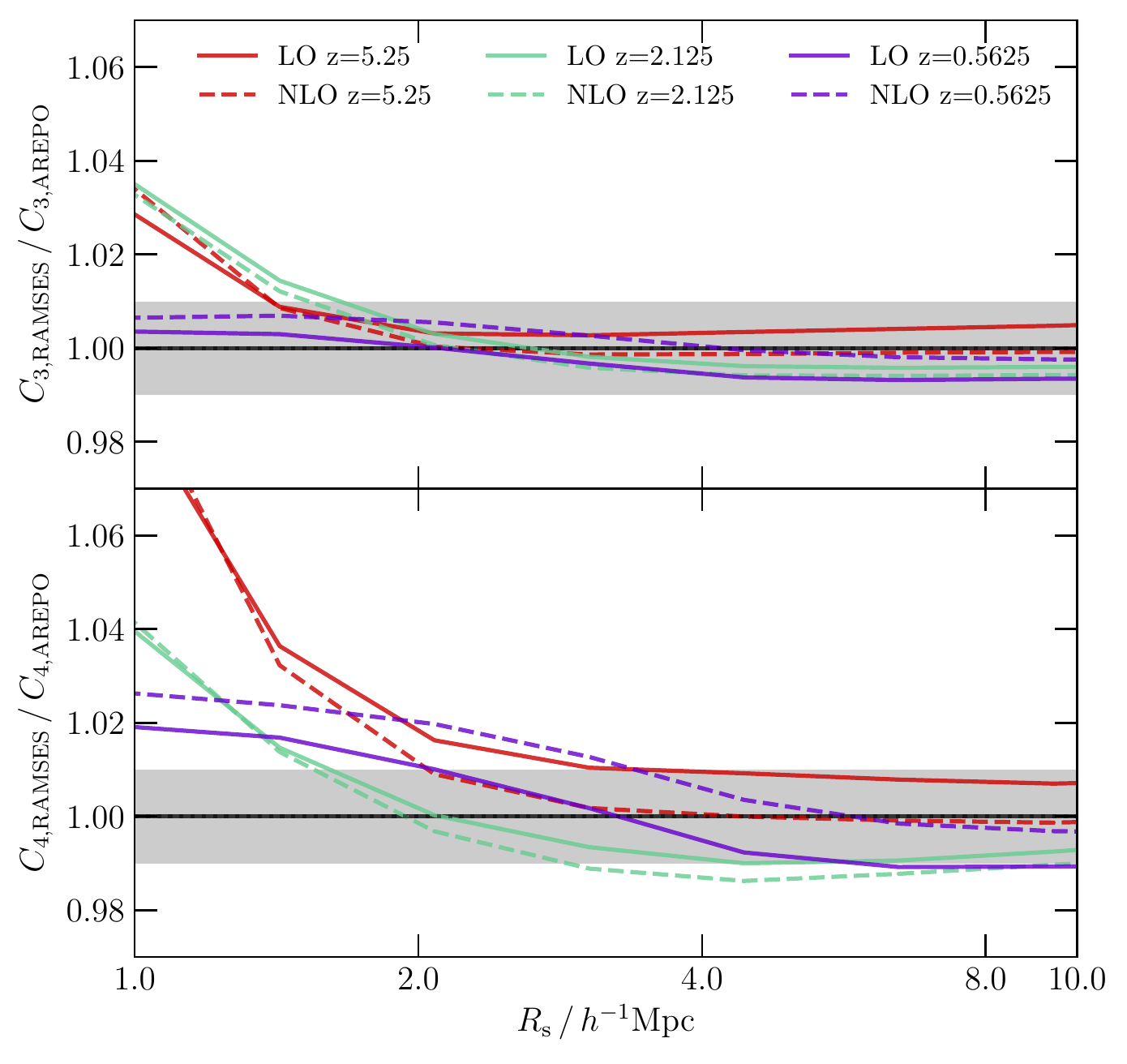}
	\caption{\label{fig:stats_cumulants_codes} Ratio of the third and fourth cumulant, $C_3$ (top panel) and $C_4$ (bottom panel) of the baryon density field between the {\sc Ramses} and the {\sc Arepo} simulations as a function of the scale of the applied top hat filter $R_{\rm TH}$. Results from the LO runs (1LPT/1PPT, $z_{\rm start}=49$) are shown as solid lines, those from the NLO runs (2LPT/2PPT, $z_{\rm start}=24$) are shown as dashed lines for three output times indicated by the different colours. Agreement between the two codes is at the few per cent level.  }
\end{figure}

\begin{figure*}
	\centering
	\includegraphics[type=pdf, ext=.pdf, read=.pdf, width=0.9\textwidth]{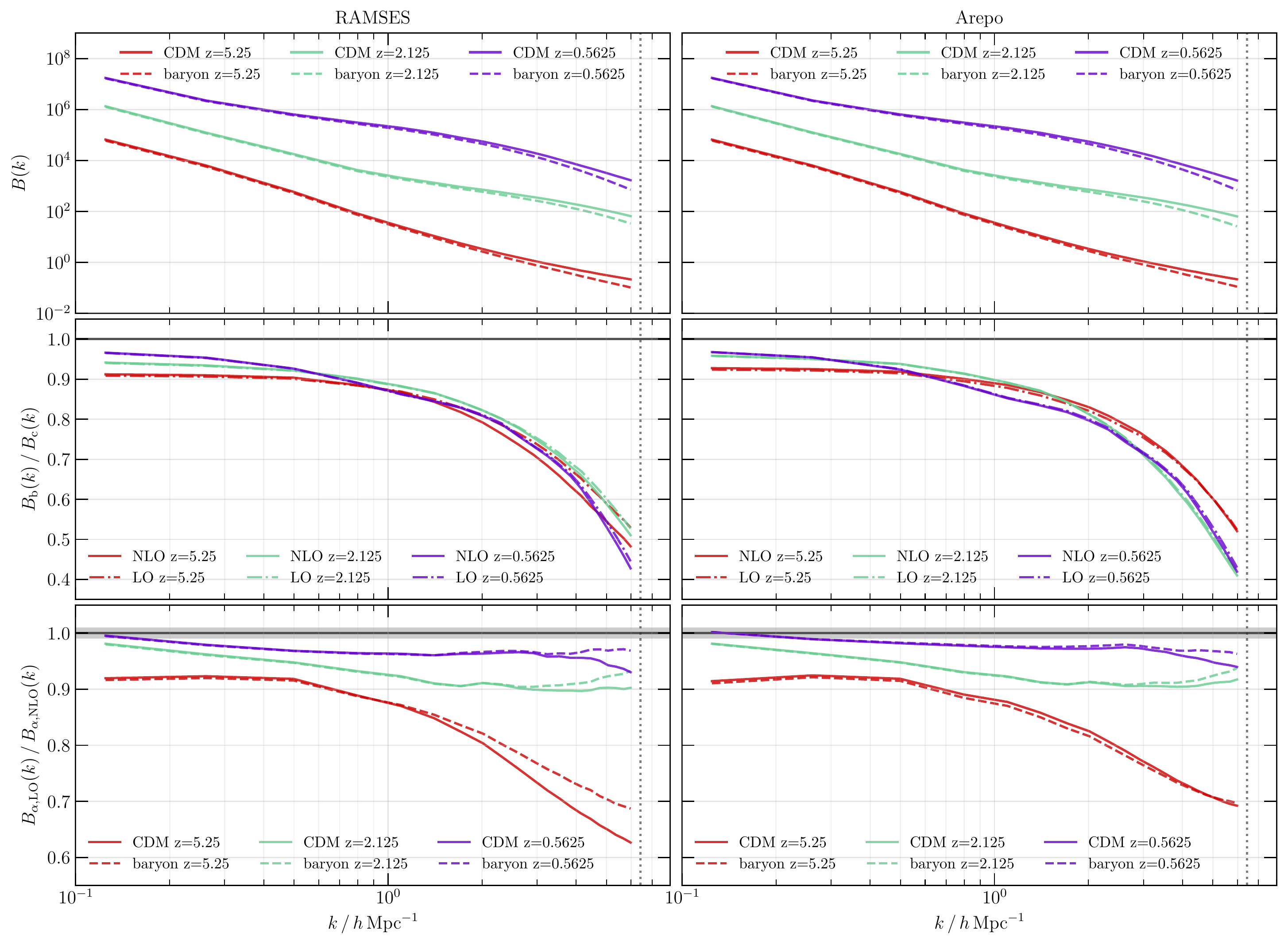}
	\caption{\label{fig:bispectrum} Equilateral bispectra of the baryon and CDM density fields for the {\sc Ramses} (left panels) and {\sc Arepo} (right panels) simulations. {\bf Top panels:} the bispectra at $z=5.25$, $2.125$ and $0.5625$ for the CDM (solid lines) and baryon (dashed lines) density fields. {\bf Middle panels:} Ratio between the baryon and CDM bispectra for the three different redshifts for the LO (i.e., 1LPT/1PPT, $z_{\rm start}=49$) and NLO (i.e., 2LPT/2PPT, $z_{\rm start}=24$) simulations. {\bf Bottom panels:} ratio of the LO and NLO bispectra for CDM (solid) and baryons (dashed) for the three redshifts. }
\end{figure*}

\begin{figure}
	\centering
\includegraphics[type=pdf, ext=.pdf, read=.pdf, width=0.95\columnwidth]{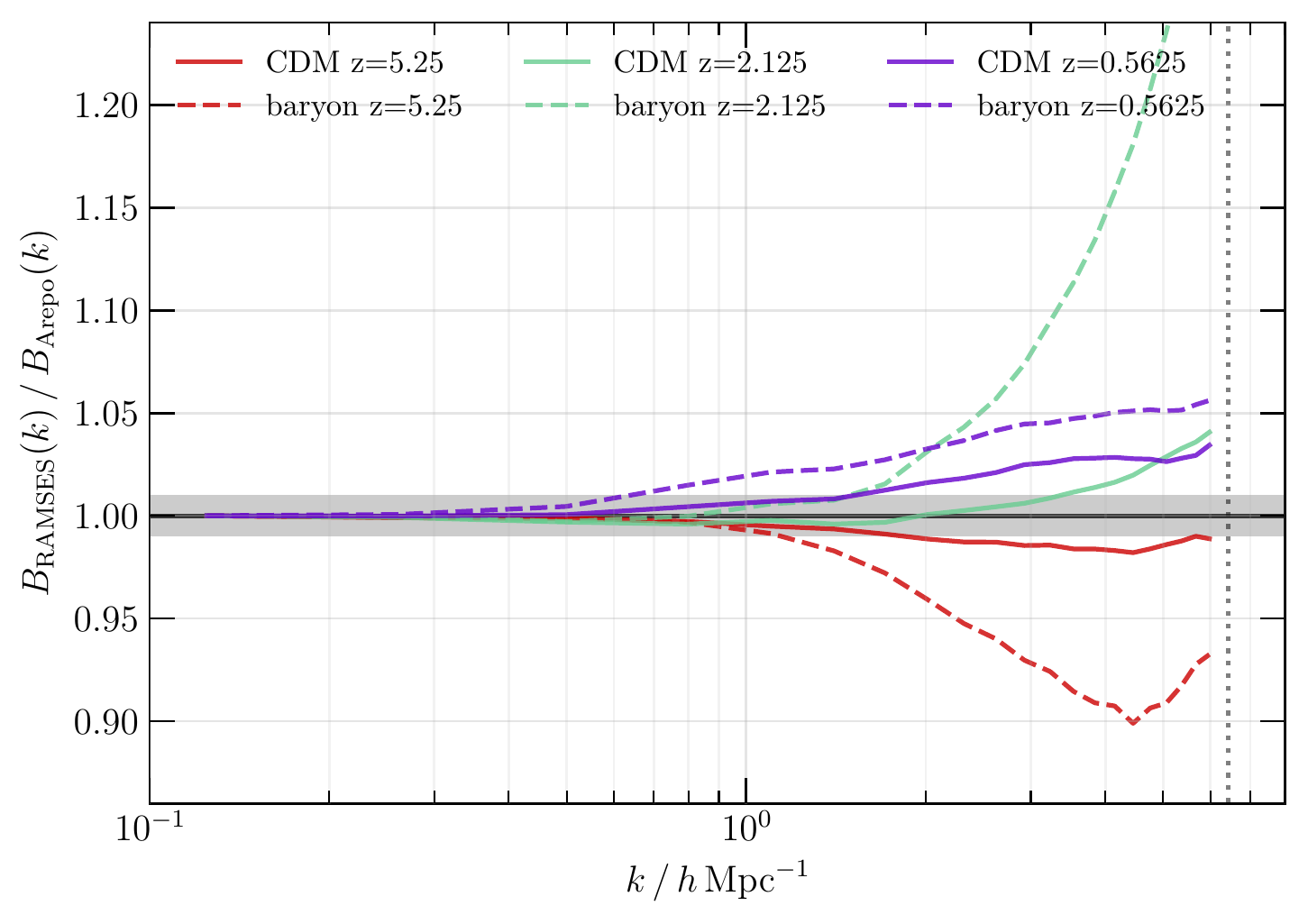}
	\caption{\label{fig:bispectrum_codes} Ratio of the equilateral bispectra of the baryon and CDM density fields between the {\sc Ramses} and the {\sc Arepo} simulations as a function of the triangle scale $k$. We show only results from the NLO simulations indicating a significantly boosted bispectrum amplitude on small scales in {\sc Ramses} compared to {\sc Arepo}.  Note that in this plot, we have re-normalized the respective amplitudes so that the ratio is unity in the smallest $k$-bin to suppress contribution from slightly different output times and growth in the two codes (cf.\ Fig.~\ref{fig:stats_cumulants_codes}).}
\end{figure}

\subsection{Evolution of non-Gaussianity -- cumulants and bispectrum}
\label{sec:evolution_nongaussian}
Finally, we quantify the evolution of non-Gaussianity due to gravitational instability in the density fields. It is well known that convergence in higher-order statistics depends sensitively on both the starting time of simulation and the truncation order in the PT expansion \citep[e.g.][]{Munshi:1994,Scoccimarro:1998,Crocce:2006,Michaux:2020}. Due to the absence of higher-order PT schemes for baryon-CDM simulations, such tests have not been made in the two-fluid case to our knowledge. Here, we specifically consider two summary statistics: (1) the third and fourth cumulants (i.e., skewness and kurtosis) of the baryon density field as a function of filtering scale covering the range between the non-linear and the linear scales of our simulations, and (2) the equilateral bispectrum, i.e., the harmonic version of the three-point correlation function, of the baryon and the CDM density field.

\paragraph*{Cumulant statistics of the baryon density field.} \ To quantify the improvement brought about by going to higher order PT when generating initial conditions, we first study the influence of first vs. second order PT on the one-point statistics of the smoothed baryon density field. Specifically, we investigate ratios of the $C_3$ and $C_4$ cumulants, as defined in Eq.\,\eqref{eq:cumulants}, between the LO runs (i.e., using either  1LPT or 1PPT with a starting time of $z_{\rm start}=49$) and the NLO runs (which use 2LPT/2PPT with a starting time of $z_{\rm start}=24$). Note that we vary both the starting time and the order since the NLO start is too late for first order PT. An even earlier start might improve agreement, but usually comes at the cost of larger numerical errors in the solution
(see e.g. \citealt{Michaux:2020}). 

The results are shown for $C_3$ in the top panel, and for $C_4$ in the bottom panel of Figure~\ref{fig:stats_cumulants} for simulations run with both {\sc Ramses} (solid lines) and {\sc Arepo} (dashed lines). In agreement with previous studies for the total matter density field (e.g., \citealt{Crocce:2006} see in particular their Fig.~5), we find that first-order ICs underestimate the degree of non-Gaussianity also in the baryon field. Errors are systematically larger at higher redshift ($z\gtrsim2$), approaching more than 10 per cent on all scales at $z\gtrsim 5$ for our simulation set-up. The overall improvement brought about by going from LO to NLO is virtually identical (within better than one per cent) for both {\sc Ramses} and {\sc Arepo}, which is a clear validation of our PPT approach for Eulerian finite volume methods. It is particularly interesting to see that advection errors of the Eulerian code, that are relatively prominent in the power spectrum on small scales, are not reflected here.

To compare more accurately the results obtained with the two codes, we show in Figure~\ref{fig:stats_cumulants_codes} explicitly the ratio of the cumulants measured in the {\sc Ramses} and the {\sc Arepo} simulations. While we find that both $C_3$ and $C_4$ agree when smoothed on large scales to about one per cent, {\sc Ramses} shows a consistently larger amount of non-Gaussianity on small scales $R\lesssim 2h^{-1}{\rm Mpc}$. One can speculate that on small scales particle noise or poor sampling in underdense regions in the Lagrangian code could impact these results, and/or that advection errors in the Eulerian code could lead to decreased variance, while possibly higher order cumulants are better retained, so that these normalized cumulants appear boosted. While we only note this systematic discrepancy here, it is certainly worthwhile to investigate its origin and detailed dependence on resolution and/or code parameters in future work in light of precision predictions of the Lyman-$\alpha$ forest.


\paragraph*{Bispectrum evolution in the two-fluid system.} \ Finally, we analyse the evolution of the baryon and CDM density bispectrum, and its dependence on PT order and simulation code. We focus here for simplicity  only on the equilateral bispectrum. Our main results are shown in Figure~\ref{fig:bispectrum}, which presents results for the NLO simulations obtained with {\sc Ramses} in the left panels, and for {\sc Arepo} in the right panels. The CDM (solid line) and baryon (dashed line) bispectra are shown for our usual three redshifts in the top most panel. Similarly as for the baryon and CDM power spectra, one observes that the baryon bispectrum is suppressed relative to the CDM one, particularly so on small scales.  To make this more explicit, the middle panels of the figure show the ratio of the baryon to CDM bispectra, $B_{\rm b}(k)/B_{\rm c}(k)$  revealing a relatively time independent suppression comparable in amplitude to the suppression in the power spectrum (cf. second panels from top in Figure~\ref{fig:stats_ramses_arepo}).  Note that for the power ratio $P_{\rm b}(k)/P_{\rm c}(k)$, we observed a significant evolution of the suppression in the {\sc Arepo} run, growing from smaller to larger scales over time. 

We also compare the bispectra obtained from the NLO IC runs against those with only LO (first order) initial conditions in the bottom panels of Figure~\ref{fig:bispectrum}. The impact on the component bispectra is as one would expect similar to the single fluid case, where one observes also a strongly suppressed bispectrum at high redshifts for first order ICs except when very early starting times are used to initialize the simulations \citep[e.g.][]{Crocce:2006,McCullagh:2015} -- the well-known transient. The bispectra from early start and with low-order ICs are however more impacted by the accumulation of discreteness errors \citep{Michaux:2020}, which one would expect to be even more dramatic in the baryon component if a diffusive Eulerian scheme is used. Since we do not disentangle starting redshift and PT order, discreteness and truncation errors are somewhat convoluted here. Comparing the {\sc Arepo} and {\sc Ramses} results for NLO vs.\ LO ICs, one can still draw a few interesting conclusions (and we leave more thorough investigations to future work): The LO/NLO bispectrum ratios for baryons are very similar for {\sc Arepo} and {\sc Ramses}, on all scales, showing that numerical diffusion due to the longer integration time in LO vs.\ NLO is not important. In the particle component, we see however a stronger suppression in {\sc Ramses}, which could be a consequence of the effectively lower force resolution at early times due to the AMR scheme. At the same time, particle noise might induce spurious effects. Note that in the absence of exact solutions or at least a full convergence study, all such conclusions are speculative. 

In summary, the excellent agreement between the improvement between LO and NLO for both 2LPT and 2PPT ICs clearly validates the PPT approach for higher-order ICs for baryons for Eulerian codes. \cite{Foreman:2019} have previously studied the baryon bispectrum in a full ``physics'' galaxy formation simulation including cooling and AGN feedback, however starting from ICs where baryons trace CDM perfectly. It will be interesting to compare these results with our adiabatic runs as well as with a more realistic astrophysical simulation that takes our new ICs into account.

As a last comparison, we show the explicit ratios of bispectra between the NLO IC simulations performed with {\sc Ramses} and with {\sc Arepo}. Due to the slight difference in snapshot times, we divide all bispectra by the value in the first $k$-bin. The result is shown in Figure~\ref{fig:bispectrum_codes}. We caution that in order to establish which results are converged, one would have to conduct a rigorous resolution test. Here, we are more interested at the level of typical differences due to the different methods. We find for the earliest snapshot, at $z=5.25$, a $\sim2$ per cent suppression of the CDM bispectrum close to the Nyquist wave number in the {\sc Ramses} run compared to {\sc Arepo}, and a much larger suppression of the baryon bispectrum by up to 10 per cent close to the Nyquist wave number in the baryon bispectrum. This difference is consistent with a similar suppression visible also in the power spectrum and owed to an effectively smoother IC in the baryons and arguably also additional advection errors. What is more curious is that at late times, this is reversed, and we observe a higher amplitude in the bispectrum in the {\sc Ramses} run, particularly so for baryons, where the effect is very significant at intermediate redshifts $z\sim 2$. This behaviour is consistent with a similar difference in the cumulants when the baryon density field is smoothed on relatively small scales reported above (cf. Figure~\ref{fig:stats_cumulants_codes}) -- and we have already speculated about possible reasons there. Still, the results for the {\sc Ramses} and {\sc Arepo} appear consistent at the few per cent level, which is remarkable in light of the very different approaches to evolving baryons that these codes adopt. The often mentioned advection errors incurred by Eulerian schemes seem to have much less influence on higher order statistics.


\section{Summary and conclusions}
\label{sec:summary}
With the increasing precision of current and upcoming cosmological observations, the long standing problem of how to generate accurate initial conditions for cosmological simulations that model the distinct non-linear evolution of both CDM and baryons has become more pressing. In this paper, we present the numerical implementation of a novel approach to set up initial conditions for two-fluid cosmological simulations, and validate our implementation and its performance based on various summary statistics. In brief, our new approach
\begin{enumerate}
\item provides higher-order Lagrangian (`$n$LPT') ICs  for two gravitationally coupled fluids in the cold limit, by restricting to a generalization of the `growing-mode' solutions of standard LPT;
\item applies a field-theoretic approach to LPT to initialize Eulerian simulations using PPT \citep{Uhlemann:2019,Rampf:2020};
\item relies on backscaling the late-times input fields to initialization time (instead of a forward approach), thereby improving accuracy at low $z$, and having sub-per cent errors at $z\lesssim 24$ on most scales of interest; and
\item prevents the typical problematic excitation of spurious growth well known for two-fluid $N$-body systems, even when very small gravitational softening is used.
\end{enumerate}
The theoretical foundations are presented in the companion paper \cite{Rampf:2020}, and summarized in Sections~\ref{sec:eulerian_theory}, \ref{sec:Lagrange} and~\ref{sec:ppt} in this article. All methods are implemented in the IC generator {\sc Monofonic Music-2} (i.e., the single resolution, non-``zoom'' version of {\sc Music-2}), 
which we make publicly available\footnote{Available from \url{https://bitbucket.org/ohahn/monofonic}.}.

We validate the quality of our initial conditions in two steps, first by considering the purely gravitational collisionless evolution of the two-fluid system using the $N$-body method (specifically {\sc Gadget-2}, Section~\ref{sec:two-fluid-gravity-evolution}), and in a second step using two commonly used cosmological $N$-body + collisional hydrodynamics codes, specifically the Eulerian {\sc Ramses} code and the {\sc Arepo} moving mesh code (Section~\ref{sec:two-fluid-full-evolution}). The respective main results are as follows.

\paragraph*{Collisionless Simulations.} \ Using a suite of collisionless two-fluid simulations, including up to 3LPT ICs, our conclusions based on an extensive analysis of power spectra are that
\begin{enumerate}
\item  Erroneous growth due to discreteness, previously observed in two-fluid simulations, is absent when initial mass variations instead of displacement perturbations are used (see Fig.~\ref{fig:delta_r_nbody}). These mass variations  are small (per cent level), vary mostly on large scales, and are independent of the starting redshift.
\item Residual discreteness and truncation (LPT transient) errors in two-fluid systems are similar to those in single-fluid systems and confined to scales close to the particle Nyquist wave number \citep[cf.][]{Michaux:2020}.
\item Therefore, late starting times with high-order LPT yield the best accuracy (before shell-crossing), by optimizing the impact of perturbative truncation errors vs.\ discreteness errors. 
\end{enumerate}
Furthermore, we confirm a previously reported non-linear suppression in the baryon to CDM power ratio in collisionless simulations \citep{Somogyi:2010,Angulo:2013} also in the absence of different gravitational softening for baryons and CDM.

\paragraph*{Hydrodynamic Simulations.} \ For the fully hydrodynamic simulations, we use our novel PPT approach up to second order, to set up the baryon initial conditions on a regular mesh for the Eulerian finite volume code {\sc Ramses}, while {\sc Arepo} is initialized with our novel two-fluid LPT, identical to the collisionless simulations. The CDM $N$-body particles in {\sc Ramses} are of course also initialized using LPT. Using an analysis of CDM, baryon and total matter density power spectra, cumulants of the baryon density field, and CDM and baryon density bispectra, we validate the performance of both the $n$LPT and $n$PPT ICs between $z\sim 5-0.5$. Since the main purpose of this paper is to present and validate the numerical implementation of our novel PT approaches, our conclusions based on hydrodynamic simulations do no include a rigorous resolution study. Also a study of the impact of the numerous parameters of each code on the results is beyond the scope of this work. Several of the conclusions below should therefore be followed up with more rigorous convergence tests in future work. Our main findings based on the hydrodynamic simulations are
\begin{enumerate}
\item the PPT approach to set up Eulerian baryon ICs leads to a natural suppression of power on the smallest scales due to the finite effective `Jeans'-scale associated with the $\hbar$-parameter of this method (which is set mostly by resolution). In contrast, two-fluid LPT is perfectly cold and has no such scale.
\item Despite this initial suppression, the late time ($z\lesssim 2.5$) baryon evolution and all power spectra agree well (i.e., within a few per cent) between the Eulerian and the moving mesh runs.
\item The improvement brought about by second-order over first-order PT in setting up ICs is virtually identical for the LPT and the PPT initial conditions, as demonstrated by our study of higher-order cumulants and bispectra. The impact is similar to the improvements seen for single-fluid total-matter PT \citep{Crocce:2006}, and most important for higher-order statistics.
\item Finite pressure in our non-radiative two fluid simulations leads to a very similar suppression of the total matter spectrum compared to the collisionless simulations at late times, independent of the simulation code, and in broad agreement with previous results based on ``full-physics'' simulations \citep{Chisari:2019}.
\item We find some interesting differences in the amount of small-scale non-Gaussianity between the {\sc Ramses} and {\sc Arepo} simulations that possibly warrant further investigation.
\end{enumerate}

In conclusion, we presented the numerical implementation of the `growing mode' two-fluid LPT/PPT approach discussed in detail in the companion paper \cite{Rampf:2020}, and validated its performance for both Eulerian and Lagrangian hydrodynamics codes. We believe that the presented improvements are on par with the necessary increase in the precision of cosmological simulations, in particular when probing baryons at increasingly higher redshifts. 

\section*{Acknowledgements}

We thank our referee St\'ephane Colombi for various valuable suggestions that helped to improve the presentation of our results. We thank Micha\"el Michaux for his bispectrum scripts, Raul Angulo, Bruno Marcos, Fabian Schmidt, and Romain Teyssier for useful discussions, and Volker Springel for comments on an earlier version of the manuscript. O.H.\ acknowledges funding from the European Research Council (ERC) under the European Union's Horizon 2020 research and innovation programme, Grant agreement No.\ 679145 (COSMO-SIMS). C.R.\ is a Marie Sk\l odowska-Curie Fellow and acknowledges funding from the People Programme (Marie Curie Actions) of the European Union Horizon 2020 Programme under Grant Agreement No.\ 795707 (COSMO-BLOW-UP). This work was granted access to the HPC resources of TGCC/CINES under the allocation A0060410847 attributed by GENCI (Grand Equipement National de Calcul Intensif). We thank the authors of {\sc Ramses}, {\sc Arepo}, {\sc Class}, and {\sc BSkit} for making their software publicly available.

\section*{Data Availability}
The software implementing all described techniques to generate initial conditions is freely available at \url{https://bitbucket.org/ohahn/monofonic}. 
The data underlying this article will be shared on reasonable request to the corresponding author.



\bibliographystyle{mnras}

\begin{thebibliography}{}
\makeatletter
\relax
\def\mn@urlcharsother{\let\do\@makeother \do\$\do\&\do\#\do\^\do\_\do\%\do\~}
\def\mn@doi{\begingroup\mn@urlcharsother \@ifnextchar [ {\mn@doi@}
  {\mn@doi@[]}}
\def\mn@doi@[#1]#2{\def\@tempa{#1}\ifx\@tempa\@empty \href
  {http://dx.doi.org/#2} {doi:#2}\else \href {http://dx.doi.org/#2} {#1}\fi
  \endgroup}
\def\mn@eprint#1#2{\mn@eprint@#1:#2::\@nil}
\def\mn@eprint@arXiv#1{\href {http://arxiv.org/abs/#1} {{\tt arXiv:#1}}}
\def\mn@eprint@dblp#1{\href {http://dblp.uni-trier.de/rec/bibtex/#1.xml}
  {dblp:#1}}
\def\mn@eprint@#1:#2:#3:#4\@nil{\def\@tempa {#1}\def\@tempb {#2}\def\@tempc
  {#3}\ifx \@tempc \@empty \let \@tempc \@tempb \let \@tempb \@tempa \fi \ifx
  \@tempb \@empty \def\@tempb {arXiv}\fi \@ifundefined
  {mn@eprint@\@tempb}{\@tempb:\@tempc}{\expandafter \expandafter \csname
  mn@eprint@\@tempb\endcsname \expandafter{\@tempc}}}

\bibitem[\protect\citeauthoryear{{Ahn}}{{Ahn}}{2016}]{Ahn:2016}
{Ahn} K.,  2016, \mn@doi [\apj] {10.3847/0004-637X/830/2/68}, \href
  {https://ui.adsabs.harvard.edu/abs/2016ApJ...830...68A} {830, 68}

\bibitem[\protect\citeauthoryear{{Almgren}, {Bell}, {Lijewski}, {Luki{\'c}}  \&
  {Van Andel}}{{Almgren} et~al.}{2013}]{Almgren:2013}
{Almgren} A.~S.,  {Bell} J.~B.,  {Lijewski} M.~J.,  {Luki{\'c}} Z.,   {Van
  Andel} E.,  2013, \mn@doi [\apj] {10.1088/0004-637X/765/1/39}, \href
  {https://ui.adsabs.harvard.edu/abs/2013ApJ...765...39A} {765, 39}

\bibitem[\protect\citeauthoryear{{Angulo} \& {Pontzen}}{{Angulo} \&
  {Pontzen}}{2016}]{Angulo:2016}
{Angulo} R.~E.,  {Pontzen} A.,  2016, \mn@doi [\mnras] {10.1093/mnrasl/slw098},
  \href {https://ui.adsabs.harvard.edu/abs/2016MNRAS.462L...1A} {462, L1}

\bibitem[\protect\citeauthoryear{{Angulo}, {Hahn}  \& {Abel}}{{Angulo}
  et~al.}{2013}]{Angulo:2013}
{Angulo} R.~E.,  {Hahn} O.,   {Abel} T.,  2013, \mn@doi [\mnras]
  {10.1093/mnras/stt1135}, \href
  {https://ui.adsabs.harvard.edu/abs/2013MNRAS.434.1756A} {434, 1756}

\bibitem[\protect\citeauthoryear{{Aric{\`o}}, {Angulo},
  {Hern{\'a}ndez-Monteagudo}, {Contreras}, {Zennaro}, {Pellejero-Iba{\~n}ez}
  \& {Rosas-Guevara}}{{Aric{\`o}} et~al.}{2020}]{Arico:2020}
{Aric{\`o}} G.,  {Angulo} R.~E.,  {Hern{\'a}ndez-Monteagudo} C.,  {Contreras}
  S.,  {Zennaro} M.,  {Pellejero-Iba{\~n}ez} M.,   {Rosas-Guevara} Y.,  2020,
  \mn@doi [\mnras] {10.1093/mnras/staa1478}, \href
  {https://ui.adsabs.harvard.edu/abs/2020MNRAS.495.4800A} {495, 4800}

\bibitem[\protect\citeauthoryear{{Bernardeau}, {Colombi}, {Gazta{\~n}aga}  \&
  {Scoccimarro}}{{Bernardeau} et~al.}{2002}]{Bernardeau:2002}
{Bernardeau} F.,  {Colombi} S.,  {Gazta{\~n}aga} E.,   {Scoccimarro} R.,  2002,
  \mn@doi [\physrep] {10.1016/S0370-1573(02)00135-7}, \href
  {https://ui.adsabs.harvard.edu/abs/2002PhR...367....1B} {367, 1}

\bibitem[\protect\citeauthoryear{{Bernardeau}, {van de Rijt}  \&
  {Vernizzi}}{{Bernardeau} et~al.}{2012}]{Bernardeau:2012}
{Bernardeau} F.,  {van de Rijt} N.,   {Vernizzi} F.,  2012, \mn@doi [\prd]
  {10.1103/PhysRevD.85.063509}, \href
  {https://ui.adsabs.harvard.edu/abs/2012PhRvD..85f3509B} {85, 063509}

\bibitem[\protect\citeauthoryear{Binney \& Tremaine}{Binney \&
  Tremaine}{2008}]{Binney:2008}
Binney J.,  Tremaine S.,  2008, Galactic Dynamics: Second Edition.
Princeton Series in Astrophysics, Princeton University Press

\bibitem[\protect\citeauthoryear{{Bird}, {Feng}, {Pedersen}  \&
  {Font-Ribera}}{{Bird} et~al.}{2020}]{Bird:2020}
{Bird} S.,  {Feng} Y.,  {Pedersen} C.,   {Font-Ribera} A.,  2020, \mn@doi
  [\jcap] {10.1088/1475-7516/2020/06/002}, \href
  {https://ui.adsabs.harvard.edu/abs/2020JCAP...06..002B} {2020, 002}

\bibitem[\protect\citeauthoryear{{Blas}, {Lesgourgues}  \& {Tram}}{{Blas}
  et~al.}{2011}]{Blas:2011}
{Blas} D.,  {Lesgourgues} J.,   {Tram} T.,  2011, \mn@doi [\jcap]
  {10.1088/1475-7516/2011/07/034}, \href
  {https://ui.adsabs.harvard.edu/abs/2011JCAP...07..034B} {2011, 034}

\bibitem[\protect\citeauthoryear{{Blazek}, {McEwen}  \& {Hirata}}{{Blazek}
  et~al.}{2016}]{Blazek:2016}
{Blazek} J.~A.,  {McEwen} J.~E.,   {Hirata} C.~M.,  2016, \mn@doi [\prl]
  {10.1103/PhysRevLett.116.121303}, \href
  {https://ui.adsabs.harvard.edu/abs/2016PhRvL.116l1303B} {116, 121303}

\bibitem[\protect\citeauthoryear{{Brandbyge}, {Rampf}, {Tram}, {Leclercq},
  {Fidler}  \& {Hannestad}}{{Brandbyge} et~al.}{2017}]{2017MNRAS.466L..68B}
{Brandbyge} J.,  {Rampf} C.,  {Tram} T.,  {Leclercq} F.,  {Fidler} C.,
  {Hannestad} S.,  2017, \mn@doi [\mnras] {10.1093/mnrasl/slw235}, \href
  {https://ui.adsabs.harvard.edu/abs/2017MNRAS.466L..68B} {466, L68}

\bibitem[\protect\citeauthoryear{{Bryan} et~al.,}{{Bryan}
  et~al.}{2014}]{Bryan:2014}
{Bryan} G.~L.,  et~al., 2014, \mn@doi [\apjs] {10.1088/0067-0049/211/2/19},
  \href {https://ui.adsabs.harvard.edu/abs/2014ApJS..211...19B} {211, 19}

\bibitem[\protect\citeauthoryear{{Buehlmann} \& {Hahn}}{{Buehlmann} \&
  {Hahn}}{2019}]{Buehlmann:2019}
{Buehlmann} M.,  {Hahn} O.,  2019, \mn@doi [\mnras] {10.1093/mnras/stz1243},
  \href {https://ui.adsabs.harvard.edu/abs/2019MNRAS.487..228B} {487, 228}

\bibitem[\protect\citeauthoryear{Chen, Castorina  \& White}{Chen
  et~al.}{2019}]{Chen:2019}
Chen S.-F.,  Castorina E.,   White M.,  2019, \mn@doi [Journal of Cosmology and
  Astroparticle Physics] {10.1088/1475-7516/2019/06/006}, 2019, 006

\bibitem[\protect\citeauthoryear{{Chisari} et~al.,}{{Chisari}
  et~al.}{2019}]{Chisari:2019}
{Chisari} N.~E.,  et~al., 2019, \mn@doi [The Open Journal of Astrophysics]
  {10.21105/astro.1905.06082}, \href
  {https://ui.adsabs.harvard.edu/abs/2019OJAp....2E...4C} {2, 4}

\bibitem[\protect\citeauthoryear{{Colombi}, {Jaffe}, {Novikov}  \&
  {Pichon}}{{Colombi} et~al.}{2009}]{Colombi:2009}
{Colombi} S.,  {Jaffe} A.,  {Novikov} D.,   {Pichon} C.,  2009, \mn@doi
  [\mnras] {10.1111/j.1365-2966.2008.14176.x}, \href
  {https://ui.adsabs.harvard.edu/abs/2009MNRAS.393..511C} {393, 511}

\bibitem[\protect\citeauthoryear{{Crocce}, {Pueblas}  \&
  {Scoccimarro}}{{Crocce} et~al.}{2006}]{Crocce:2006}
{Crocce} M.,  {Pueblas} S.,   {Scoccimarro} R.,  2006, \mn@doi [\mnras]
  {10.1111/j.1365-2966.2006.11040.x}, \href
  {https://ui.adsabs.harvard.edu/abs/2006MNRAS.373..369C} {373, 369}

\bibitem[\protect\citeauthoryear{{Dalal}, {Pen}  \& {Seljak}}{{Dalal}
  et~al.}{2010}]{Dalal:2010}
{Dalal} N.,  {Pen} U.-L.,   {Seljak} U.,  2010, \mn@doi [\jcap]
  {10.1088/1475-7516/2010/11/007}, \href
  {https://ui.adsabs.harvard.edu/abs/2010JCAP...11..007D} {2010, 007}

\bibitem[\protect\citeauthoryear{{Dubois} et~al.,}{{Dubois}
  et~al.}{2014}]{Dubois:2014}
{Dubois} Y.,  et~al., 2014, \mn@doi [\mnras] {10.1093/mnras/stu1227}, \href
  {https://ui.adsabs.harvard.edu/abs/2014MNRAS.444.1453D} {444, 1453}

\bibitem[\protect\citeauthoryear{{Durrer}}{{Durrer}}{2008}]{Durrer:2008}
{Durrer} R.,  2008, {The Cosmic Microwave Background}.
Cambridge University Press, Cambridge

\bibitem[\protect\citeauthoryear{{Efstathiou}, {Davis}, {White}  \&
  {Frenk}}{{Efstathiou} et~al.}{1985}]{Efstathiou:1985}
{Efstathiou} G.,  {Davis} M.,  {White} S.~D.~M.,   {Frenk} C.~S.,  1985,
  \mn@doi [\apjs] {10.1086/191003}, \href
  {https://ui.adsabs.harvard.edu/abs/1985ApJS...57..241E} {57, 241}

\bibitem[\protect\citeauthoryear{{Emberson}, {Frontiere}, {Habib}, {Heitmann},
  {Larsen}, {Finkel}  \& {Pope}}{{Emberson} et~al.}{2019}]{Emberson:2019}
{Emberson} J.~D.,  {Frontiere} N.,  {Habib} S.,  {Heitmann} K.,  {Larsen} P.,
  {Finkel} H.,   {Pope} A.,  2019, \mn@doi [\apj] {10.3847/1538-4357/ab1b31},
  \href {https://ui.adsabs.harvard.edu/abs/2019ApJ...877...85E} {877, 85}

\bibitem[\protect\citeauthoryear{{Fialkov}, {Barkana}, {Tseliakhovich}  \&
  {Hirata}}{{Fialkov} et~al.}{2012}]{Fialkov:2012}
{Fialkov} A.,  {Barkana} R.,  {Tseliakhovich} D.,   {Hirata} C.~M.,  2012,
  \mn@doi [\mnras] {10.1111/j.1365-2966.2012.21318.x}, \href
  {https://ui.adsabs.harvard.edu/abs/2012MNRAS.424.1335F} {424, 1335}

\bibitem[\protect\citeauthoryear{Fidler, Tram, Rampf, Crittenden, Koyama  \&
  Wands}{Fidler et~al.}{2017a}]{Fidler:2017ebh}
Fidler C.,  Tram T.,  Rampf C.,  Crittenden R.,  Koyama K.,   Wands D.,  2017a,
  \mn@doi [\jcap] {10.1088/1475-7516/2017/06/043}, 1706, 043

\bibitem[\protect\citeauthoryear{{Fidler}, {Tram}, {Rampf}, {Crittenden},
  {Koyama}  \& {Wands}}{{Fidler} et~al.}{2017b}]{2017JCAP...06..043F}
{Fidler} C.,  {Tram} T.,  {Rampf} C.,  {Crittenden} R.,  {Koyama} K.,   {Wands}
  D.,  2017b, \mn@doi [\jcap] {10.1088/1475-7516/2017/06/043}, \href
  {https://ui.adsabs.harvard.edu/abs/2017JCAP...06..043F} {2017, 043}

\bibitem[\protect\citeauthoryear{{Foreman}, {Coulton}, {Villaescusa-Navarro}
  \& {Barreira}}{{Foreman} et~al.}{2020}]{Foreman:2019}
{Foreman} S.,  {Coulton} W.,  {Villaescusa-Navarro} F.,   {Barreira} A.,  2020,
  \mn@doi [\mnras] {10.1093/mnras/staa2523}, \href
  {https://ui.adsabs.harvard.edu/abs/2020MNRAS.498.2887F} {498, 2887}

\bibitem[\protect\citeauthoryear{{Greif}, {White}, {Klessen}  \&
  {Springel}}{{Greif} et~al.}{2011}]{Greif:2011}
{Greif} T.~H.,  {White} S. D.~M.,  {Klessen} R.~S.,   {Springel} V.,  2011,
  \mn@doi [\apj] {10.1088/0004-637X/736/2/147}, \href
  {https://ui.adsabs.harvard.edu/abs/2011ApJ...736..147G} {736, 147}

\bibitem[\protect\citeauthoryear{{Guth}, {Hertzberg}  \&
  {Prescod-Weinstein}}{{Guth} et~al.}{2015}]{Guth:2015}
{Guth} A.~H.,  {Hertzberg} M.~P.,   {Prescod-Weinstein} C.,  2015, \mn@doi
  [\prd] {10.1103/PhysRevD.92.103513}, \href
  {https://ui.adsabs.harvard.edu/abs/2015PhRvD..92j3513G} {92, 103513}

\bibitem[\protect\citeauthoryear{{Hahn} \& {Abel}}{{Hahn} \&
  {Abel}}{2011}]{Hahn:2011}
{Hahn} O.,  {Abel} T.,  2011, \mn@doi [\mnras]
  {10.1111/j.1365-2966.2011.18820.x}, \href
  {https://ui.adsabs.harvard.edu/abs/2011MNRAS.415.2101H} {415, 2101}

\bibitem[\protect\citeauthoryear{{Hockney} \& {Eastwood}}{{Hockney} \&
  {Eastwood}}{1981}]{Hockney:1981}
{Hockney} R.~W.,  {Eastwood} J.~W.,  1981, {Computer Simulation Using
  Particles}.
Computer Simulation Using Particles, New York: McGraw-Hill, 1981

\bibitem[\protect\citeauthoryear{{Hopkins}}{{Hopkins}}{2015}]{Hopkins:2015}
{Hopkins} P.~F.,  2015, \mn@doi [\mnras] {10.1093/mnras/stv195}, \href
  {https://ui.adsabs.harvard.edu/abs/2015MNRAS.450...53H} {450, 53}

\bibitem[\protect\citeauthoryear{{Hu} \& {Dodelson}}{{Hu} \&
  {Dodelson}}{2002}]{Hu:2002}
{Hu} W.,  {Dodelson} S.,  2002, \mn@doi [\araa]
  {10.1146/annurev.astro.40.060401.093926}, \href
  {https://ui.adsabs.harvard.edu/abs/2002ARA&A..40..171H} {40, 171}

\bibitem[\protect\citeauthoryear{{Huang}, {Eifler}, {Mandelbaum}  \&
  {Dodelson}}{{Huang} et~al.}{2019}]{Huang:2019}
{Huang} H.-J.,  {Eifler} T.,  {Mandelbaum} R.,   {Dodelson} S.,  2019, \mn@doi
  [\mnras] {10.1093/mnras/stz1714}, \href
  {https://ui.adsabs.harvard.edu/abs/2019MNRAS.488.1652H} {488, 1652}

\bibitem[\protect\citeauthoryear{{Joyce} \& {Marcos}}{{Joyce} \&
  {Marcos}}{2007}]{Joyce:2007}
{Joyce} M.,  {Marcos} B.,  2007, \mn@doi [\prd] {10.1103/PhysRevD.75.063516},
  \href {https://ui.adsabs.harvard.edu/abs/2007PhRvD..75f3516J} {75, 063516}

\bibitem[\protect\citeauthoryear{{Joyce}, {Marcos}, {Gabrielli}, {Baertschiger}
   \& {Sylos Labini}}{{Joyce} et~al.}{2005}]{Joyce:2005}
{Joyce} M.,  {Marcos} B.,  {Gabrielli} A.,  {Baertschiger} T.,   {Sylos Labini}
  F.,  2005, \mn@doi [\prl] {10.1103/PhysRevLett.95.011304}, \href
  {https://ui.adsabs.harvard.edu/abs/2005PhRvL..95a1304J} {95, 011304}

\bibitem[\protect\citeauthoryear{{Klypin} \& {Shandarin}}{{Klypin} \&
  {Shandarin}}{1983}]{Klypin:1983}
{Klypin} A.~A.,  {Shandarin} S.~F.,  1983, \mn@doi [\mnras]
  {10.1093/mnras/204.3.891}, \href
  {http://adsabs.harvard.edu/abs/1983MNRAS.204..891K} {204, 891}

\bibitem[\protect\citeauthoryear{{Kravtsov}, {Klypin}  \&
  {Khokhlov}}{{Kravtsov} et~al.}{1997}]{Kravtsov:1997}
{Kravtsov} A.~V.,  {Klypin} A.~A.,   {Khokhlov} A.~M.,  1997, \mn@doi [\apjs]
  {10.1086/313015}, \href
  {https://ui.adsabs.harvard.edu/abs/1997ApJS..111...73K} {111, 73}

\bibitem[\protect\citeauthoryear{{Lewis}, {Challinor}  \& {Lasenby}}{{Lewis}
  et~al.}{2000}]{Lewis2000}
{Lewis} A.,  {Challinor} A.,   {Lasenby} A.,  2000, \mn@doi [\apj]
  {10.1086/309179}, \href
  {http://adsabs.harvard.edu/cgi-bin/nph-bib_query?bibcode=2000ApJ...538..473L&db_key=AST}
  {538, 473}

\bibitem[\protect\citeauthoryear{{Ludlow}, {Schaye}, {Schaller}  \&
  {Bower}}{{Ludlow} et~al.}{2020}]{Ludlow:2020}
{Ludlow} A.~D.,  {Schaye} J.,  {Schaller} M.,   {Bower} R.,  2020, \mn@doi
  [\mnras] {10.1093/mnras/staa316}, \href
  {https://ui.adsabs.harvard.edu/abs/2020MNRAS.493.2926L} {493, 2926}

\bibitem[\protect\citeauthoryear{Marcos, Baertschiger, Joyce, Gabrielli  \&
  Sylos~Labini}{Marcos et~al.}{2006}]{Marcos:2006cn}
Marcos B.,  Baertschiger T.,  Joyce M.,  Gabrielli A.,   Sylos~Labini F.,
  2006, \mn@doi [Phys. Rev.] {10.1103/PhysRevD.73.103507}, D73, 103507

\bibitem[\protect\citeauthoryear{Matsubara}{Matsubara}{2015}]{Matsubara:2015ipa}
Matsubara T.,  2015, \mn@doi [\prd] {10.1103/PhysRevD.92.023534}, 92, 023534

\bibitem[\protect\citeauthoryear{McCullagh, Jeong  \& Szalay}{McCullagh
  et~al.}{2015}]{McCullagh:2015}
McCullagh N.,  Jeong D.,   Szalay A.~S.,  2015, \mn@doi [\mnras]
  {10.1093/mnras/stv2525}, 455, 2945

\bibitem[\protect\citeauthoryear{{Michaux}, {Hahn}, {Rampf}  \&
  {Angulo}}{{Michaux} et~al.}{2020}]{Michaux:2020}
{Michaux} M.,  {Hahn} O.,  {Rampf} C.,   {Angulo} R.~E.,  2020, \mn@doi
  [\mnras] {10.1093/mnras/staa3149}, \href
  {https://ui.adsabs.harvard.edu/abs/2020MNRAS.500..663M} {500, 663}

\bibitem[\protect\citeauthoryear{{Munshi}, {Sahni}  \& {Starobinsky}}{{Munshi}
  et~al.}{1994}]{Munshi:1994}
{Munshi} D.,  {Sahni} V.,   {Starobinsky} A.~A.,  1994, \mn@doi [\apj]
  {10.1086/174925}, \href
  {https://ui.adsabs.harvard.edu/abs/1994ApJ...436..517M} {436, 517}

\bibitem[\protect\citeauthoryear{{Naoz} \& {Barkana}}{{Naoz} \&
  {Barkana}}{2005}]{Naoz:2005}
{Naoz} S.,  {Barkana} R.,  2005, \mn@doi [\mnras]
  {10.1111/j.1365-2966.2005.09385.x}, \href
  {https://ui.adsabs.harvard.edu/abs/2005MNRAS.362.1047N} {362, 1047}

\bibitem[\protect\citeauthoryear{{Naoz}, {Yoshida}  \& {Barkana}}{{Naoz}
  et~al.}{2011}]{Naoz:2011}
{Naoz} S.,  {Yoshida} N.,   {Barkana} R.,  2011, \mn@doi [\mnras]
  {10.1111/j.1365-2966.2011.19025.x}, \href
  {https://ui.adsabs.harvard.edu/abs/2011MNRAS.416..232N} {416, 232}

\bibitem[\protect\citeauthoryear{{O'Leary} \& {McQuinn}}{{O'Leary} \&
  {McQuinn}}{2012}]{OLeary:2012}
{O'Leary} R.~M.,  {McQuinn} M.,  2012, \mn@doi [\apj]
  {10.1088/0004-637X/760/1/4}, \href
  {https://ui.adsabs.harvard.edu/abs/2012ApJ...760....4O} {760, 4}

\bibitem[\protect\citeauthoryear{{Planck Collaboration} et~al.,}{{Planck
  Collaboration} et~al.}{2020}]{Planck:2018a}
{Planck Collaboration} et~al., 2020, \mn@doi [\aap]
  {10.1051/0004-6361/201833910}, \href
  {https://ui.adsabs.harvard.edu/abs/2020A&A...641A...6P} {641, A6}

\bibitem[\protect\citeauthoryear{{Porqueres}, {Hahn}, {Jasche}  \&
  {Lavaux}}{{Porqueres} et~al.}{2020}]{Porqueres:2020}
{Porqueres} N.,  {Hahn} O.,  {Jasche} J.,   {Lavaux} G.,  2020, \mn@doi [\aap]
  {10.1051/0004-6361/202038482}, \href
  {https://ui.adsabs.harvard.edu/abs/2020A&A...642A.139P} {642, A139}

\bibitem[\protect\citeauthoryear{Rampf}{Rampf}{2012}]{Rampf:2012b}
Rampf C.,  2012, \mn@doi [\jcap] {10.1088/1475-7516/2012/12/004}, 1212, 004

\bibitem[\protect\citeauthoryear{{Rampf} \& {Hahn}}{{Rampf} \&
  {Hahn}}{2020}]{Rampf:2020b}
{Rampf} C.,  {Hahn} O.,  2020, preprint (arXiv:2010.12584)

\bibitem[\protect\citeauthoryear{{Rampf}, {Villone}  \& {Frisch}}{{Rampf}
  et~al.}{2015}]{Rampf:2015}
{Rampf} C.,  {Villone} B.,   {Frisch} U.,  2015, \mn@doi [\mnras]
  {10.1093/mnras/stv1365}, \href
  {https://ui.adsabs.harvard.edu/abs/2015MNRAS.452.1421R} {452, 1421}

\bibitem[\protect\citeauthoryear{Rampf, Uhlemann  \& Hahn}{Rampf
  et~al.}{2020}]{Rampf:2020}
Rampf C.,  Uhlemann C.,   Hahn O.,  2020, preprint (arXiv:2008.09123)

\bibitem[\protect\citeauthoryear{{Schaller}, {Gonnet}, {Chalk}  \&
  {Draper}}{{Schaller} et~al.}{2016}]{Schaller:2016}
{Schaller} M.,  {Gonnet} P.,  {Chalk} A. B.~G.,   {Draper} P.~W.,  2016,
  preprint (arXiv:1606.02738)

\bibitem[\protect\citeauthoryear{{Schaye} et~al.,}{{Schaye}
  et~al.}{2015}]{Schaye:2015}
{Schaye} J.,  et~al., 2015, \mn@doi [\mnras] {10.1093/mnras/stu2058}, \href
  {https://ui.adsabs.harvard.edu/abs/2015MNRAS.446..521S} {446, 521}

\bibitem[\protect\citeauthoryear{{Schmidt}}{{Schmidt}}{2016}]{Schmidt:2016}
{Schmidt} F.,  2016, \mn@doi [\prd] {10.1103/PhysRevD.94.063508}, \href
  {https://ui.adsabs.harvard.edu/abs/2016PhRvD..94f3508S} {94, 063508}

\bibitem[\protect\citeauthoryear{{Schneider} \& {Teyssier}}{{Schneider} \&
  {Teyssier}}{2015}]{Schneider:2015}
{Schneider} A.,  {Teyssier} R.,  2015, \mn@doi [\jcap]
  {10.1088/1475-7516/2015/12/049}, \href
  {https://ui.adsabs.harvard.edu/abs/2015JCAP...12..049S} {2015, 049}

\bibitem[\protect\citeauthoryear{{Schneider}, {Teyssier}, {Stadel}, {Chisari},
  {Le Brun}, {Amara}  \& {Refregier}}{{Schneider}
  et~al.}{2019}]{Schneider:2019}
{Schneider} A.,  {Teyssier} R.,  {Stadel} J.,  {Chisari} N.~E.,  {Le Brun} A.
  M.~C.,  {Amara} A.,   {Refregier} A.,  2019, \mn@doi [\jcap]
  {10.1088/1475-7516/2019/03/020}, \href
  {https://ui.adsabs.harvard.edu/abs/2019JCAP...03..020S} {2019, 020}

\bibitem[\protect\citeauthoryear{{Scoccimarro}}{{Scoccimarro}}{1998}]{Scoccimarro:1998}
{Scoccimarro} R.,  1998, \mn@doi [\mnras] {10.1046/j.1365-8711.1998.01845.x},
  \href {https://ui.adsabs.harvard.edu/abs/1998MNRAS.299.1097S} {299, 1097}

\bibitem[\protect\citeauthoryear{{Scoccimarro}}{{Scoccimarro}}{2000}]{Scoccimarro:2000}
{Scoccimarro} R.,  2000, \mn@doi [\apj] {10.1086/317248}, \href
  {https://ui.adsabs.harvard.edu/abs/2000ApJ...544..597S} {544, 597}

\bibitem[\protect\citeauthoryear{{Sefusatti}, {Crocce}, {Scoccimarro}  \&
  {Couchman}}{{Sefusatti} et~al.}{2016}]{Sefusatti:2016}
{Sefusatti} E.,  {Crocce} M.,  {Scoccimarro} R.,   {Couchman} H.~M.~P.,  2016,
  \mn@doi [\mnras] {10.1093/mnras/stw1229}, \href
  {https://ui.adsabs.harvard.edu/abs/2016MNRAS.460.3624S} {460, 3624}

\bibitem[\protect\citeauthoryear{{Shandarin} \& {Zeldovich}}{{Shandarin} \&
  {Zeldovich}}{1989}]{Shandarin:1989}
{Shandarin} S.~F.,  {Zeldovich} {\relax Ya. B}.,  1989, \mn@doi [Reviews of
  Modern Physics] {10.1103/RevModPhys.61.185}, \href
  {https://ui.adsabs.harvard.edu/abs/1989RvMP...61..185S} {61, 185}

\bibitem[\protect\citeauthoryear{{Slepian} \& {Eisenstein}}{{Slepian} \&
  {Eisenstein}}{2015}]{Slepian:2015}
{Slepian} Z.,  {Eisenstein} D.~J.,  2015, \mn@doi [\mnras]
  {10.1093/mnras/stu2627}, \href
  {https://ui.adsabs.harvard.edu/abs/2015MNRAS.448....9S} {448, 9}

\bibitem[\protect\citeauthoryear{{Slepian} et~al.,}{{Slepian}
  et~al.}{2018}]{2018MNRAS.474.2109S}
{Slepian} Z.,  et~al., 2018, \mn@doi [\mnras] {10.1093/mnras/stx2723}, \href
  {https://ui.adsabs.harvard.edu/abs/2018MNRAS.474.2109S} {474, 2109}

\bibitem[\protect\citeauthoryear{{Somogyi} \& {Smith}}{{Somogyi} \&
  {Smith}}{2010}]{Somogyi:2010}
{Somogyi} G.,  {Smith} R.~E.,  2010, \mn@doi [\prd]
  {10.1103/PhysRevD.81.023524}, \href
  {https://ui.adsabs.harvard.edu/abs/2010PhRvD..81b3524S} {81, 023524}

\bibitem[\protect\citeauthoryear{{Springel}}{{Springel}}{2005}]{Springel:2005}
{Springel} V.,  2005, \mn@doi [\mnras] {10.1111/j.1365-2966.2005.09655.x},
  \href {http://adsabs.harvard.edu/abs/2005MNRAS.364.1105S} {364, 1105}

\bibitem[\protect\citeauthoryear{{Springel}}{{Springel}}{2010}]{Springel:2010}
{Springel} V.,  2010, \mn@doi [\mnras] {10.1111/j.1365-2966.2009.15715.x},
  \href {https://ui.adsabs.harvard.edu/abs/2010MNRAS.401..791S} {401, 791}

\bibitem[\protect\citeauthoryear{{Springel} et~al.,}{{Springel}
  et~al.}{2018}]{Springel:2018}
{Springel} V.,  et~al., 2018, \mn@doi [\mnras] {10.1093/mnras/stx3304}, \href
  {https://ui.adsabs.harvard.edu/abs/2018MNRAS.475..676S} {475, 676}

\bibitem[\protect\citeauthoryear{{Taruya}, {Nishimichi}  \& {Jeong}}{{Taruya}
  et~al.}{2018}]{Taruya:2018}
{Taruya} A.,  {Nishimichi} T.,   {Jeong} D.,  2018, \mn@doi [\prd]
  {10.1103/PhysRevD.98.103532}, \href
  {https://ui.adsabs.harvard.edu/abs/2018PhRvD..98j3532T} {98, 103532}

\bibitem[\protect\citeauthoryear{{Teyssier}}{{Teyssier}}{2002}]{Teyssier:2002}
{Teyssier} R.,  2002, \mn@doi [\aap] {10.1051/0004-6361:20011817}, \href
  {http://adsabs.harvard.edu/abs/2002A%26A...385..337T} {385, 337}

\bibitem[\protect\citeauthoryear{{Tomlinson}, {Jeong}  \& {Kim}}{{Tomlinson}
  et~al.}{2019}]{Tomlinson:2019}
{Tomlinson} J.,  {Jeong} D.,   {Kim} J.,  2019, \mn@doi [\aj]
  {10.3847/1538-3881/ab3223}, \href
  {https://ui.adsabs.harvard.edu/abs/2019AJ....158..116T} {158, 116}

\bibitem[\protect\citeauthoryear{{Tseliakhovich} \& {Hirata}}{{Tseliakhovich}
  \& {Hirata}}{2010}]{Tseliakhovich:2010}
{Tseliakhovich} D.,  {Hirata} C.,  2010, \mn@doi [\prd]
  {10.1103/PhysRevD.82.083520}, \href
  {https://ui.adsabs.harvard.edu/abs/2010PhRvD..82h3520T} {82, 083520}

\bibitem[\protect\citeauthoryear{{Uhlemann}, {Rampf}, {Gosenca}  \&
  {Hahn}}{{Uhlemann} et~al.}{2019}]{Uhlemann:2019}
{Uhlemann} C.,  {Rampf} C.,  {Gosenca} M.,   {Hahn} O.,  2019, \mn@doi [\prd]
  {10.1103/PhysRevD.99.083524}, \href
  {http://adsabs.harvard.edu/abs/2019PhRvD..99h3524U} {99, 083524}

\bibitem[\protect\citeauthoryear{{Valkenburg} \&
  {Villaescusa-Navarro}}{{Valkenburg} \&
  {Villaescusa-Navarro}}{2017}]{Valkenburg:2017}
{Valkenburg} W.,  {Villaescusa-Navarro} F.,  2017, \mn@doi [\mnras]
  {10.1093/mnras/stx376}, \href
  {https://ui.adsabs.harvard.edu/abs/2017MNRAS.467.4401V} {467, 4401}

\bibitem[\protect\citeauthoryear{{Wadsley}, {Keller}  \& {Quinn}}{{Wadsley}
  et~al.}{2017}]{Wadsley:2017}
{Wadsley} J.~W.,  {Keller} B.~W.,   {Quinn} T.~R.,  2017, \mn@doi [\mnras]
  {10.1093/mnras/stx1643}, \href
  {https://ui.adsabs.harvard.edu/abs/2017MNRAS.471.2357W} {471, 2357}

\bibitem[\protect\citeauthoryear{{Weinberger}, {Springel}  \&
  {Pakmor}}{{Weinberger} et~al.}{2020}]{Weinberger:2020}
{Weinberger} R.,  {Springel} V.,   {Pakmor} R.,  2020, \mn@doi [\apjs]
  {10.3847/1538-4365/ab908c}, \href
  {https://ui.adsabs.harvard.edu/abs/2020ApJS..248...32W} {248, 32}

\bibitem[\protect\citeauthoryear{Weltman et~al.}{Weltman
  et~al.}{2020}]{Bull:2018lat}
Weltman A.,  et~al., 2020, \mn@doi [\pasa] {10.1017/pasa.2019.42}, 37, e002

\bibitem[\protect\citeauthoryear{{White}}{{White}}{1996}]{White:1996}
{White} S.~D.~M.,  1996, in {Schaeffer} R.,  {Silk} J.,  {Spiro} M.,
  {Zinn-Justin} J.,  eds, Cosmology and Large Scale Structure. p.~349

\bibitem[\protect\citeauthoryear{{Yoo}, {Dalal}  \& {Seljak}}{{Yoo}
  et~al.}{2011}]{Yoo:2011}
{Yoo} J.,  {Dalal} N.,   {Seljak} U.,  2011, \mn@doi [\jcap]
  {10.1088/1475-7516/2011/07/018}, \href
  {https://ui.adsabs.harvard.edu/abs/2011JCAP...07..018Y} {2011, 018}

\bibitem[\protect\citeauthoryear{{Yoshida}, {Sugiyama}  \&
  {Hernquist}}{{Yoshida} et~al.}{2003}]{Yoshida:2003}
{Yoshida} N.,  {Sugiyama} N.,   {Hernquist} L.,  2003, \mn@doi [\mnras]
  {10.1046/j.1365-8711.2003.06829.x}, \href
  {https://ui.adsabs.harvard.edu/abs/2003MNRAS.344..481Y} {344, 481}

\bibitem[\protect\citeauthoryear{{Zel'dovich}}{{Zel'dovich}}{1970}]{Zeldovich:1970}
{Zel'dovich} {\relax Ya}.~B.,  1970, \aap, \href
  {https://ui.adsabs.harvard.edu/abs/1970A&A.....5...84Z} {500, 13}

\bibitem[\protect\citeauthoryear{{Zennaro}, {Bel}, {Villaescusa-Navarro},
  {Carbone}, {Sefusatti}  \& {Guzzo}}{{Zennaro}
  et~al.}{2017}]{2017MNRAS.466.3244Z}
{Zennaro} M.,  {Bel} J.,  {Villaescusa-Navarro} F.,  {Carbone} C.,  {Sefusatti}
  E.,   {Guzzo} L.,  2017, \mn@doi [\mnras] {10.1093/mnras/stw3340}, \href
  {https://ui.adsabs.harvard.edu/abs/2017MNRAS.466.3244Z} {466, 3244}

\bibitem[\protect\citeauthoryear{{Zheligovsky} \& {Frisch}}{{Zheligovsky} \&
  {Frisch}}{2014}]{Zheligovsky:2014}
{Zheligovsky} V.,  {Frisch} U.,  2014, \mn@doi [J. Fluid Mech.]
  {10.1017/jfm.2014.221}, \href
  {https://ui.adsabs.harvard.edu/abs/2014JFM...749..404Z} {749, 404}

\bibitem[\protect\citeauthoryear{{van Leer}}{{van Leer}}{1979}]{VanLeer:1979}
{van Leer} B.,  1979, \mn@doi [Journal of Computational Physics]
  {10.1016/0021-9991(79)90145-1}, \href
  {https://ui.adsabs.harvard.edu/abs/1979JCoPh..32..101V} {32, 101}

\makeatother
\end{thebibliography}
\input{twofluidics.bbl}


\appendix

\section{Code accuracy parameters}
We carried out all simulations with the parameter settings listed in Table~\ref{tab:ramses_params} for {\sc Ramses}, and in Table~\ref{tab:arepo_params} for {\sc Arepo} (and {\sc Gadget-2} where applicable).

\begin{table}
\begin{center}
\begin{tabular}{ll}
\hline
\texttt{RUN\_PARAMS}&\\
\texttt{nsubcycle} & \texttt{1, 2} \\
\hline
\texttt{AMR\_PARAMS}&\\
\texttt{levelmin} & \texttt{9} \\
\texttt{levelmax} & \texttt{17} \\
\hline
\texttt{HYDRO\_PARAMS}&\\
\texttt{courant\_factor} & \texttt{0.8} \\
\texttt{slope\_type} & \texttt{2} \\
\texttt{pressure\_fix} & \texttt{.true.} \\
\texttt{scheme} & \texttt{'muscl'} \\
\texttt{riemann} & \texttt{hllc} \\
\hline
\texttt{REFINE\_PARAMS}&\\
\texttt{m\_refine} & \texttt{4.,10*8.} \\
\texttt{interpol\_var} & \texttt{1} \\
\texttt{interpol\_type} & \texttt{0}\\
\hline
\end{tabular}
\end{center}
\caption{\label{tab:ramses_params}{\sc Ramses} code parameter values used in this paper.}
\vspace{1cm}
\begin{center}
\begin{tabular}{ll}
\hline
\texttt{TypeOfTimestepCriterion}          & \texttt{0} \\
\texttt{ErrTolIntAccuracy}                & \texttt{0.025} \\
\texttt{CourantFac}                       & \texttt{0.8} \\
\texttt{MaxSizeTimestep}                  & \texttt{0.01} \\
\texttt{TypeOfOpeningCriterion}           & \texttt{1} \\
\texttt{ErrTolTheta}                      & \texttt{0.7} \\
\texttt{ErrTolForceAcc}                   & \texttt{0.0025} \\
\texttt{SofteningComovingType0}           & \texttt{0.025} \\
\texttt{SofteningComovingType1}           & \texttt{0.025} \\
\texttt{SofteningMaxPhysType0}            & \texttt{0.25} \\
\texttt{SofteningMaxPhysType1}            & \texttt{0.25} \\
\texttt{GasSoftFactor}                    & \texttt{2.5} \\
\texttt{SofteningTypeOfPartType0}         & \texttt{0} \\
\texttt{SofteningTypeOfPartType1}         & \texttt{1} \\
\texttt{MinimumComovingHydroSoftening}    & \texttt{0.025} \\
\texttt{CellShapingSpeed}                 & \texttt{0.5} \\
\texttt{CellMaxAngleFactor}               & \texttt{2.25} \\
\texttt{ReferenceGasPartMass}             & \texttt{0} \\
\texttt{TargetGasMassFactor}              & \texttt{1} \\
\texttt{RefinementCriterion}              & \texttt{1} \\
\texttt{DerefinementCriterion}            & \texttt{1} \\
\hline
\end{tabular}
\end{center}
\caption{\label{tab:arepo_params}{\sc Arepo} code parameter values used in this paper. The timestep and force accuracy/softening parameters also apply to the collisionless {\sc Gadget-2} runs, where baryons are however treated as \texttt{Type2} particles instead of \texttt{Type0}. Also for the $2\times256^3$ {\sc Gadget-2} run, the softening used was twice larger.}
\end{table}

\bsp	
\label{lastpage}
\end{document}